\documentclass[a4paper,12pt]{article}
\usepackage{amssymb}
\usepackage{amsmath}
\usepackage{graphicx,multicol}
\usepackage{color}
\definecolor{rosso}{cmyk}{0,1,1,0.4}
\definecolor{rossos}{cmyk}{0,1,1,0.55}
\definecolor{rossoc}{cmyk}{0,1,1,0.2}
\definecolor{blu}{cmyk}{1,1,0,0.3}
\definecolor{blus}{cmyk}{1,1,0,0.6}
\definecolor{bluc}{cmyk}{1,1,0,0.1}
\definecolor{verde}{cmyk}{0.92,0,0.59,0.25}
\definecolor{verdec}{cmyk}{0.92,0,0.59,0.15}
\definecolor{verdes}{cmyk}{0.92,0,0.59,0.7}

\newcommand{\fig}[1]{~{\rm\ref{fig:#1}}}

\def\circa#1{\,\raise.3ex\hbox{$#1$\kern-.75em\lower1ex\hbox{$\sim$}}\,}

\font\tenrsfs=rsfs10 at 12pt
\font\sevenrsfs=rsfs7
\font\fiversfs=rsfs5
\newfam\rsfsfam
\textfont\rsfsfam=\tenrsfs
\scriptfont\rsfsfam=\sevenrsfs
\scriptscriptfont\rsfsfam=\fiversfs
\def\mathscr#1{{\fam\rsfsfam\relax#1}}
\def\Lag{\mathscr{L}}

\newcommand{\Usl}{U\hspace{-1.6ex}/\,}
\newcommand{\Psl}{P\hspace{-1.5ex}/}
\newcommand{\Ksl}{K\hspace{-1.5ex}/}

\newcommand{\eV}{\,{\rm eV}}

\newcommand{\GeV}{\,{\rm GeV}}
\newcommand{\TeV}{\,{\rm TeV}}

\makeatletter
\def\art{\@ifnextchar[{\eart}{\oart}}
\def\eart[#1]#2#3#4#5#6{{\rm #2}, {#3 #4} {\rm (#6) #5 [#1]}}
\def\hepart[#1]#2{{\rm #2, #1}}
\newcommand{\oart}[5]{{\rm #1}, {#2 \rm #3} {\rm (#5) #4}}

\newcommand{\NP}{Nucl. Phys.}
\newcommand{\Jhep}{{\rm JHEP}}
\newcommand{\PRL}{Phys. Rev. Lett.}
\newcommand{\PL}{Phys. Lett.}
\newcommand{\PR}{Phys. Rev.}

\newcommand{\Nt}{{\tilde{N}}}
\newcommand{\Lt}{{\tilde{L}}}
\newcommand{\Ht}{{\tilde{H}}}

\newcommand{\beq}{\begin{equation}}
\newcommand{\eeq}{\end{equation}}
\newcommand{\bea}{\begin{eqnarray}}
\newcommand{\eea}{\end{eqnarray}}
\newcommand{\ba}{\begin{array}}
\newcommand{\ea}{\end{array}}
\newcommand{\bi}{\begin{itemize}}
\newcommand{\ei}{\end{itemize}}
\newcommand{\bn}{\begin{enumerate}}
\newcommand{\en}{\end{enumerate}}
\newcommand{\bc}{\begin{center}}
\newcommand{\ec}{\end{center}}

\newcommand{\nn}{\nonumber\\}

\newcommand{\eq}[1]{eq.~(\ref{#1})}

\newcommand{\gsim}{\lower.7ex\hbox{$\;\stackrel{\textstyle>}{\sim}\;$}}
\newcommand{\lsim}{\lower.7ex\hbox{$\;\stackrel{\textstyle<}{\sim}\;$}}

\newcounter{alphaequation}[equation]
\def\thealphaequation{\theequation\hbox to
0.6em{\hfil\alph{alphaequation}\hfil}}
\def\eqnsystem#1{
\def\@eqnnum{{\rm (\thealphaequation)}}
\def\@@eqncr{\let\@tempa\relax \ifcase\@eqcnt \def\@tempa{& & &} \or
  \def\@tempa{& &}\or \def\@tempa{&}\fi\@tempa
  \if@eqnsw\@eqnnum\refstepcounter{alphaequation}\fi
\global\@eqnswtrue\global\@eqcnt=0\cr}
\refstepcounter{equation} \let\@currentlabel\theequation \def\@tempb{#1}
\ifx\@tempb\empty\else\label{#1}\fi
\refstepcounter{alphaequation}
\let\@currentlabel\thealphaequation
\global\@eqnswtrue\global\@eqcnt=0 \tabskip\@centering\let\\=\@eqncr
$$\halign to \displaywidth\bgroup \@eqnsel\hskip\@centering
$\displaystyle\tabskip\z@{##}$&\global\@eqcnt\@ne
\hskip2\arraycolsep\hfil${##}$\hfil& \global\@eqcnt\tw@\hskip2\arraycolsep
$\displaystyle\tabskip\z@{##}$\hfil
\tabskip\@centering&\llap{##}\tabskip\z@\cr}

\def\endeqnsystem{\@@eqncr\egroup$$\global\@ignoretrue} \makeatother

\begin{document}
\tolerance=100000
\thispagestyle{empty}
\setcounter{page}{0}

{hep-ph/0310123\hfill IFUP--TH/2003--37\hfill
CERN-TH/2003-240}
\vspace{1cm}

\begin{center}
{\LARGE \bf \color{rossos}
Towards a complete theory of thermal 
 \\[0.3cm]
leptogenesis  in the  SM and MSSM
}\\[2.cm]

{
{\large\bf G.F. Giudice}$^1$,
{\large\bf A. Notari}$^2$,
 {\large\bf M. Raidal}$^{3}$,
 {\large\bf A. Riotto}$^{4}$,  {\large\bf A. Strumia}$^{5}$
}  
\\[7mm]
{\it $^1$ Theoretical Physics Division, CERN, CH-1211
Geneva 23, 
Switzerland} \\[3mm]
{\it $^2$ Scuola Normale Superiore, Piazza dei Cavalieri 7,
Pisa, I-56126, 
Italy } \\[3mm]
{\it $^3$ National Institute of Chemical Physics and
Biophysics, 
Tallinn 10143, Estonia}  \\[3mm]
{\it $^4$ INFN, Sezione di Padova, via Marzolo 8, Padova
I-35131, 
Italy} \\[3mm]
{\it $^5$ Dipartimento di Fisica dell'Universit\`a di Pisa
and INFN, 
Italia}\\[1cm]
\vspace{1cm}
{\large\bf\color{blus} Abstract}
\end{center}
\begin{quote}
{\noindent\color{blus}
We perform a thorough study of thermal leptogenesis
adding finite temperature effects, 
RGE corrections,
scatterings involving gauge bosons and by 
properly avoiding overcounting on-shell processes.
Assuming hierarchical right-handed neutrinos with arbitrary abundancy,
successful leptogenesis can be achieved if 
left-handed neutrinos are lighter than $ 0.15\eV$
and right-handed neutrinos heavier than 
$2\times 10^7\GeV$  (SM case, $3\sigma$ C.L.).
MSSM results are similar.
Furthermore, we study how reheating after inflation affects thermal leptogenesis.
Assuming that the inflaton reheats SM particles but not directly  right-handed neutrinos,
we derive the lower bound on the reheating
temperature to be 
$T_{\rm RH} \circa{>}  2\times 10^9\GeV$.
This bound conflicts with
the cosmological gravitino bound present in supersymmetric theories.
We study some scenarios that avoid this conflict: `soft leptogenesis',
 leptogenesis in presence of a large right-handed (s)neutrino abundancy or of
a  sneutrino condensate.}

\end{quote}

\newpage

\setcounter{page}{1}



\section{Introduction}
\label{intro}
If some new physics violates lepton number ${\cal L}$
at an energy scale $\Lambda_{\cal L}$, neutrinos
get small Majorana masses via
the dimension-5 effective operator
$(LH)^2/\Lambda_{\cal L}$.
Experiments suggest $\Lambda_{\cal L}\sim 10^{14}\GeV$.
Indeed the solar and atmospheric data 
can be explained by neutrino oscillations
induced by the following neutrino masses and mixings~\cite{oscdata}
\begin{equation}\begin{array}{cc}
|\Delta m^2_{\rm atm}| = (2.0^{+ 0.4}_{-0.3}) \times 10^{-3}\eV^2,
&
\sin^22\theta_{\rm atm} = 1.00 \pm 0.04,      \\[2mm]
\Delta m^2_{\rm sun} = (7.2 \pm 0.7)\times 10^{-5}\eV^2,&
\tan^2\theta_{\rm sun} = 0.44 \pm  0.05.
\end{array}\label{eq:oscdata}
\end{equation}
Experiments will make further progress towards measuring
effects accessible at low energy,
completely described by 9 Majorana parameters:
3 neutrino masses, 3 mixing angles, 3 CP-violating phases.

One possible mechanism to generate the dimension-5 operator
$(LH)^2/\Lambda_{\cal L}$
is known as `see-saw' mechanism~\cite{seesaw}.
Adding three right-handed neutrinos $N_{1,2,3}$
with heavy Majorana masses $m_{N_3}> m_{N_3}> m_{N_1}\gg M_Z$
and Yukawa couplings $Y^\nu_{ij}$
\begin{equation}
\label{eq:L}
\Lag = \Lag_{\rm SM} + \left(
\frac{m_{N_i}}{2} N_i^2 + Y^\nu_{ij}   L_i N_j H + \hbox{h.c.}\right) ,
\end{equation}
one obtains light neutrino states with
Majorana masses $m_\nu = - (v Y^\nu)^T  ~m_N^{-1} ~ (v Y^\nu)$.
 The see-saw is a simple and elegant mechanism, 
 but hard to test experimentally.
It predicts no relation between
the 9 low-energy parameters, just reproducing them
in terms of 18 high-energy ones.
The right-handed neutrinos which constitute the essence of the see-saw
are too heavy or too weakly coupled to be experimentally observed.
There are few possible indirect probes. For instance, 
in some supersymmetric models the Yukawa couplings $Y^\nu_{ij}$ might induce
sizable rates for lepton flavour violating processes such as $\mu\to e\gamma$.
On the cosmological side, 
{\it thermal leptogenesis} \cite{fuk} 
provides an attractive scenario
for the generation of the baryon asymmetry of the universe 
\cite{reviewbau}. The
three necessary conditions for the 
generation of the baryon asymmetry \cite{sak} are satisfied in the 
Standard Model (SM) with additional, heavy singlet right-handed neutrinos:
the  
baryon number is violated by sphaleron processes which convert the 
lepton asymmetry induced by the Majorana nature of the right-handed neutrino 
masses into baryon asymmetry; CP-violation is due to
the 
Yukawa interactions of the right-handed neutrinos
with the SM  lepton doublets and out-of-equilibrium is induced by the
right-handed neutrino decays.
For some recent analyses see~\cite{k-sm,BRS,vari, mBound}.

The goal of this paper is to perform a thorough analysis of 
thermal leptogenesis within the SM and the Minimal Supersymmetric
Standard Model (MSSM). 
We improve the computation of baryon asymmetry
generated through the mechanism of thermal leptogenesis by
\begin{itemize}
\item[{\em i)}] including finite temperature corrections to
propagators, masses, decay and scattering processes, and to the
CP-asymmetry; 
\item[{\em ii)}] renormalizing couplings at the relevant scale 
($\sim 2\pi T$, where $T$ is the
relevant temperature,  rather than $\sim M_Z$);
\item[{\em iii)}] adding $\Delta L=1$ scatterings
 involving gauge bosons 
which turn out to be  
comparable or larger than the ones involving the top quark 
included in previous computations;
\item[{\em iv)}] performing a proper subtraction 
of  washout  scatterings mediated by intermediate on-shell particles
(once this is correctly done, they are no longer resonantly enhanced);

\item[{\em v)}] extending the analysis
to situations where right-handed
(s)neutrinos give a sizable contribution to the total energy density;

\item[{\em vi)}] discussing how the predictions
of thermal leptogenesis depend upon the 
cosmological assumptions. In particular, 
we study the 
effects of reheating after inflation
and compute the lowest value  of the reheating temperature
$T_{\rm RH}$ for successful leptogenesis.
\end{itemize}
The paper is organized as follows. 
In section~\ref{qft} we briefly summarize some general results of
field theory at finite temperature.
In section~\ref{thermal} we discuss
how we include thermal corrections and how they affect
the different ingredients of leptogenesis.
Details can be found in a series of appendices:
Boltzmann equations in appendix~\ref{Boltz},
scattering rates in appendix~\ref{gamma},
CP-asymmetries in appendices~\ref{AppEpsilon} (SM) and~\ref{CPMSSM} (MSSM).
In section~\ref{SM} we combine all ingredients to get the final baryon
asymmetry predicted within 
SM leptogenesis, and study which thermal corrections turn out to
be numerically important.
On the basis of the lesson learned for the SM,
in section~\ref{MSSM} we address the more involved case of 
supersymmetric leptogenesis.
We apply our improved computation also to the `soft leptogenesis' scenario~\cite{softl2, softl}.
In section~\ref{reh} we discuss how the baryon asymmetry changes
when the maximal temperature
 reached by the universe after inflation is
 not much higher than $m_{N_1}$.
 The variation depends on one extra parameter,
 the reheating temperature $T_{\rm RH}$. Finally our results are
summarized in section~\ref{secconc}.

\section{Finite-temperature propagators}\label{qft}
In order to consider finite temperature effects in the
plasma, we work in the
so-called real time formalism (RTF) \cite{bellac}
of thermal field theory. The Green's functions computed in
this formalism
are directly time-ordered ones. The RTF requires the
introduction
of a ghost field dual to each physical field leading to the
doubling
of degrees of freedom. The thermal propagator has therefore
a $2\times 2$ 
structure: the $(11)$ component refers to the physical
field, the
$(22)$ component to the corresponding ghost field, with the
off-diagonal
components 
$(12)$ and $(21)$ mixing them.

\subsubsection{Scalars}
The complete propagator of a scalar particle 
({\it e.g.} the Higgs) in momentum space is
\begin{equation}
G(K)=\left(
\begin{array}{cc}
G^{11}(K)& G^{12}(K)\\
G^{21}(K) & G^{22}(K)
\end{array}\right)
=U\left(T,K\right)
\left(
\begin{array}{cc}
\Delta_B(K)& 0\\
0 & \Delta_B^*(K)
\end{array}\right)U\left(T,K\right)\, ,
\label{pb}
\end{equation}
\begin{equation}
U\left(T,K\right)=\left(
\begin{array}{cc}
{\rm cosh}\theta_K & {\rm sinh}\theta_K \\
{\rm sinh}\theta_K     & {\rm cosh}\theta_K
\end{array}\right) ,
\end{equation}
\begin{equation}
{\rm cosh}\theta_K=\sqrt{1+f_B(\omega )} \, ,
\,\,\,\, {\rm sinh}\theta_K=
\sqrt{f_B(\omega )},\, \qquad f_B\equiv
\frac{1}{e^{|\omega|/T}-1},
\end{equation}
where $K$ is the particle four-momentum. In a general 
frame where the thermal bath has four-velocity $U^\mu$
($U^\mu U_\mu =1$),
we define the Lorentz-invariant quantities
\begin{equation}
\omega \equiv K^\mu U_\mu ,~~~k \equiv \sqrt{ \left( K^\mu
U_\mu
\right)^2-K^\mu K_\mu }.
\end{equation}
These coincide with the particle energy and momentum in
the rest frame
of the thermal bath, $U^\mu =(1,0,0,0)$. 

In Eq. (\ref{pb}), $\Delta_B(K)$ is the resummed propagator
\begin{equation}
\Delta_B(K)=\frac{i}{K^2-m_B^2-\Sigma_B(K)+i\epsilon}\, ,
\end{equation}
where $m_B$ is the bare mass and 
$\Sigma_B(K)$ is the finite-temperature self-energy of the
scalar 
boson field. $\Sigma_B(K)$ describes the continuous
interactions
with the heat bath altering the propagation of the boson,
and it 
modifies the 
dispersion relation, substituting particles with
quasiparticles.
At one-loop the self-energy takes the form $\Sigma_B(K)= 
{\rm Re}\,\Sigma_B(K)+
i\, {\rm Im}\,\Sigma_B(K)$, where ${\rm
Re}\,\Sigma_B(K)=m_B^2(T)$ is the 
effective 
plasma mass squared
 and ${\rm Im}\,\Sigma_B(K)=-2\,\omega\,\Gamma_B$ is
proportional
to the damping rate $\Gamma_B$ of the boson in the plasma.
Since 
$\left|{\rm Im}\,\Sigma_B(K)\right|$ is suppressed compared
to
$\left|{\rm Re}\,\Sigma_B(K)\right|$ \cite{enqvist}, in the
following we
will work with resummed propagators for scalar 
bosons neglecting the absorptive
part. Notice also that since we are considering one-loop
thermal corrections
to processes where
all external fields are physical, we only need 
the $(11)$ component of the bosonic propagator
\begin{equation}
 G^{11}(K)={\rm cosh}^2\theta_K
\left(\frac{i}{K^2-m_B(T)^2+i 
\epsilon}\right)+
 {\rm sinh}^2\theta_K
\left(\frac{-i}{K^2-m_B(T)^2-i\epsilon}\right) \ ,
\end{equation}
where we have included the bare mass in $m_B(T)$.
Using the property 
\begin{equation}
\frac{1}{x+i\epsilon}=P\left( \frac{1}{x}\right) - 
i \pi \delta (x)\label{iepsilon}  \ ,
\end{equation} 
where $P$ denotes the principal value,
the propagator can be rewritten as:
\begin{equation}
G^{11}(K)=\frac{i}{K^2-m_B(T)^2+i \epsilon} + 2 
\pi f_B(\omega) \delta \left[ K^2-m_B(T)^2\right] \ .   
\label{bosonpropagator}
\end{equation}

\subsubsection{Fermions}
The (one-loop) resummed propagators for fermion fields can
be
written in RTF as
\begin{equation}
S(K)=\left(
\begin{array}{cc}
S^{11}(K)& S^{12}(K)\\
S^{21}(K) & S^{22}(K)
\end{array}\right)
=M\left(T,K\right)
\left(
\begin{array}{cc}
\Delta_F(K)& 0\\
0 & \Delta_F^*(K)
\end{array}\right)M\left(T,K\right)\, ,
\label{pf}
\end{equation}
where

\begin{equation}
M\left(T,K\right)=\left(
\begin{array}{cc}
{\rm cos}\phi_K & -{\rm sin}\phi_K \\
{\rm sin}\phi_K     & {\rm cos}\phi_K
\end{array}\right)
\end{equation}
and
\begin{equation}
{\rm cos}\phi_K=\left[\theta(\omega)-\theta(-\omega)\right]
\sqrt{1-f_F(\omega)}\, ,
\,\,\,\, {\rm sin}\phi_K=
\sqrt{f_F(\omega)}, \, \qquad
f_F(\omega)\equiv \frac{1}{e^{|\omega|/T}+1}.
\end{equation}
In Eq. (\ref{pf}) $\Delta_F(K)$ is the resummed propagator

\begin{equation}
\Delta_F(K)=\frac{i}{\gamma^\mu K_\mu -m_0
-\Sigma_F(K)+i\epsilon}\, 
,\label{deltaFermion}
\end{equation}
where $m_0$ is the fermion bare mass and
$\Sigma_F(K)$ is the self-energy of the boson field at
finite temperature.
At one-loop the fermionic self-energy is given by
\cite{weldon}
\begin{equation}
\Sigma_F(K)=-a(K)\,\gamma^\mu K_\mu -b(K)\,\gamma^\mu
 U_\mu \, ,
\end{equation}
where $U^\mu$ is the four-velocity of the thermal bath as
seen from a general
frame. Neglecting the zero-temperature mass $m_0$, the
coefficients $a(K)$ and $b(K)$
are given by
\begin{eqnarray}
a(K)&=&\frac{m_F^2(T)}{k^2}\left[ 1+\frac{\omega}{2k}\ln
\frac{\omega
-k}{\omega +k}
+\frac{\omega k}{T^2}I+\frac{k^2}{T^2}J
\right], \label{aaa}\\
b(K)&=&-\frac{m_F^2(T)}{k}\left[
\frac{\omega}{k}+\frac{1}{2}\left(
\frac{\omega^2}{k^2}-1\right)\ln \frac{\omega
-k}{\omega +k}+ \frac{(\omega^2-k^2)}{T^2}I\right]
,\label{bbb}
\end{eqnarray}
\begin{eqnarray}
I&=& \int_0^\infty
\frac{dp}{\pi^2k^2}\left[f_B(p)+f_F(p)\right]
\left[ \frac{\omega}{2}\ln\frac{4p(p+k)+k^2-
\omega^2}{4p(p-k)+k^2-\omega^2}+p \ln 
\frac{4p^2-(k+\omega)^2}{4p^2-(k-\omega)^2}\right] \\
J&=& \int_0^\infty
\frac{dp}{2\pi^2k}\left[f_B(p)-f_F(p)\right]
\ln\frac{4p(p+k)+k^2-
\omega^2}{4p(p-k)+k^2-\omega^2}.
\end{eqnarray}
Here
$f_{B,F}$ are the Bose-Einstein and Fermi-Dirac
distributions, respectively,
and $m_F(T)$ is the effective mass of the fermion in the
plasma. Notice
that this mass is not a ``true'' fermionic mass since it
does not
affect the chiral symmetry. The thermal mass is generated
radiatively
and comes from terms which are chiral-symmetric in the
Lagrangian.
The integrals $I$ and $J$ are gauge-dependent and have been
computed here
in the Feynman gauge.
In the high-temperature limit ($T\gg k$), the terms in
eqs.~(\ref{aaa})
and (\ref{bbb}) proportional to the integrals $I$ and $J$
can be neglected,
as they only give subleading contributions, leaving a
result for the
coefficients $a$ and $b$ which is gauge
independent\footnote{Our subsequent
computations involve weakly coupled particles, so that
$m_F^2/T^2$ turns out to be small and
neglecting $I$ and $J$ is an excellent approximation.}.

For a fermion charged under an ${\rm SU}(N)$ gauge group
with coupling $g$ and
having Yukawa coupling $Y$, the thermal fermionic mass
squared is
$m_F^2(T)=g^2 T^2 C(R)/8+ N_f\left|Y\right|^2T^2/16$, where
$C(R)$ is the quadratic Casimir of the fermionic
representation
({\it e.g.} $C(R)=(N^2-1)/(2N)$, when $R$ is a fundamental
of $SU(N)$)
and $N_f$ is the particle multiplicity flowing in the loop.

\begin{figure}
$$\includegraphics[width=17cm]{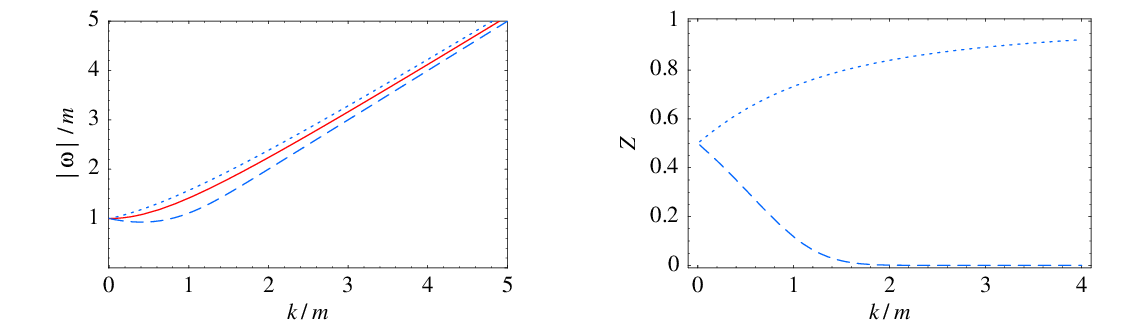}$$
\caption{\label{fig:mT}\em 
The dispersion relation $|\omega(k)|$ (fig.\fig{mT}a)
and the residue (fig.\fig{mT}b) of
particle (dotted line) and `hole' (dashed line)
excitations of a fermion with thermal mass $m$ at temperature $T$
for $m\ll T$.
The solid line shows the approximation $\omega^2 = m^2 + k^2$.}
\end{figure}

Interactions of the fermions with the thermal bath
modify the fermionic dispersion relation~\cite{weldon}
leading to two different types
of excitations with positive energy:
`particles'  and  `holes' with the wrong correlation
between chirality and helicity in the bare massless limit.
The names `particles' and `holes' are suggested by an
analogy with superconductors~\cite{weldon}. 
This is because the propagator in
eq.~(\ref{deltaFermion})
has poles at $ \omega =\pm k -b/(1+a)$,
an equation with two different positive-energy
solutions $E_p$ and $E_h$ ($E=|\omega|$, see fig.~\ref{fig:mT}a).
At low-momentum, $k\ll m_F$,
\begin{equation}
E_p=m_F+\frac{k}{3}+\frac{k^2}{3m_F}+\cdots,\qquad
E_h= m_F-\frac{k}{3}+\frac{k^2}{3m_F}+\cdots \, ,
\end{equation}
while at larger momenta $T\gg k\gg m_F$
\begin{equation}
E_p =
k+\frac{m_F^2}{k}-\frac{m_F^4}{2k^3}\ln\frac{2k^2}{m_F^2}+\cdots
,\qquad
E_h = k+2k \exp(-1-2k^2/m_F^2)+\cdots  .
\end{equation}
The most remarkable property of hole dispersion relation is that
its minimum occurs at the nonzero momentum $k_*\simeq 0.4\,m_F$.
The residues of the particle and hole propagators
 are given by~\cite{weldon}
\beq
Z = \bigg[\frac{d}{d\omega} b-(1+a)(k-\omega) \bigg]^{-1}= \frac{E^2-k^2}{2m_F^2}
\eeq
which differs from the standard
value $Z = 1$.
This reduces to $Z_p=Z_h =1/2$ at $k\ll m_F$
and
$Z_p=1$ and $Z_h=0$ at $k\gg m_F$  (see fig.~\ref{fig:mT}b):
holes interact only at low momentum. 
Notice that the dispersion relation
and the residues of the holes can be directly obtained
from those of the particles, $E_h(k)=E_p(-k)$ and $Z_h(k)=Z_p(-k)$.
The functions $a(K)=a(k,\omega)$ and $b(K)=b(k,\omega)$ defined
 in eq.~(\ref{aaa})--(\ref{bbb}),
when evaluated
for on-shell particles or holes are functions, for instance, only of $k$.
Correspondingly, a useful
property is $a(k,E_p(k))=a(-k,E_h(-k))$ and 
$|b(k,E_p(k))|=|b(-k,E_h(-k))|$.

Despite these subtleties, since $Z_p + Z_h \approx 1$
the fermionic dispersion relation
can be well approximated by
$\omega=[|\vec{k}|^2+m_F^2(T)]^{1/2}$ 
(see fig.~\ref{fig:mT})
and this is the form we have used in our calculation,
verifying that it is accurate enough when its validity
seems doubtful.
Notice that temperature-dependent masses enter into the
kinematics and into the
dispersion relations, but spinor functions must be taken
identical
to vacuum spinors, although with a modified dispersion
relation \cite{weldon}.

As for the bosonic case, we are interested only in the
$(11)$ component
of the resummed fermionic propagator
\begin{equation}
S^{11}(K)=
{\rm cos}^2\phi_K\,\Delta_F(K)-{\rm
sin}^2\phi_K\,\Delta^*_F(K),
\label{pinin}
\end{equation}
\begin{equation}
\Delta_F(K)=i\frac{(1+a) \gamma^\mu K_\mu + b \gamma^\mu
U_\mu 
}{(1+a)^2 K^2 + 2 (1+a)b K \cdot U +b^2 + i\epsilon}   .
\end{equation}
Assuming that the self-energy $\Sigma$ is real (which is 
always true in the high-$T$ limit), so that $a$ and $b$ are
real, with
the help of eq.~(\ref{iepsilon}), we can rewrite
eq.~(\ref{pinin}) as
\begin{eqnarray}
S^{11}(K)=\bigg[(1+a) \gamma^\mu K_\mu + b \gamma^\mu U_\mu
\bigg]
\bigg[\frac{i }{[(1+a)\omega +b]^2-(1+a)^2k^2  + 
i\epsilon} \nonumber \\
- 2\pi f_F(\omega)\delta \left([(1+a)\omega
+b]^2-(1+a)^2k^2 
\right) \bigg] .
\label{fermionpropagator}
\end{eqnarray}
Before applying these results to the computation of the
baryon asymmetry, 
let
us briefly discuss the issue of infrared singularities
at finite temperature \cite{yaffe}.
It is well-known that perturbation theory at finite
temperature is
afflicted by infrared problems which are worse than the
ones
appearing at zero-temperature field theory where the
Kinoshita-Lee-Nauenberg (KLN) theorem \cite{Kin,LN}
demonstrates that singularities appearing at intermediate
stages of the
calculation cancel out in the final physical result.
In an interacting scalar
theory the plasma mass may receive large two-loop infrared
contributions
and an all-loop  resummation is needed to get a final
result. 
In this
case the plasma mass obtained at one-loop is used as 
an infrared regulator. In order to deal with 
infrared
divergences one can perform the so-called {\it hard thermal
loop} (HTL)
resummation \cite{BPR}. 
In the 
HTL resummation one makes a distinction between
hard momenta of order $T$ and  soft momenta of order $gT$
and performs a resummation  only for the soft lines.
Indeed,
the corrections to the bare propagator $K^2$, being of
order $g^2T^2$,
start to be relevant only when $K\sim gT$, which is the
soft scale.
Therefore, if the internal momentum is
hard, then ordinary bare propagators and vertices are
sufficient, but if
the momentum is soft then effective propagators and
vertices must be used.
In this way, one obtains an improved perturbation theory, 
in which the
HTLs become generic analytic
functions of the external momenta, $g^n T^2
f(\omega_i,k_i)$. Using the HTL technique, we have
for instance explicitly checked that 
the top Yukawa coupling constant $\lambda_t$ entering the 
vertex $HQ_3U_3$  (which is involved in the
$\Delta L=1$ scattering $LH\rightarrow Q_3U_3$) gets renormalized at 
finite temperature by the exchange of gauge bosons between the 
Higgs and the top lines and that the correction is tiny, $\delta\lambda_t/
\lambda_t\simeq 10^{-1}\,g_2$.
Furthermore
one can  extend the KLN theorem at finite temperature,
and the infrared divergences in the corrections from
virtual gauge-boson
exchange cancel once we include absorption and emission of
real particles
that are too soft to escape the thermal
bath~\cite{weldonir}.

\begin{figure}[t]
$$\includegraphics[width=17cm,height=6cm]{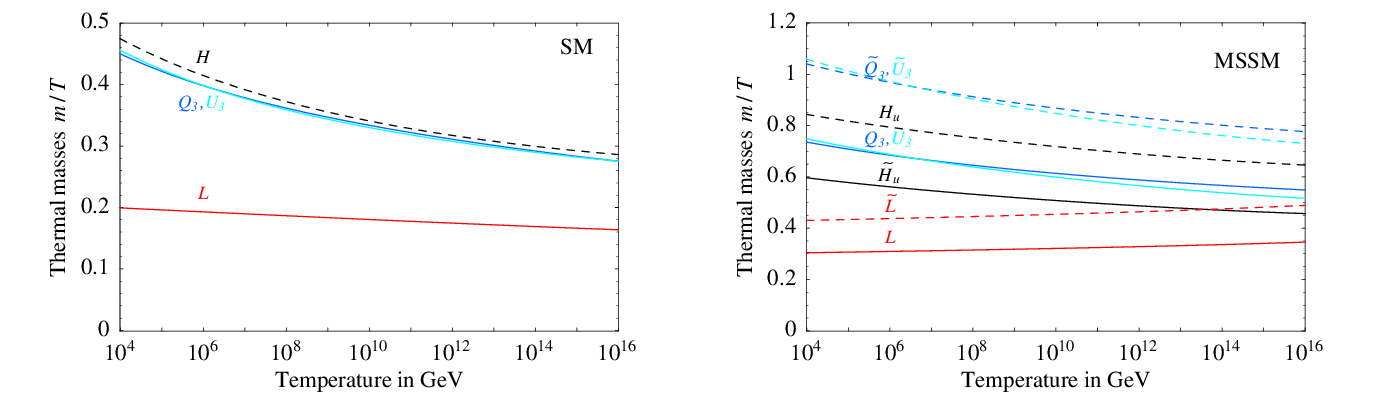}$$
\caption{\label{fig:thermalmasses}\em
Thermal masses in the SM (left) and in the MSSM (right), in units
of the temperature $T$.}
\end{figure}

\section{Including thermal corrections}\label{thermal}
Performing a complete study of leptogenesis
including all finite-temperature corrections is a very
difficult task. 
In our computation we include the leading
finite-temperature 
corrections which
are quantitatively relevant for the computation of the 
final baryon asymmetry.
These are given by: {\it i)} thermal corrections to gauge
and Yukawa
couplings; {\it ii)} thermal corrections to lepton, quark, 
and Higgs propagators;
{\it iii)} thermal corrections to the CP-violating
asymmetry.

In this section we discuss how we implement these three
kinds 
of corrections and how large they are.
Explicit formul\ae{} can be found in the appendices.
In the next section we 
indicate how the finite-temperature computation of the
baryon asymmetry
can be significantly simplified
by including only those effects which a posteriori turn out
to be 
numerically most relevant.

\subsection{Corrections to couplings}\label{couplings}
Renormalization of gauge and Yukawa couplings in the
thermal plasma
has been extensively
studied (see {\it e.g.} ref.~\cite{mikko}). 
A very good approximation is to renormalize the couplings
at the first Matsubara mode,
\begin{equation}
E_{r}=2 \pi T,
\label{scale}
\end{equation}
using the zero-temperature renormalization group equations
(RGE).
This result can be understood by recalling that
the average particle energy in the thermal plasma 
is larger than the temperature,
and so must be the renormalization scale.
For our purposes, the important couplings are the
gauge and top Yukawa, which we evaluate using the
appropriate RGE
at the scale in \eq{scale}.
Therefore those couplings are always functions of
temperature, even 
if not explicitly indicated.

Leptogenesis also depends on neutrino couplings and masses.
The neutrino mass matrix can be renormalized from 
low energy (where it is measured)
up to the high-energy scale relevant for leptogenesis using
the well known RGE
reported in appendix~\ref{gamma}.
They are often solved with a `diagonalize and run' approach~\cite{diagrun}
which focuses on 
the
neutrino masses and mixings probed by oscillation
experiments. 
We instead purse a `run and diagonalize'  strategy, as {\it
e.g.}\ in 
ref.~\cite{BRS},
which makes it easier to see how the combinations of
neutrino masses 
relevant for leptogenesis renormalize.
The solution can be trivially written as
$m(\bar \mu') = r~I\cdot m(\bar \mu) \cdot I $ where
$r$ is an overall rescaling factor and
\begin{equation}
I\simeq  1\hspace{-0.4ex}\hbox{I} +
\hbox{diag}(y_e^2,y_\mu^2,y_\tau^2)\frac{\ln(\bar\mu'/\bar\mu)}{(4\pi
)^2}\times\left\{\begin{array}{ll}
  -3/2    & \hbox{in the SM}   \\
  1/\cos^2\beta    & \hbox{in the MSSM} 
\end{array}\right. 
\end{equation}(higher powers of  $\ln(\bar\mu'/\bar\mu)$
can be easily resummed).
Here $y_\ell = m_\ell/v$ with $v=174\GeV$.
Unless one considers large values of $\tan\beta$ 
the flavour dependent term $I$ is very close to the unity
matrix.
In such a case, $I$ can be relevant only if one considers
special
neutrino mass matrices, fine-tuned such that 2 or 3
eigenvalues are
almost degenerate.
We can neglect $I$, as long as we do not consider such
cases.

\begin{figure}
$$\includegraphics[width=12cm]{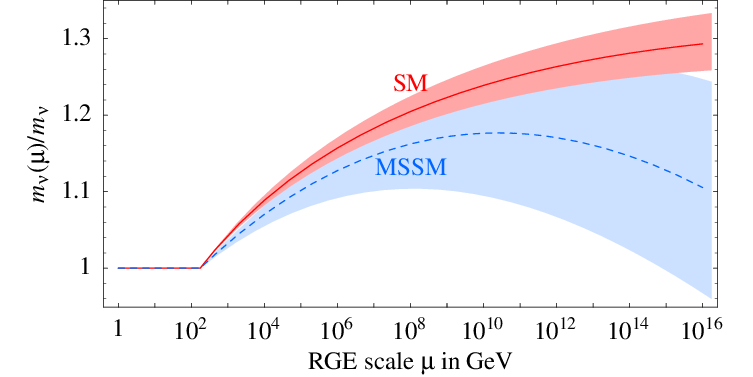}$$
\caption{\label{fig:RGE}\em Universal running of 
$m_\nu$ in the SM and in the MSSM.
The bands give an indication of the uncertainties,
as explained in the text.}
\end{figure}

The relevant correction to neutrino masses is therefore
given by the  overall rescaling factor  $r$.
Its numerical value is plotted in fig.\fig{RGE}.
The SM value has been computed assuming 
$\alpha_3 (M_Z)= 0.118\pm 0.003$,
a pole top mass of $m_t =175\pm 5\GeV$ and
a Higgs mass $m_h = 115\GeV$.
The band indicates the present uncertainty induced by
the errors on $m_t$ and $\alpha_3$.
Varying the Higgs mass has a negligible impact,
unless $m_h$ is close to the triviality bound, $m_h\sim 180\GeV$,
where the quartic Higgs coupling becomes non-perturbative at high scales
inducing arbitrarily large values of $r$.
This issue was also discussed in~\cite{SMRGE2}.

The MSSM central value has been computed assuming also
moderately large $\tan\beta\sim 10$,
unification of gauge couplings at $M_{\rm GUT} = 2\times 10^{16}\GeV$,
and $\lambda_t(M_{\rm GUT}) = 0.6$.
Ref.~\cite{heavym0}  explains why $\lambda_t(M_{\rm GUT})$ is
still significantly uncertain, about between $0.5\div0.7$,
giving rise to a correspondingly large uncertainty in RGE effects,
illustrated by the shaded area in fig.\fig{RGE}.

\subsection{Thermal corrections to decay and scattering
processes}\label{rates}
To calculate the generated baryon  asymmetry of the
universe in the 
thermal leptogenesis scenario we have to solve a set of
Boltzmann 
equations, discussed in appendix~\ref{Boltz},
which take into account processes that create or wash-out
the asymmetry.
We have recalculated the relevant reaction densities taking
into account 
the propagation of particles inside the thermal plasma.

As discussed in section~\ref{qft}, the wave-functions that 
describe external fermion states
in decay and scattering amplitudes are the same as in the
zero-temperature case~\cite{weldon}.
Temperature corrections appear only in internal fermion
lines and in the kinematics. 
In the SM, for all processes except those involving gauge bosons (which
will be discussed later), the only fermion which mediates an interaction
relevant for leptogenesis is $N_1$. 
Since $N_1$ has no gauge interactions and 
its Yukawa coupling is small in all the relevant region of
parameters,
thermal corrections to ${N_1}$ propagation can be
neglected\footnote{We do
not consider the case of quasi-degenerate right-handed
neutrinos.
In such a case small corrections which break degeneracy
could not be neglected.}.

Therefore we only have to include thermal corrections to
the dispersion 
relations of lepton doublets, third-generation quarks and
Higgs bosons (and
their supersymmetric partners).
Fig.\fig{mT}a shows the dispersion relation $\omega(k)$ 
satisfied by a fermion with thermal mass $m$ at temperature
$T$.
For simplicity, we approximate it with a Lorentz-invariant
relation
$\omega^2 = m^2 + k^2$, shown by the solid line in
fig.\fig{mT}a.
As discussed later, this approximation has a negligible
impact on  our results.

\begin{figure}[t]
$$\hspace{-5mm}\includegraphics[width=15.5cm]{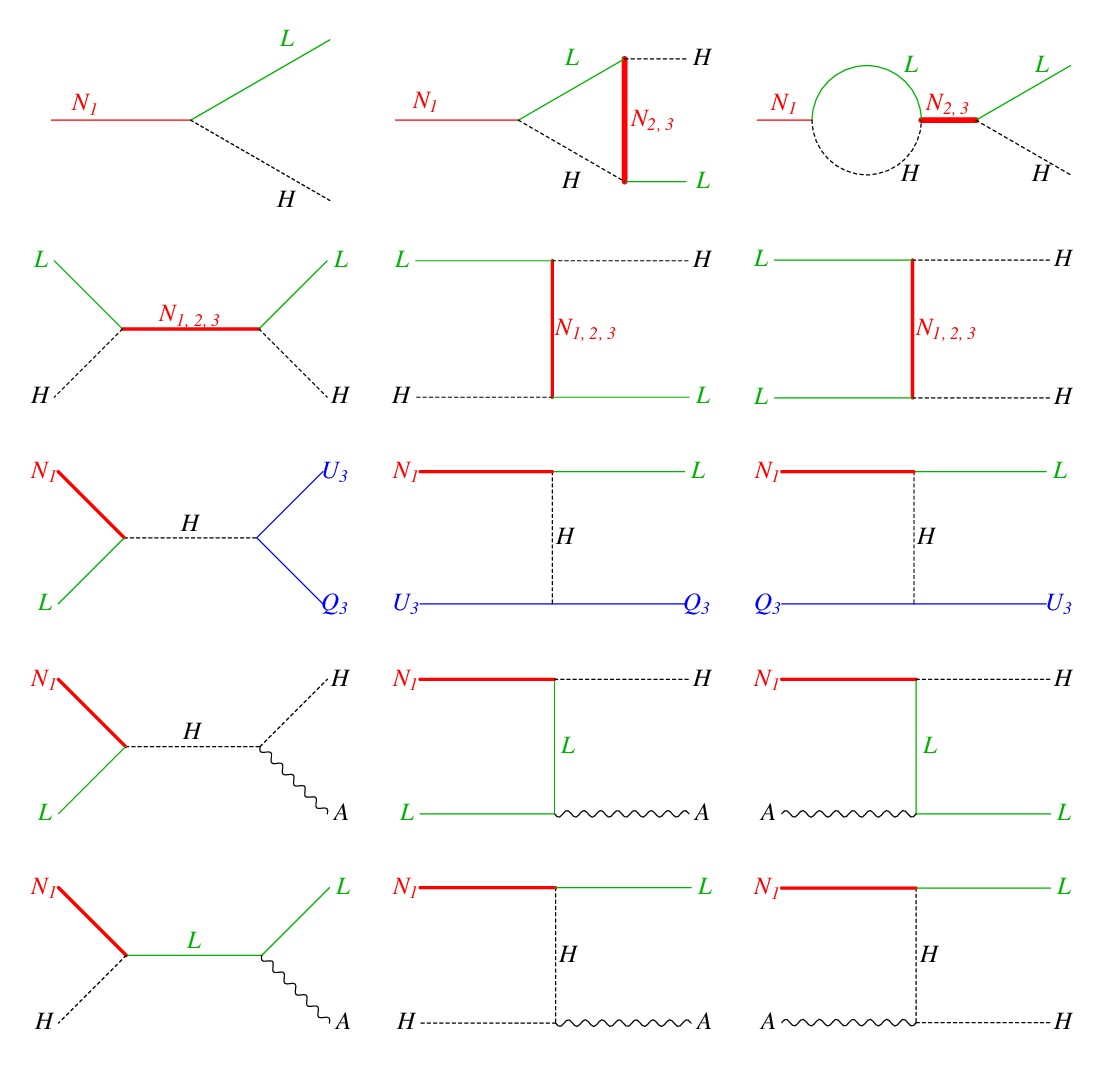}$$
\caption{\label{fig:leptogDiags}\em Feynman diagrams
contributing to SM thermal leptogenesis.}
\end{figure}

Temperature corrections to the SM and MSSM particle masses
are well 
known~\cite{thermalmasses}. 
The relevant formul\ae{} are collected in
appendix~\ref{gamma} and the
numerical values of
$m/T$ are plotted in fig.\fig{thermalmasses} as function of the
temperature (assuming a Higgs mass $m_h = 115\GeV$).
The computation of thermally corrected SM decay rates and
reduced cross sections is 
performed in appendix~\ref{gamma}.
Here we wish to discuss the most important features of the
results.

The processes that affect SM thermal leptogenesis are 
(Feynman diagrams are plotted in fig.\fig{leptogDiags})
\begin{itemize}
\item the
decays
$N\to H L$ and (at very high temperature when the Higgs
becomes
heavier than $N_1$)  $H\to N L$
(the relative reaction density is 
denoted as $\gamma_D$,
see appendix~\ref{gamma});
\item the $\Delta L = 2$ scatterings $LH \to \bar L\bar H$ and $LL\to \bar H \bar H$ 
($\gamma_{N}$);

\item  the $\Delta L = 1$ scatterings involving the top quark
$N_1\bar L\to Q_3 U_3$ ($\gamma_{Ss}$)
and $LQ_3\to N_1 U_3$, $LU_3\to N_1 Q_3$ ($\gamma_{St}$)
as well as their inverse reactions
(which have the same reaction densities, up to small
CP-violating corrections).
We introduce $\Delta L = 1$ scatterings involving ${\rm SU}(2)\otimes{\rm U}(1)$
gauge bosons $A$:
$N_1\bar L\to HA$  ($\gamma_{As}$),
$LH\to N_1A$, $\bar{L}A\to N_1 H$ ($\gamma_{At}$)
and define the total scattering rates 
 $\gamma_{Ss,t}=\gamma_{Hs,t}+\gamma_{As,t}$.
 \end{itemize}
Fig.\fig{fig0}a shows the reaction densities 
$\gamma_D$, $\gamma_{Ss}$, $\gamma_{St}$, $4\gamma_{N}$,
all normalized in units of $H n_{N_1}$,
as function of the temperature and for
$\tilde m_1= r\cdot |\Delta m_{\rm atm}^2|^{1/2}=0.06$~eV 
and $m_{N_1}=10^{10}\GeV$.

Fig.\fig{fig0}b (c) show the full set of reaction densities
computed including (not including) the effects added in this paper.
In these figures we use conventions adopted in previous papers, and plot
$\gamma_D$, 
$\gamma_{Hs}$, $\gamma_{Ht}$, $\gamma_{As}$, $\gamma_{At}$
normalized in units of $H n_{N_1}$
and the `subtracted $\Delta L=2$ scattering rate' (see appendix~\ref{Boltz})
$\gamma_{N}^{\rm sub}$ normalized in units of $H n_\gamma$.

\subsubsection{Decays}
\begin{figure}[t]
$$\hspace{-7mm}\includegraphics[width=17.5cm]{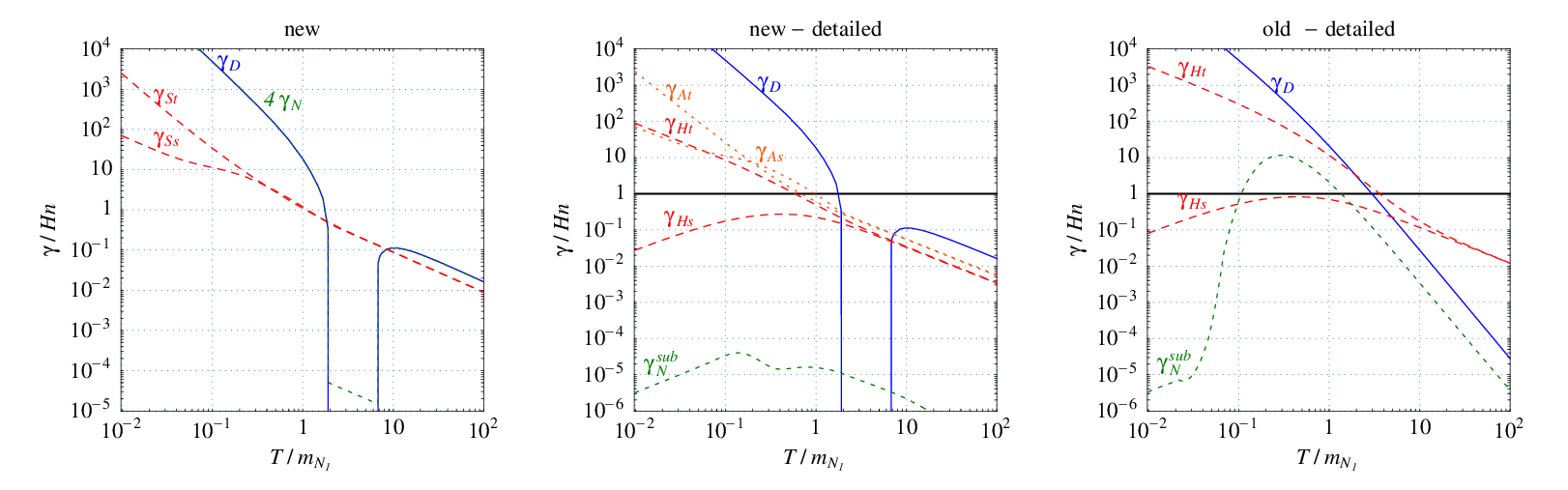}$$
\caption{\em {\bf The SM reaction densities} for $\tilde m_1
\equiv (Y_\nu Y_\nu^\dagger)_{11} v^2/m_{N_1}
= 0.06\eV$ 
and $m_{N_1}=10^{10}\GeV$.
{\color{blus} Blue line: decays ($\gamma_D$)}. 
{\color{rossos} Red long-dashed lines:
$\Delta L = 1$ scatterings  ($\gamma_{Ss,t}=\gamma_{Hs,t}+\gamma_{As,t}$)}.
{\color{verdes} Green dashed lines: $\Delta L = 2$ scatterings ($\gamma_N$)}.
\label{fig:fig0}}
\end{figure}

The modification in $\gamma_D$ is probably the most
apparent feature
of a comparison between fig.\fig{fig0}b and \fig{fig0}c,
and it occurs because at sufficiently high temperature, the
Higgs becomes
heavier than $N_1$ and the decay
$N_1\to H L$ becomes kinematically forbidden.
For temperatures in the range where $m_H - m_L<m_{N_1} <
m_H + m_L$,
there are no two-body decays involving $N_1$ at all.
At higher temperatures the Higgs becomes  
heavy enough for the  $H\to N_1L$ decays to be allowed,
and the heavy neutrinos are produced in the process
$H\leftrightarrow N_1 L$
rather than in $N_1\leftrightarrow HL$.
Including thermal masses  we get
$\gamma_D\propto T^4$ at $T\gg m_{N_1}$.
Neglecting thermal masses gave a much smaller decay rate,
$\gamma_D\propto T^2$, so that
higher order $\Delta L=1$ scatterings $\gamma_{Hs,t}$ were dominant.

\subsubsection{$\Delta L=1$ scatterings}
A comparison of fig.\fig{fig0}b,c reveals
other important numerical differences.
There is a significant reduction of $\Delta L = 1$
scattering rates (red dashed curves), mainly $\gamma_{H_t}$, over the full
temperature range. 
This comes from two different effects: 1) the top Yukawa
coupling at high temperatures
is smaller than at the electroweak scale, {\it e.g.}\ 
$y_t(10^{10}\GeV)/y_t(m_t)\approx 0.6$;
2) Higgs boson exchange  in the $t$ channel mediates
a long-range force, giving rise to cross section enhanced
by $\ln m_{N_1}/m_H$.
This enhancement disappears when the thermal Higgs mass,
$m_H\sim 0.4~ T$,
is used in place of the zero-temperature Higgs mass,
$m_H\sim 100\GeV$.
Such reduction of the $\Delta L = 1$ scattering rates turns
out to be the most important modification
and affects leptogenesis in two different ways:
less washout of the leptonic asymmetry at $T\ll m_{N_1}$
and slower thermalization of  $N_1$ at $T\gg m_{N_1}$.
It is partially compensated by the inclusion of $\Delta L = 1$ scatterings involving gauge bosons.

\subsubsection{$\Delta L=2$ scatterings}
There is a new resonant enhancement of 
$\Delta L =2$ scatterings
$L H\leftrightarrow \bar L \bar H$ mediated by $N_1$.\footnote{Scatterings mediated by
the heavier $N_{2,3}$ are included as described in section~\ref{SM}.}
At low temperatures the virtual $N_1$ can be on-shell
when exchanged in the $s$-channel, so that
$\gamma_{N}$ is enhanced by the
$s$-channel resonance at
$s=m^2_{N_1}$.
 At high temperatures the virtual $N_1$ can be on-shell
when exchanged in the $u$-channel, so that
$\gamma_{N}$ is enhanced by the
$u$-channel resonance at $u = m^2_{N_1}$.
For intermediate temperatures there is no resonance.

The $s$-channel resonance is regulated, as usual, by the $N_1$ width.
On the contrary the new $u$-channel resonance occurs when $N_1$ is stable,
because its decays are forbidden by thermal masses.
The divergence in the cross section is eliminated by the
presence of an imaginary part in the $N_1$ propagator, corresponding to
the thermal damping rate caused by the interactions with the plasma. 
Although this effect depends in a complicated way 
on $N_1$ thermal motion with respect to the plasma,
in narrow width approximation (i.e.\ for small $N_1$ Yukawa couplings)
resonant enhancements give $\gamma_N\simeq \gamma_D/4$
(see appendix~\ref{gamma}).

Therefore we performed a precise computation of the decay rate
(including also Pauli-blocking and 
stimulated emission factors, as discussed in appendix~\ref{gamma})
and computed $\gamma_{N}^{\rm sub} = \gamma_N-\gamma_D/4$,
the contribution to the $\Delta L = 2$ scattering rate
due only to off-shell scatterings.
Indeed, this quantity enters the Boltzmann equations, 
because the contribution 
of on-shell $N_1$ exchange is already taken into account by
successive decays and inverse decay processes,
$LH\leftrightarrow N_1 \leftrightarrow \bar{L}\bar{H}$, 
and has to be subtracted in order to avoid double
counting. In appendix~\ref{Boltz} we show how to properly 
perform the subtractions to the $N_1$ propagator. 
Our result differs from the one of ref.~\cite{k-sm,mBound}. 
Indeed, the subtraction method used in ref.~\cite{k-sm,mBound} 
leaves a spurious contribution which effectively
double counts the decay process.
This is why our  $\gamma_N^{\rm sub}$ in fig.\fig{fig0}b
no longer has the ``off-shell resonance'' found by 
previous computations, shown in fig.\fig{fig0}c.
At leading order in the neutrino Yukawa coupling $Y_\nu$ 
$\gamma_D,\gamma_{Ss},\gamma_{St} \propto (Y_\nu Y_\nu^\dagger)_{11}$ and
$\gamma_{N}^{\rm sub}\propto  (Y_\nu Y_\nu^\dagger)_{11}^2$.
Therefore, we find that the off-shell contribution is relevant only  when $Y_\nu\sim 1$,
which is not the case in our example of fig.\fig{fig0}.
When $Y_\nu\sim 1$ a fully precise computation should include also $\Delta L=0$
$L\bar L\to N_1N_1$ scatterings,
which would play a minor role, affecting the $N_1$ abundancy.
More importantly, in such a case off-shell $\Delta L=2$ scatterings suppress $n_B$ exponentially,
because (unlike $\Delta L=2$ mediated by on-shell $N_1$)
at $T\circa{<} m_{N_1}$ they are not suppressed
by the $N_1$ abundancy.

\medskip

Furthermore, fig.\fig{fig0}b allows to get the rates for other values of $\tilde{m}_1$
by applying the appropriate rescaling.
For other values of $m_{N_1}$ the 
rescaling of the Yukawa couplings
$(Y_\nu Y_\nu^\dagger)_{11} = \tilde{m}_1 m_{N_1}/v^2$
is the dominant effect, but is not the only one.
One needs to recompute the rates taking into account 
the running of the couplings.

\subsection{CP violation at finite temperature\label{cp}}
In this section we investigate the finite-temperature
effects on
the CP asymmetries
\begin{equation}
\epsilon_i=\frac{\gamma^{\rm eq}(i\rightarrow
f)-\gamma^{\rm eq}(\bar{\hbox{\em \i}}
\rightarrow\bar{f})}{\gamma^{\rm eq}(i \rightarrow
f)+\gamma^{\rm eq}(\bar{\hbox{\em \i}}\rightarrow
\bar{f})}  \ .
  \label{DefEpsilon}\end{equation}
  where $\gamma^{\rm eq}$ are the thermally averaged decay rates.
The decay processes relevant to our analysis are
$N_1\rightarrow L H$,
which is allowed for
$m_{N_1}>m_L(T)+m_H(T)$ and $H\rightarrow L N_1$, which is
allowed at higher
$T$, when $m_H(T)>m_L(T)+m_{N_1}$ and is CP-violating only
because
of purely finite-$T$ effects.

The issue of CP-violating decays at finite $T$ was already
investigated
in ref.~\cite{CoviTh}, although neglecting thermal masses.
As we will see, the effect of the masses is crucial, giving
a non-trivial $T$ dependence of the CP violation.
The effect of thermal masses is taken into account by using
the one-loop
finite temperature resummed propagators, and by using
modified dispersion
relations, as discussed in section~\ref{qft}.

We choose to work
in the rest frame of the plasma, where
$U_\mu=(1,0,0,0)$.
In this way the finite-$T$ Feynman rules are simplified,
while the kinematics
is more complicated, since we have to consider decaying
particles
in motion.
Although we have performed our calculation in the general
case,
neglecting the thermal motion of the decaying particles
allows to write
reasonably accurate analytical approximations.
As already said,  we approximate the complicated dispersion
relation for fermions with
$\omega^2=k^2+m_F^2$.

%

The CP asymmetries in the relevant decays
come from interference between the tree-level decay
amplitude with
the one-loop contributions. There are two relevant
one-loop diagrams: the so-called vertex and wave-function
contributions
(shown in fig.~\ref{Ndecay}).
We compute the imaginary part of the one-loop graphs
using the Cutkosky cutting rules at finite
 temperature~\cite{Kobes},
which are more complicated than at $T=0$ (even in the
absence of type-$2$ vertices).
While at $T=0$ most cuttings (in our case two of the three
possible cuttings)
give no contribution due to energy-conservation,
this is no longer true at $T\neq 0$,
since particles may absorb energy from the plasma.
Formally, this means that we must also consider cut
lines with negative-energy on-shell particles.

Nevertheless, we ignore cuts which involve heavy
right-handed
neutrinos $N_{2,3}$, because they are
suppressed by a  Boltzmann factor $\exp(-m_{N_{2,3}}/T)$,
which is negligibly small since we assume a hierarchical
spectrum $m_{N_{2,3}}\gg m_{N_1}$, and we work at $T$ much smaller than $m_{N_{2,3}}$.
Therefore, as illustrated in fig.~\ref{Ndecay}
we can restrict ourselves to the standard cutting of the
Higgs and
lepton lines, but with energy flows in both directions.

\subsubsection{CP-asymmetry in $N_1$ decay}\label{cpN1}
We first consider the CP asymmetry in $N_1$ decay,
\begin{equation}
\epsilon_{N_1} \equiv \frac{\gamma^{\rm eq}(N_1\rightarrow H L)-\gamma^{\rm eq}(
N_1\rightarrow \bar{H} \bar{L})}{\gamma^{\rm eq}(N_1\rightarrow H
L)+\gamma^{\rm eq}
(N_1\rightarrow \bar{H} \bar{L})}.
\end{equation}
The full result employed in our leptogenesis code
 is presented in appendix \ref{AppEpsilon}.
Here we present a simple analytic approximation obtained neglecting
the thermal motion of $N_1$ with respect to the plasma,
which is justified at $T \ll m_{N_1}$ and still reasonably accurate
at higher temperatures (see fig.~\ref{figepsilon}a).
The result is
\begin{equation}
\epsilon_{N_1}(T) = \epsilon_{N_1}(0)R_\epsilon(T)\, ,\qquad
 \epsilon_{N_1}(0)=\frac{1}{8\pi}\sum_{j\neq 1}
\frac{\textrm{Im}
\left[ (Y^{\dagger} Y)_{j 1}^2\right] }{\left[Y^{\dagger}
Y\right]_{11}}
f\left(\frac{m_{N_j}^2}{m_{N_1}^2}\right)  . \label{eps0}
\end{equation}
The function  $f$ describes the usual result at $T=0$
which, in the SM,
is given by
\begin{equation}
f(x)=\sqrt{x}\left[ \frac{x-2}{x-1}-(1+x)\ln
\left(\frac{1+x}{x} \right)
 \right] \stackrel{x\gg 1}{\longrightarrow} - \frac{3}{2
\sqrt{x}} \ . \label{f(x)}
\end{equation}
The thermal correction is described by the function $R_\epsilon$, given by\footnote{If $N_{2,3}$ are not much heavier
than $N_1$ one can easily include effects suppressed by higher powers of
$m_{N_1}/m_{N_{2,3}}$, obtaining a more lengthy analytical expression.}
\begin{equation}
R_\epsilon= {16 \frac{k^2}{m_{N_1}} [\omega(1+a_L)+b_L]}J
[1+f_H-f_L-2 f_H f_L]\qquad
J = \left|\left|
\begin{matrix}
\partial \delta_H/\partial\omega &\partial
\delta_L/\partial\omega \\
\partial \delta_H/\partial k &\partial \delta_L/\partial k
\end{matrix}\right|\right|^{-1} \label{eq:g1}
\end{equation}
where $||M||\equiv |\det M|$,
$K_L=(\omega,k)$ is the energy-momentum quadri-vector of $L$,
and the functions $a_L\equiv a(K_L)$ and $b_L\equiv b(K_L)$ are defined in eq.s~(\ref{aaa},\ref{bbb}).
The Fermi-Dirac and Bose-Einstein distributions
$$ f_L = (e^{E_L/T}+ 1)^{-1}\qquad
f_H = (e^{E_H/T}- 1)^{-1}$$
in the third factor
are evaluated at the fermion energy $E_L = \omega$ and at the
boson energy $E_H=m_{N_1}-\omega$.
The first term is obtained by computing the relevant
Feynman graph in the limit $m_{N_{2,3}}\gg m_{N_1}$ and dividing by the tree level
rate.
The Jacobian $J$ is obtained when imposing the on-shell
conditions for the cut particles $H$ and $L$
\begin{equation}
\delta_H \equiv (m_{N_1}-\omega)^2 - k^2 - m_H^2 = 0
\qquad
\delta_L \equiv [(1+a_L)\omega +b]^2-(1+a_L)^2k^2  = 0
\end{equation}
which fixes the values of $\omega$ and $k$ in terms of
$m_{N_1},\, m_L,\, m_H$.
As discussed in section~\ref{qft} the equation for $L$ has
two different solutions:  `particles' and `holes'.
A numerical computation shows
that the `hole' contribution to the CP-asymmetry is
negligible because,
as explained in section~\ref{qft}, relativistic holes have
negligible interactions.
The `particle' contribution is well approximated
by inserting in eq.~(\ref{eq:g1})
the values of $\omega$ and $k$
\begin{equation}\label{eq:omegak}
\omega = \frac{m_{N_1}^2 + m_L^2 -
m_H^2}{2m_{N_1}},\qquad
k = \sqrt{\omega^2 - m_L^2}\end{equation}
obtained approximating
$\delta_L\approx \omega^2 -k^2 - m_L^2=0$.


The main feature shown in fig.~\ref{figepsilon}
is  that  $\epsilon(T)$ goes to zero as
the
temperature increases and the
process becomes kinematically forbidden.
This happens because the particles in the final state
coincide with the
cut particles in the loop: $L$ and $H$.
Therefore the threshold at which the cut particles can no
longer be on the mass-shell
is the same at which the decay becomes kinematically
forbidden, i.e.\ when
$m_{N_1}\approx m_H+m_L$.

There is another important effect which gives an additional
suppression.
The $1+f_B-f_F-2 f_B f_F$ factor in \eq{eq:g1} was first
derived by the authors of ref.~\cite{CoviTh}
who, neglecting the $L$ and $H$ thermal masses and thus
setting
$\omega = m_{N_1}/2$, found it to be equal to 1.
However, only if the arguments of the Bose-Einstein and
Fermi-Dirac
distributions are the same, there is a peculiar
cancellation: $f_B-f_F-2 f_B f_F=0$.
Physically this cancellation can be understood as a
compensation between
stimulated emission and Pauli blocking. Only if bosons and
fermions
enter with the same energy,  an exact cancellation holds.

\subsubsection{CP-asymmetry in $H$ decay}
\begin{figure}[t]
\centerline{
\raisebox{-4mm}{\includegraphics[height=6.3cm]{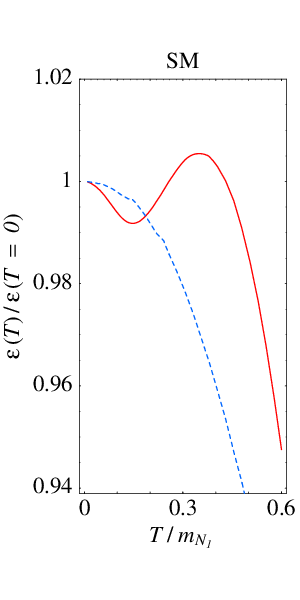}}
\includegraphics[height =5.4cm]{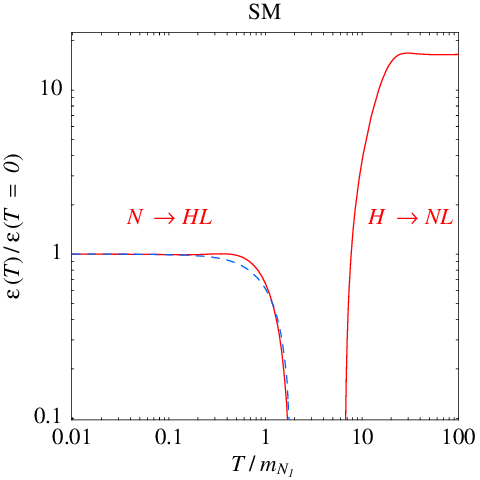}~~
\raisebox{-3mm}{\includegraphics[height =6cm,width=8cm]{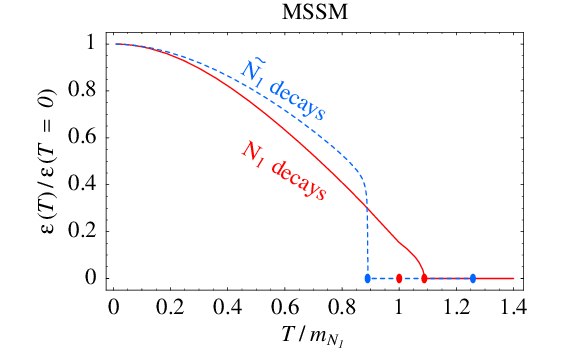}}}
\caption{\em  {\bf Thermal corrections to the CP asymmetry}.
$\epsilon(T)/\epsilon(T=0)$ as a function of
temperature
for $m_{N_{2,3}}\gg m_{N_1} = 10^{10}\GeV$
in the SM (left plots) and in the MSSM (right plot, the dots indicate
the various thresholds for $N_1$ and $\tilde{N}_1$ decays).
In the SM plot
the solid line shows our more accurate result derived in 
appendix~\ref{AppEpsilon},
while the dashed line shows the approximate result described in the main text,
obtained neglecting the $N_1$ thermal motion.
\label{figepsilon}}
\end{figure}

The computation of the CP asymmetry in Higgs decay,
\begin{equation}
\epsilon_H \equiv \frac{\gamma^{\rm eq}(H\rightarrow N_1 L)-\gamma^{\rm eq}(
\bar{H}\rightarrow N_1 \bar{L})}{\gamma^{\rm eq}(H\rightarrow N_1
L)+\gamma^{\rm eq}
(\bar{H}\rightarrow N_1 \bar{L})},
\end{equation}
is similar to the previous one, although
with some important differences.
Also in this case there are two relevant cuts (in wave and
vertex one loop diagrams,
see fig.~\ref{Hdecay}) which involve the Higgs and lepton
lines.
The difference is that such graphs would have no imaginary
parts
with the usual Feynman rules at $T=0$, as cuttings of $H$
and $L$ would be
kinematically forbidden.
On the contrary, at finite $T$ absorption of particles by the
plasma allows also negative energies in the cuts.
The asymmetry
$\epsilon_H$ turns out to be proportional to
the purely thermal factor
$
f_H-f_L-2 f_Hf_L
 \label{nBnF2}$.
The non-standard cut (see fig.~\ref{Hdecay}) implies a more complicated kinematics which
does not allow us to obtain an
analytic result for $\epsilon_H$. 
The computation is presented in appendix \ref{AppEpsilon}, where we
neglect effects due to particle motion and due to non-trivial fermion
dispersion relation, since eventually 
$\epsilon_H$ turns out to have a negligible effect
on the final results for leptogenesis.
$\epsilon_H$ approaches a constant value at high $T\gg m_{N_1}$
(see fig.~\ref{figepsilon}).

\subsubsection{CP-asymmetries in the MSSM}
The situation in the MSSM becomes more complicated than in the SM because
we must now study both $N_1$ and $\tilde{N_1}$ decays,
each having two possible decay channels, with
each channel having more diagrams.
The CP-asymmetries at zero temperature were first computed correctly 
in ref.~\cite{Covi}.
We do not study $H$ and $\tilde{H}$ decays, which appear only 
at high temperature.

The computations at finite temperature are analogous to the SM case and described in appendix~\ref{CPMSSM}.
The result is different because the cut particles in the loop are different from the  final-state particles:
the threshold at which the cut particles can no more go on-shell is different
from the one at which the decay becomes forbidden.
Therefore, in the single decay modes, CP-violation disappears either
before or after the decay mode becomes kinematically forbidden.

The source of CP-asymmetry which appears in the MSSM Boltzmann equations
are the CP-asymmetries $\epsilon_{N_1}$ and $\epsilon_{\Nt_1}$
in $N_1$ and $\tilde{N}_1$ decays,
averaged over the different decay channels
$N_1\to HL,\Ht\Lt$ and $\Nt_1\to H\Lt,\Ht L$.
As illustrated in fig.~\ref{figepsilon}b $\epsilon_{N_1}$ and $\epsilon_{\Nt_1}$
behave in rather different ways.

Like in the SM, $\epsilon_{N_1}$ goes to zero at the same temperature threshold
at which $N_1$ decays become kinematically forbidden,
as a consequence of the fact that both decay channels
contribute to the imaginary part of both decay modes.

On the contrary, for $\epsilon_{\Nt_1}$, only one-loop diagrams 
with internal bosons $H,\tilde{L}$ contribute to the CP-asymmetry
of decays into fermions $\tilde{H},L$ and viceversa:
since thermal corrections make bosons heavier than the corresponding fermions (see appendix~\ref{gamma})
$\epsilon_{\Nt_1}$ vanishes when $\tilde{N}_1$ decays into fermions are still kinematically allowed.
The dashed line in fig.~\ref{figepsilon}b shows the final result.
Thermal corrections are significant at $T\sim m_{N_1}$, but
give almost no effect at lower temperatures.
This happens in a non-trivial way.
Decays into scalars have a rate significantly enhanced by stimulated emission
and a CP-asymmetry significantly suppressed by Pauli blocking,
while the opposite happens for decays into fermions.
If thermal masses can be neglected, the two effects compensate each other,
as noticed in ref.~\cite{CoviTh}.
This cancellation no longer takes place when thermal masses become sizable,
giving rise to the behavior of $\epsilon_{\Nt_1}(T)$ shown in fig.~\ref{figepsilon}b.

%

\section{Leptogenesis in the Standard Model}\label{SM}
We assume that right-handed neutrinos are hierarchical,
$m_{N_{2,3}}\gg m_{N_1}$ so that we have to study the
evolution of
the number density of  $N_1$ only.
In such a case the final amount of ${\cal B}-{\cal L}$ asymmetry
$Y_{{\cal B}-{\cal L}}=n_{{\cal B}-{\cal L}}/s$
generated by $N_1$
assuming no pre-existing asymmetry
can be conveniently parameterized as
\begin{equation}
Y_{{{\cal B}-{\cal L}}}=-\epsilon_{N_1} ~ \eta ~ Y^{\rm eq}_{N_1}
(T\gg m_{N_1}) .
\end{equation}
Here $\epsilon_{N_1}$ is the CP-asymmetry parameter in $N_1$
decays
{\em at zero temperature}, and $Y^{\rm eq}_{N_1}(T\gg
m_{N_1}) =
135 \zeta(3)/(4\pi^4g_*)$, where $g_*$ counts the effective
number of
spin-degrees of freedom in thermal equilibrium ($g_*= 106.75$ in the SM
with no right-handed neutrinos)\footnote{The formula 
used in our numerical code includes leading order thermal effects from
quarks, leptons and gauge bosons (neglecting Yukawa couplings):
$$\rho =\rho_R 
+ \rho_{N_1},\qquad
\rho_R=\bigg[\frac{427}{4}\frac{\pi^2}{30}
- \frac{7}{4}g^2_3-\frac{19}{32}g^2_2-\frac{25}{96}g^2_Y \bigg]T^4
\qquad s=\frac{4\rho_R}{3T}.$$}.
 $\eta $ is an efficiency factor that measures the number
density of
$N_1$ with respect to the equilibrium value, the
out-of-equilibrium
condition at decay, and the thermal corrections to $\epsilon_{N_1}$. 
Recalling that, after reprocessing by
sphaleron 
transitions, the baryon asymmetry is related to the ${\cal B}-{\cal L}$
asymmetry by 
\begin{equation}\label{eq:YB}
\frac{n_{\cal B}}{s}=\frac{24+4n_H}{66+13n_H}\frac{n_{{\cal B} - {\cal L}}}{s},
\end{equation}
where $n_H$ is the number of Higgs doublets, for the SM we
find
\begin{equation}
\label{uusm}
\frac{n_{\cal B}}{s} =-1.38\times 10^{-3} \epsilon_{N_1} \eta.
\end{equation}
Assuming the  `standard' $\Lambda$CDM cosmological model,
BBN and WMAP measurements imply
\begin{equation}
\label{eq:exp}
\frac{n_{\cal B}}{n_\gamma} = (6.15\pm 0.25)\times 10^{-10}\qquad
\hbox{with}\qquad s= 7.04~ n_\gamma.
\end{equation}

Computing $\eta$ is the most difficult part of the
calculation, since
it is obtained from numerical solution of Boltzmann
equations.
In general the result depends on how the lepton asymmetry
is distributed in the three lepton flavours.
For simplicity one usually ignores flavour issues and
solves the approximated Boltzmann equation
for the total lepton asymmetry described in
appendix~\ref{Boltz}.
If all mixing angles of left and right-handed neutrinos are large, 
as data might suggest,
 this `one-flavour approximation'  is accurate up to ${\cal O}(1)$ corrections.
In order to include flavour factors one must 
solve the Boltzmann equations for the $3\times 3$ density matrix of
 lepton doublets,  as discussed in ref.~\cite{bcst}.

The $\Delta L=2$ scatterings mediated by heavier right-handed neutrinos
$m_{N_{2,3}}\gg m_{N_1}$ are relevant at
$m_{N_1}\circa{>} 10^{14}\GeV$.
Below, they can have ${\cal O}(1)$ effects if  neutrinos are quasi-degenerate.
Although these scatterings are produced by the same effective dimension 5 operator
that generates neutrino masses, their contribution to $\hat\sigma_{Ns,t}$ depends
on unknown high-energy parameters: 
the flavour composition of the neutrino coupled to $N_1$.
Therefore, following ref.~\cite{bcst}, we introduce
a parameter $\xi $ that, in eqs.~\ref{sigNs}--\ref{sigNt}, parameterizes the
unknown contribution of $N_{2,3}$,
which is important only if $m_{N_1}\circa{>} 10^{14}\GeV$.
If there were only one neutrino flavour with mass $m_\nu$,
the value of $\xi $ would be  $\xi  = m_\nu/\tilde{m}_1 - 1$.
With three neutrinos we cannot even write the precise definition of $\xi $,
as flavour factors cannot be correctly included
in Boltzmann equations valid in `one-flavour approximation'.
In our numerical results we assumed
$\xi =\max(1,m_{\rm atm}/\tilde{m}_1)$:
if all mixing angles are large this choice is reasonably correct in all the parameter space.


\medskip

Having fixed $\xi$, in  `one-flavour approximation' $\eta$ depends only on two
parameters:
$$\tilde{m}_1\equiv (Y_\nu Y_\nu^\dagger)_{11}
v^2/m_{N_1}\qquad\hbox{and}\qquad m_{N_1}$$
(and almost only on $\tilde{m}_1$ if $m_{N_1}\ll
10^{14}\GeV$ i.e.\ when 
$\Delta L = 2$ scatterings mediated by $N_i$ off-resonance
are negligible).
This well-known fact remains true also when thermal
corrections are included, as
can be seen from appendix~\ref{gamma}.
The parameter $\tilde{m}_1$ is
`the contribution to the neutrino mass mediated by $N_1$'.
To see what this means in practice, let us temporarily
assume that the left-handed neutrinos have a hierarchical
spectrum $m_1 \ll m_2 = m_{\rm sun} \ll m_3 = m_{\rm atm}\equiv |\Delta m_{\rm atm}^2|^{1/2}$.
In such a case $\tilde{m}_1 = m_{\rm atm}$ if
$N_1$ exchange gives rise to the atmospheric mass
splitting,
or $\tilde{m}_1\circa{>} m_{\rm sun}$  if $N_1$ gives
rise to the solar mass splitting.
A smaller  $\tilde{m}_1\ge m_1$ can be obtained if $N_1$ gives rise to $m_1$,
which can be arbitrarily small.
A $\tilde{m}_1$ larger than $m_{\rm atm}$ can be obtained if $N_{2,3}$
exchange cancels out the $N_1$ contribution to neutrino masses.
Stronger (weaker) restrictions are obtained if there are less (more) than 3 right-handed neutrinos.

In conclusion, measuring neutrino masses does not fix
$\tilde{m}_1$ and $m_{N_1}$,
which remain as free parameters.
Therefore we compute $\eta$ as function of $m_{N_1}$ and of
$\tilde{m}_1$ renormalized at the high scale $m_{N_1}$
(at high scales $\tilde{m}_1$ is about $(20\div 30)\%$ larger than at low energy).

\medskip

\begin{figure}[t]
\centerline{
\includegraphics{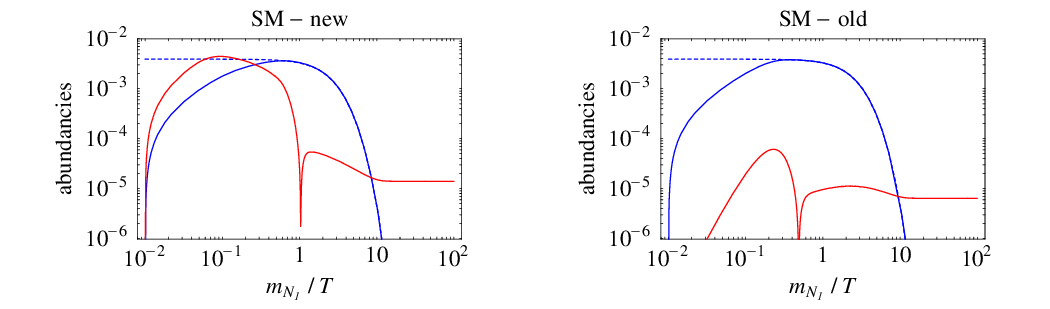}
}
\caption{\em \label{fig3}
Evolution of $Y_{N_1}$ (blue curves) and
$|Y_{B-L}/\epsilon_{N_1}|$ (red curves) with temperature
in the SM. We fix  $m_{N_1}=10^{10}$ GeV and $\tilde
m_1(m_{N_1})=0.06\eV$. The dashed line shows
the thermal abundance of $Y_{N_1}$.
Left plot: full computation, the efficiency is $\eta=0.0036$.
Right plot: no new effect included (and on-shell scatterings
incorrectly subtracted), $\eta=0.0017$. 
\vspace*{0.5cm}}
\end{figure}

\begin{figure}[t]
$$\hspace{-6mm}\includegraphics[width=17.5cm]{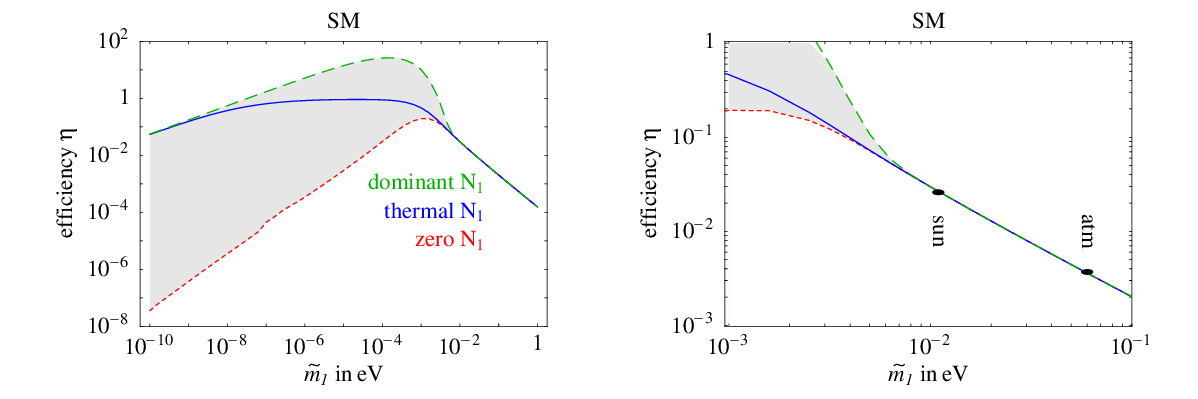}$$
$$\hspace{-8mm}\includegraphics[width=17.5cm]{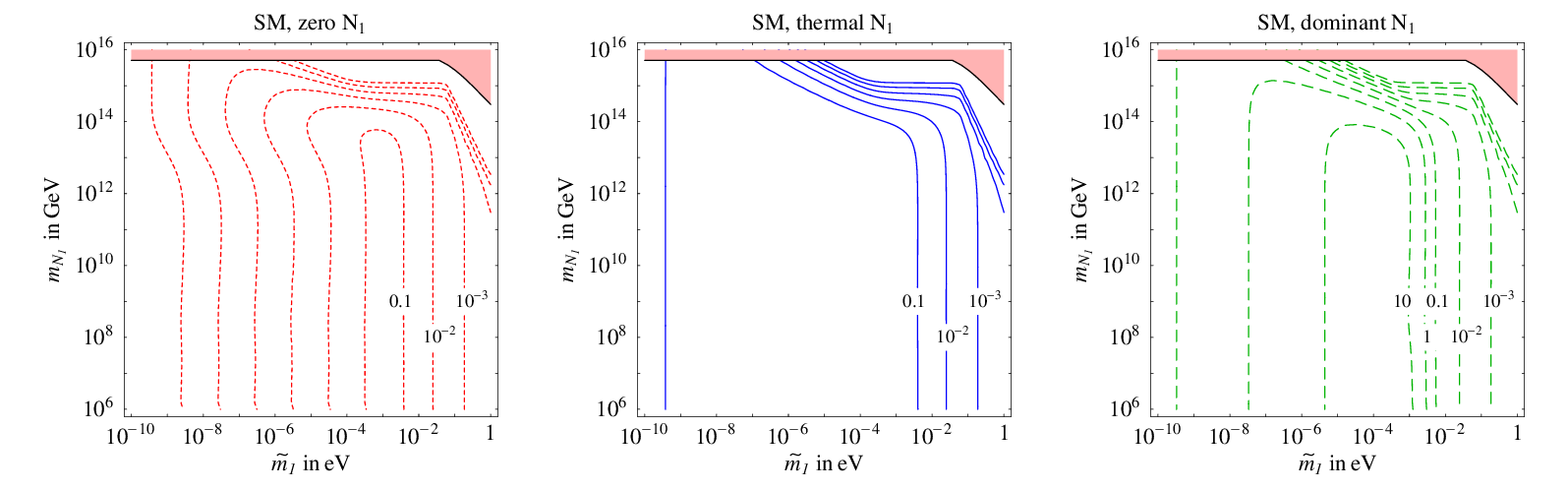}$$
\caption{\em
{\bf Efficiency $\eta$ of leptogenesis} in the SM,
 assuming zero (dashed red line), thermal (continuous blue line) or dominant (long dashed green line)
 initial $N_1$ abundancy. 
Upper plots: $\eta$ as function of
$\tilde{m}_1$ (renormalized at $m_{N_1}$) for $m_{N_1} = 10^{10}\GeV$.
Lower plots: contours of $\eta(\tilde{m}_1,m_{N_1}) = 10^{-6,-5,\ldots,0,1}$.
In the shaded regions the neutrino Yukawa couplings are non-perturbative.
\label{figSM}}
\end{figure}

\subsubsection{Results}
Figure~\ref{fig3} shows the evolution of the $N_1$ and ${\cal B}-{\cal L}$ abundances
at our sample `atmospheric' point:
$\tilde{m}_1(m_{N_1}) = r(\Delta m^2_{\rm atm})^{1/2} = 0.06\eV$
and $m_{N_1}=10^{10}\GeV$.
For these values the $N_1$ abundancy remains close to thermal equilibrium, so that 
leptogenesis is mainly determined only by the
later stages of the evolution at relatively small temperatures.
This explains why, despite the significant variations at higher temperature,
there is only a mild correction to the final baryon asymmetry.
Proper subtraction of on-shell scatterings
reduces wash-out by a $3/2$ factor.
This gives a $3/2$ increase of the efficiency,
as can be  seen from the analytical approximation of ref.~\cite{bcst}.

\medskip

We now present our results for the thermal
leptogenesis efficiency parameter $\eta$.
We assume no pre-existing ${\cal B}-{\cal L}$ asymmetry,
and we study the three cases of 
\begin{itemize}
\item[(0)] zero initial $N_1$ population, $Y_{N_1} = 0$ at $T\gg m_{N_1}$.
This case can be realized {\it e.g.}\ if an inflaton field
reheated the universe decaying mostly into SM particles.

\item[(1)]  thermal initial $N_1$ population, $Y_{N_1} = Y_{N_1}^{\rm eq}$  
at $T\gg m_{N_1}$.
This case can be realized in presence of extra interactions at $T\gg m_{N_1}$,
mediated {\it e.g.}\ by a heavy  $Z'$ boson related to SO(10) unification.

\item[($\infty$)] dominant initial $N_1$ abundancy, $\rho_{N_1} \gg \rho_R$  
at early times.
This case can be realized {\it e.g.}\ if an inflaton field
reheated the universe decaying mostly into $N_1$.
\end{itemize}
In order to study the latter case
we modified the Boltzmann equations employed
in previous analyses including the effects of $N_1$ reheating
(in the case (1) $N_1$ reheating is relevant only if
 $\tilde{m}_1\circa{<} 10^{-6}\eV$). 
 
If $m_{N_1}\ll 10^{14}\GeV$  
the efficiency parameter $\eta$ depends almost only on $\tilde{m}_1$.
In fig.~\ref{figSM} we show $|\eta|$ as 
function of $\tilde{m}_1$ for $m_{N_1}= 10^{10}\GeV$.
If $\tilde{m}_1> 10^{-2}\eV$ neutrino Yukawa couplings keep $N_1$ so 
close to  thermal equilibrium
that $\eta$ does not depend
on the unknown initial $N_1$ abundancy
(similarly, an eventual pre-existing lepton asymmetry would be washed out
if $N_1$ Yukawa couplings act on all flavours).

At smaller $\tilde{m}_1$ the efficiency $\eta$ depends on the initial $N_1$ abundancy,
ranging between the limiting cases (0) and ($\infty$), as illustrated by the gray band in fig.~\ref{figSM}.
As expected, the maximal value of $\eta\sim g_*$ 
is reached at $\tilde{m}_1\sim \tilde{m}_1^* \equiv
 {256\sqrt{g_*} v^2}/{3M_{\rm Pl}}=2.2\times 10^{-3}\eV$ in case ($\infty$).
In such a case, $\eta$ decreases at $\tilde{m}_1\ll \tilde{m}_1^*$, because
$N_1$ decays out-of-equilibrium at temperature
$T_{\rm RH}^{N_1}\sim m_{N_1} \sqrt{\tilde{m}_1/\tilde{m}_1^*} \ll m_{N_1}$ 
so that
$N_1$ reheating washes out some lepton asymmetry.
In more physical terms, the particles $H,L$ emitted in $N_1$ decays have energy larger than the temperature $T$,
and split up in $\sim m_{N_1}/T_{\rm RH}^{N_1}$ particles without correspondingly increasing the
lepton asymmetry, so that $\eta \sim g_* \sqrt{\tilde{m}_1^*/\tilde{m}_1}$.

When $\tilde{m}_1\circa{<}10^{-6}\eV$,
$N_1$ reheating starts to be significant even in case (1) giving $\eta <1$.
In fact, even if $N_1$ initially has a thermal abundancy  
$\rho_{N_1}/\rho_R\sim  g_{N_1}/g_* \ll 1$,
its contribution to the total density of the universe becomes no longer negligible,
$\rho_{N_1}/\rho_R\sim  (g_{N_1} m_{N_1})/(g_\star T)$,
if it decays strongly out of equilibrium at $T\ll m_{N_1}$.
For the reasons explained above, this effect gives a suppression of $\eta$ (rather than an enhancement), and for very small $\tilde{m}_1$ the case (1) and 
$(\infty)$ give the same result.

 \medskip

The lower panel of fig.~\ref{figSM} contains our result for
the efficiency $|\eta|$ of thermal leptogenesis computed in cases (0), (1) and $(\infty)$ as function
of both $\tilde{m}_1$ and $m_{N_1}$.
At $m_{N_1}\circa{>}10^{14}\GeV$ non-resonant $\Delta L = 2$ scatterings
enter in thermal equilibrium strongly suppressing $\eta$.
Details depend on unknown flavour factors.


Our results in fig.~\ref{figSM} can be summarized with
simple analytical fits
\begin{equation}
\frac{1}{\eta}\approx  \frac{3.3\times 10^{-3}\eV}{\tilde{m}_1} +   \bigg(\frac{\tilde{m}_1}{0.55\times 10^{-3}\eV}\bigg)^{1.16}\qquad
\hbox{in case (0)}
\end{equation}
valid for $m_{N_1}\ll 10^{14}\GeV$. 
This enables the 
reader to study leptogenesis 
in neutrino mass models without setting up and solving the
complicated 
Boltzmann equations.

\medskip

\begin{figure}[t]
$$\includegraphics[width=0.98\textwidth]{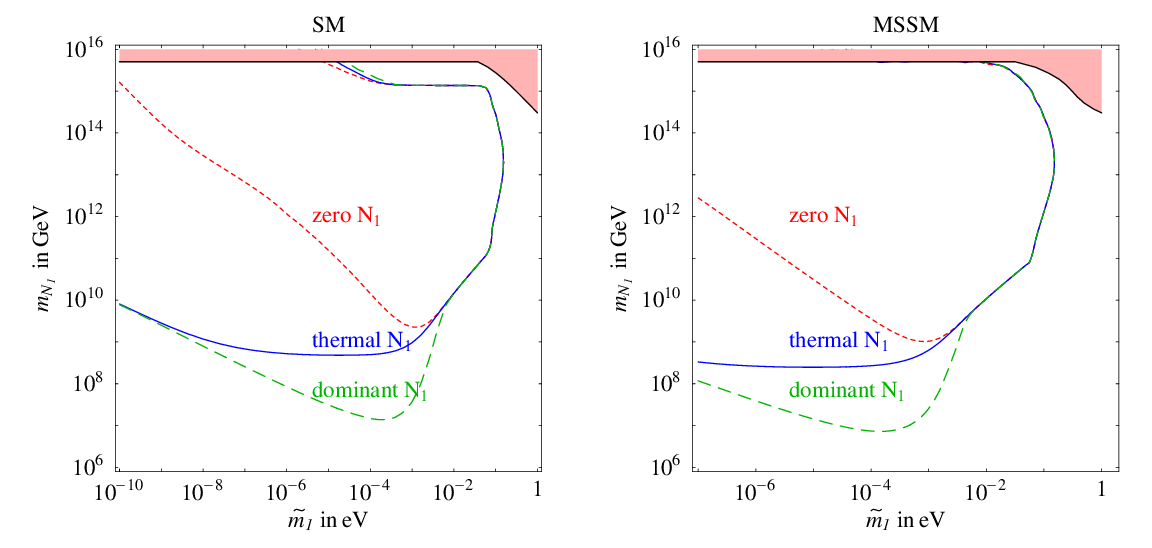}$$
\caption{\em\label{fig:Mm}
{\bf Allowed range of $\tilde m_1$ and $m_{N_1}$} for
leptogenesis in the SM and MSSM assuming $m_3 = \max(\tilde{m}_1, m_{\rm atm})$
and $\xi = m_3/\tilde{m}_1$.
Successful leptogenesis is possible in the area inside the
curves (more likely around the border).}
\end{figure}

\begin{figure}[t]
$$\includegraphics[width=0.98\textwidth]{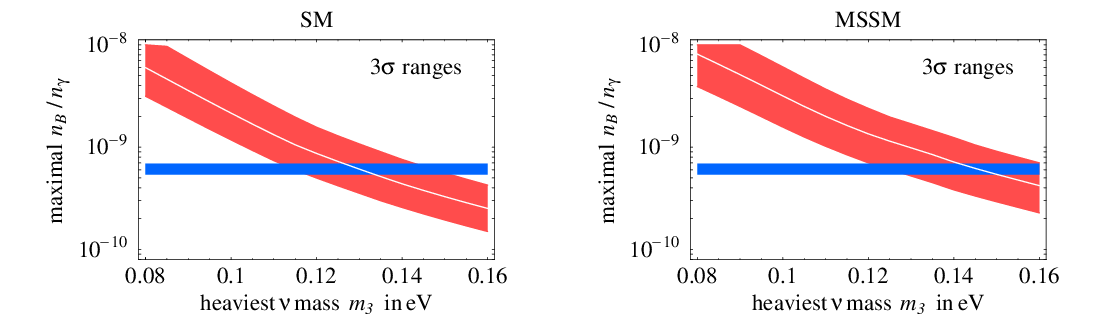} $$
\caption{\em\label{fig:m3}
{\bf Leptogenesis bound on neutrino masses}.
The plot shows the measured baryon asymmetry (horizontal line)
compared with the maximal leptogenesis value as function of
the heaviest neutrino mass $m_3$, renormalized at low energy.
Error bars are at $3\sigma$.}
\end{figure}

\subsubsection{Implications}
Experiments have not yet determined the mass $m_3$ of the heaviest mainly
left-handed neutrino.
We assume $m_3 = \max(\tilde{m}_1, m_{\rm atm})$.
Slightly different plausible assumptions are possible when $\tilde{m}_1\approx m_{\rm atm}$,
and very different fine-tuned assumptions are always possible.
 As discussed above, we assume $\xi = m_3/\tilde{m}_1$
(the parameter $\xi$ controls $\Delta L=2$ scatterings mediated by $N_{2,3}$).
These assumptions do not affect the absolute bounds 
on the masses of left-handed and right-handed neutrinos that we now discuss,
but allow to present them in one simple plot, fig.~\ref{fig:Mm}.

The crucial assumption behind fig.~\ref{fig:Mm}
is that right-handed neutrinos are very hierarchical.
Under this hypothesis the CP asymmetry is bounded by the expression given in~\cite{epsilon}
(see also~\cite{di2,mBound}), that in the 
hierarchical and  quasi-degenerate light neutrino limits simplifies to
\begin{equation}
\label{eq:di}
|\epsilon_{N_1}|\le \frac{3}{16\pi}
\frac{m_{N_1} (m_3-m_1)}{v^2}  \times
\left\{\begin{array}{ll}
1-m_1/\tilde{m}_1 & \hbox{if $m_1\ll m_3$}\\
\sqrt{1- m_1^2/\tilde{m}_1^2}&
\hbox{if $m_1\simeq m_3$}
\end{array}\right. .
\end{equation}
where all parameters are renormalized at the high-energy scale $\sim m_{N_1}$
and $m_3^2 = m_1^2 + \Delta m^2_{\rm atm} + \Delta m^2_{\rm sun}$.
The $3\sigma$ ranges of $m_{\rm atm}$ and of $n_{\cal B}/n_\gamma$ imply
the lower bound
\begin{equation}
m_{N_1} > \frac{4.5\times 10^{8}\GeV}{\eta} > \left\{\begin{array}{rl}
2.4\times 10^{9} \GeV    &   \hbox{in case (0)} \\
 4.9\times 10^{8}\GeV     & \hbox{in case (1)} \\  
1.7\times 10^{7}\GeV     & \hbox{in case ($\infty$)} 
\end{array}\right.
\label{pipp} 
\end{equation}
with the absolute bound realized in case ($\infty$).

The allowed regions are shown in fig.~\ref{fig:Mm} as function of $\tilde{m}_1$.
The bound on $m_{N_1}$ becomes stronger if left-handed neutrinos
are heavier than what suggested by oscillations, $m_3 > m_{\rm atm}$,
until thermal leptogenesis can no longer generate the observed $n_{\cal B}$ (see fig.~\ref{fig:Mm})
giving an upper bound on the mass of degenerate neutrinos
(renormalized at some unspecified scale~\cite{SMRGE2})
of about $0.1\eV$~\cite{mBound}.
This  happens because $\eta$ decreases with $\tilde{m}_1$ in the region
of interest and because the bound in \eq{eq:di} on $\epsilon_{N_1}$ 
becomes stronger when neutrinos become heavier.\footnote{The bound on neutrino masses is saturated around
$m_{N_1}\sim 10^{13}\GeV$ and $\tilde{m}_1\sim 0.1\eV$.
For these values computing leptogenesis in `one flavour approximation' is reliable~\cite{epsilon}.
We recall that we are assuming that the Higgs quartic coupling remains relatively small up to high energies.}

In order to study precisely the bound on neutrino masses we
relax our simplifying assumptions on $\xi$ and $m_3$ and compute
the maximal baryon asymmetry generated by thermal leptogenesis as function of $m_3$.
The results is shown in  fig.~\ref{fig:m3}:
including the new effects discussed in this paper
and combining errors  on $\Delta m^2_{\rm atm}$ and on the baryon asymmetry in quadrature
we get  $m_\nu <0.15\eV$ at $99.73\%$ CL (i.e.\ $3\sigma$).
The small difference with respect to previous results~\cite{mBound} is due to various factors:
correct subtraction of on-shell scatterings
(makes leptogenesis $50\%$ more efficient and the bound on neutrino masses $7\%$ weaker);
renormalization of neutrino masses (makes the bound $7\%$ stronger);
renormalization of $\lambda_t$, inclusion of gauge scatterings, of thermal corrections,
updated experimental determination of $\Delta m^2_{\rm atm}$ and of $n_{\cal B}/n_\gamma$,
revised (weaker) upper bound on the CP-asymmetry~\cite{epsilon}.

\medskip

We stress that the bound in 
\eq{eq:di} on the CP-asymmetry holds because we assumed
$m_{N_1}\ll m_{N_{2,3}}$: if
right-handed neutrinos were instead  quasi-degenerate 
CP violation in mixing can give an arbitrarily large CP-asymmetry, $\epsilon_{N_1}\sim 1$
and all bounds that we discussed evaporate.
In particular the leptogenesis bound on the neutrino masses holds under the 
dubious assumption 
that hierarchical right-handed neutrinos
give quasi-degenerate left-handed neutrinos.

\subsection{Simple approximation} \label{sappr}
While in our calculations we use our full code,
 we now also discuss how the computation can be
significantly simplified
by including only the following new ingredients 
which,  a posteriori, turn out to be numerically most relevant.
Neglecting all these effects we reproduce the results in ref.~\cite{bbp}.
Effects 2, 3, 4 were included in ref.~\cite{bcst}:
neglecting the other ones we reproduce their results.
Effect 2 was mentioned in ref.~\cite{SMRGE2}.
\begin{itemize}
\item[1] Proper subtraction of the on-shell $N_1$ propagators
in $\Delta L=2$ scattering processes, as discussed in Appendix A.
\item[2] Neutrino masses have to be renormalized at the
proper energy scale $\sim m_{N_1}$, 
both when computing $\eta$ and $\epsilon_{N_1}$.
\end{itemize}
Typically these effects give ${\cal O}(1)$ corrections.
For $\tilde m_1\gsim 10^{-3}$ eV the efficiency 
increases by up to almost a factor 2,
due to the suppression of $\Delta L = 1$ wash-out processes
caused by the following thermal effects:
\begin{itemize}
\item[3] Temperature corrections to the Higgs boson mass,
$m_H \sim 0.4~ T$, 
must be included
at least when computing the IR-enhanced $LN\to Q_3 U_3$
interaction rate.
\item[4]  The top Yukawa coupling must be renormalized at
the proper energy scale $\sim m_{N_1}$.
\end{itemize}
These variations are partially compensated by
\begin{itemize}
\item[5] Inclusion of previously neglected scatterings involving weak gauge bosons $A$,
$N_1 L\leftrightarrow HA$, $N_1 H\leftrightarrow LA$ and $N_1 A\leftrightarrow HL$. 
These extra scatterings are sizable because $g_2>\lambda_t$  at energies above $10^9\GeV$:
$\lambda_t$ is no longer the dominant coupling. These processes have been
recently considered for the time in ref.~\cite{pilaf}
\end{itemize}
Including these contributions one gets an
excellent approximation for $\tilde m_1\gsim
10^{-3}\eV$.
In order to get an approximation which is accurate also
at $\tilde m_1 \ll 10^{-3}\eV$,
in case (0) one needs to include one more effect:
\begin{itemize}
\item[6] Thermal corrections to the CP-violating parameter
$\epsilon_{N_1}$.
\end{itemize}
which turns out to have a sizable impact due to a more
subtle reason.
%


Neglecting washout scatterings (which are small at $\tilde{m}_1\ll 10^{-3}\eV$)
and the temperature dependence of the CP asymmetry,  in case (0)
the Boltzmann equations are solved by $Y_{{\cal B}-{\cal L}} = + \epsilon_{N_1} Y_{N_1}$.
In this approximation the lepton asymmetry generated in inverse-decay
processes at $T\gg m_{N_1}$
when $Y_{N_1} < Y_{N_1}^{\rm eq}$ is exactly  cancelled by
the asymmetry generated later in $N_1$ decays when 
$Y_{N_1} >Y_{N_1}^{\rm eq}$.
Consequently  the effect which dominantly breaks this
cancellation has a numerically important impact even if it is `small'.
$1)$
Washout interactions, dominated by $\gamma_D$,  erase the lepton asymmetry generated at earlier stages more than the one generated at later stages,
giving a small positive $\eta$.
$2)$ 
Thermal corrections to $\epsilon$ slightly reduce it at small temperatures $T\sim m_{N_1}/4$,
giving a small positive $\eta$.\footnote{The solid line 
in fig.~\ref{figepsilon}a
shows our most accurate result for 
 thermal corrections to $\epsilon$,
 that we employ in numerical computations.
 The enhancement at $T\sim 0.4 m_{N_1}$ comes from the quantum statistics factor,
$1+f_H-f_L-2f_Hf_L$, that can be larger than one
when both  thermal masses 
and $N_1$ motion with respect to the plasma
are taken into account.
In case (0) for $\tilde{m}_1 \ll 10^{-3}\eV$ the $N_1$ energy spectrum 
deviates from the thermal Fermi-Dirac distribution
that we assumed.
Using the slightly less accurate thermal correction to $\epsilon$,
obtained neglecting thermal motion of $N_1$
(dashed line in fig.~\ref{figepsilon}a)
typically affects the final result by a ${\cal O}(1)$ factor.
Other minor effects might be important:
the exact dispersion relation at finite temperature,
thermal corrections to couplings, higher order corrections,
and 
CP-violation in $\Delta L = 1$ scatterings.}


Finally, if one wants to study cases where $N_1$ can give a substantial
contribution to the total energy density,
one must include this correction in the Boltzmann equations,
as described in appendix~\ref{Boltz}.

\begin{figure}[t]
$$\hspace{-6mm}\includegraphics[width=17.5cm]{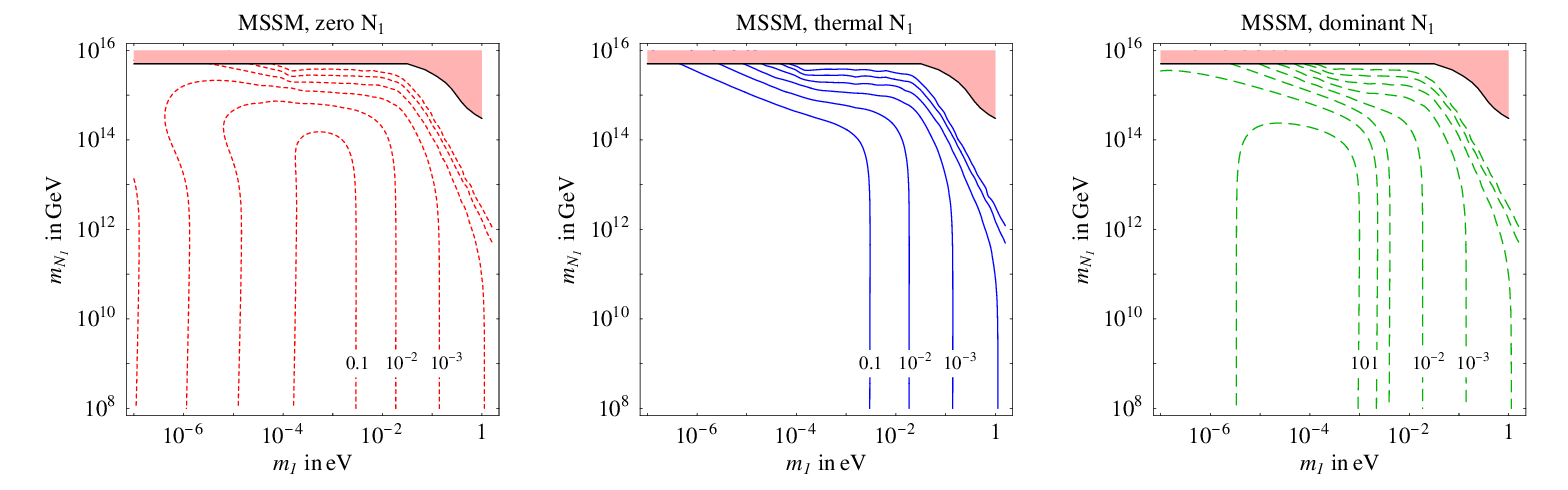}$$
\caption{\em
{\bf Efficiency ${\eta}$ of leptogenesis} in the MSSM,
 assuming zero, thermal or dominant initial $N_1,\tilde{N}_1$ abundancy. 
 In the shaded regions the neutrino Yukawa couplings are non-perturbative.
\label{figMSSM}}
\end{figure}

\section{Leptogenesis in the MSSM}\label{MSSM}
In the case of the supersymmetric extension of the SM,
the computation becomes more involved because of the presence of many
new particle
degrees of freedom.
In particular, a lepton asymmetry is generated in decays of both
the right handed neutrino $N_1$ and the right-handed
sneutrino $\tilde{N}_1$.
Since supersymmetry breaking can be ignored (except in special 
cases~\cite{softl2,softl}), the
computation
is somewhat simplified by the identities demanded by
supersymmetry,
such as
$m_{N_1} = m_{\tilde{N}_1}$ and
$\Gamma_{N_1} = \Gamma_{\tilde{N}_1}$.
Thermal corrections break supersymmetry, so that it is rather ponderous
to include them. At the moment, we do not attempt to make a full calculation
in the supersymmetric case.

On the other side, a full computation might be not
necessary.
Including only the thermal effects which in the SM
turn out to be dominant as discussed in section~\ref{sappr}, 
could be a good approximation
also for the MSSM.
Therefore we do not compute all relevant
cross sections including finite temperature effects,
but we adopt the MSSM cross sections of ref.~\cite{k-mssm}, 
inserting the
temperature-dependent
top Yukawa coupling and Higgs boson mass in IR-enhanced processes,
and performing a correct subtraction of on-shell resonances
(which again reduces $N_1$-mediated washout scatterings by a 3/2 factor).
Thermal corrections to $N_1$ and $\tilde{N}_1$ decays and
their CP-asymmetries 
are computed in appendix~\ref{CPMSSM} neglecting the thermal motion
of $N_1$, $\tilde{N}_1$
(in the SM case this would not be a very good approximation
at small $\tilde{m}_1$). 
Finally, 
we do not include $\Delta L =1$ scatterings involving
gauge bosons and gauginos.

Our MSSM results have been obtained under these approximations,
assuming moderately large values of $\tan\beta\sim 10$.
Low-energy thresholds make the top Yukawa coupling at
high energy uncertain
by about a factor of 2, as discussed in ref.~\cite{heavym0}.

Proceeding as in the case of the SM, we find that the asymmetry generated
in the MSSM is
\begin{equation}
Y_{{\cal B}-{\cal L}}=- \eta\, \epsilon_{N_1}
(Y^{\rm eq}_{N_1}+Y^{\rm eq}_{\tilde N_1})(T\gg m_{N_1}),
\label{eq:susyeff}
\end{equation}
where $\epsilon_{N_1}$ is the neutrino (or sneutrino) CP-asymmetry 
at low temperature
(equal for lepton and slepton final 
states)~\cite{Covi},
\begin{eqnarray}
\epsilon_{N_1}&=&\frac{1}{8\pi}\sum_{j\neq 1}
\frac{\textrm{Im}
\left[ (Y^{\dagger} Y)_{j 1}^2\right] }{\left[Y^{\dagger}
Y\right]_{11}}
g\left(\frac{m_{N_j}^2}{m_{N_1}^2}\right)  ,
\label{epsss}\\
g(x)&=&-\sqrt{x}\left[ \frac{2}{x-1}+\ln
\left(\frac{1+x}{x} \right)
 \right] \stackrel{x\gg 1}{\longrightarrow} - \frac{3}{
\sqrt{x}}  .
\end{eqnarray}
The number of effective degrees of freedom in the MSSM without right-handed
neutrinos is $g_*=228.75$.
Using eq. (\ref{eq:YB}) with $n_H=2$, we obtain
\begin{equation}
\label{uums}
\frac{n_{\cal B}}{s} =-1.48\times 10^{-3} \epsilon_{N_1} \eta.
\end{equation}
The MSSM results, analogous to those obtained from the SM and previously
discussed, shown in figs.~\ref{fig:Mm}--\ref{figMSSM},
 are similar to their corresponding SM results.
 It is difficult to compare with previous results,
 which have not been presented in a systematic way.
 
Extra signals may come from
lepton flavour violating decays like $\mu\to e \gamma$,
induced in supersymmetric see-saw models by the (unknown) neutrino Yukawa couplings~\cite{mueg}.
Some predictive minimal models allow to predict these rates
in terms of the measured neutrino masses and baryon asymmetry.
For example, using our revised leptogenesis computation,
 the prediction of ref.~\cite{minimalseesaw} for BR($\mu\to e\gamma$)
gets lowered by one order of magnitude.


\subsubsection{Soft leptogenesis}\label{soft}
``Soft leptogenesis''~\cite{softl2,softl} is a supersymmetric scenario of
leptogenesis which requires only one 
heavy right-handed neutrino. 
The interference between the CP-odd and
CP-even states of the heavy scalar neutrino resembles
very much the neutral kaon system. The mass splitting as well as the 
required CP violation in the heavy sneutrino system comes from the
soft supersymmetry breaking $A$ and $B$ terms,
respectively associated with the Yukawa coupling and mass term of $N_1$.
The non-vanishing value of the generated lepton asymmetry
is a pure thermal effect, since at $T=0$ the generated lepton asymmetry
in leptons exactly cancels the one in sleptons.

\begin{figure}[t]
\centerline{\includegraphics[width=0.45\textwidth]{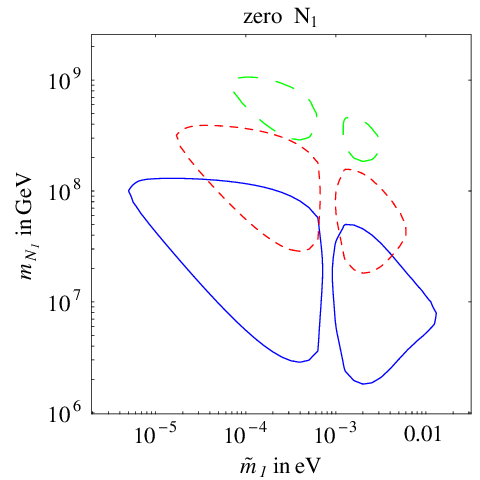}
\hfill
\includegraphics[width=0.45\textwidth]{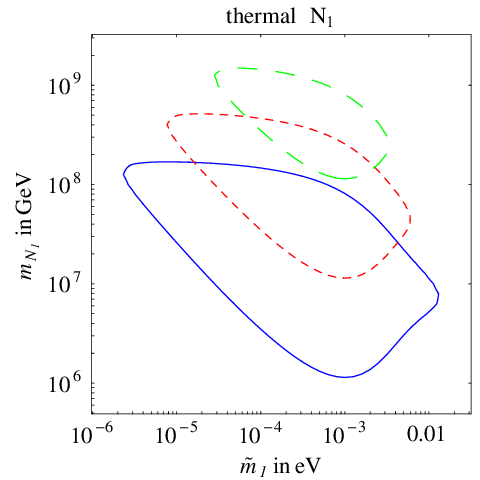}}
\caption{\em 
Regions of $(\tilde m_1,m_{N_1})$ plane where {\bf soft leptogenesis}
can produce the observed baryon asymmetry for ${\rm Im}A<{\rm TeV}$
and $\sqrt{B m_{N_1}}=100\GeV$ (solid line), $1\TeV$ (dashed),
$10\TeV$ (long-dashed)
We have assumed a vanishing 
(left) or thermal (right) initial
sneutrino density.}
\label{fig:soft}
\end{figure}

Here we improve the results of
ref.~\cite{softl}, taking into account
thermal, Pauli blocking and stimulated emission corrections to
$N_1,\tilde{N}_1$ branching ratios 
(which significantly enhance decays into bosons)
to their CP asymmetries 
(computed in appendix~\ref{CPMSSM} including thermal corrections, but
neglecting the thermal motion of $N_1,\tilde{N}_1$).
We recall that in this scenario $\epsilon_{\tilde N\to L \tilde H}$
has an opposite sign to  $\epsilon_{\tilde N\to \tilde L H}$, and~\cite{softl}
\beq
\epsilon_{\tilde N\to \tilde L H}(T=0) =
-\epsilon_{\tilde N\to L \tilde H}(T=0) = 
\frac{4\Gamma_{{\tilde N}_1} B }{4 B ^2 +\Gamma_{{\tilde N}_1}^2}\frac{{\rm Im} A}{m_{N_1}}
\label{epsilonsoft}
\eeq
We refer to~\cite{softl} for further details, and   present our improved results in fig.~\ref{fig:soft},
where we plot the regions in $(\tilde m_1,m_{N_1})$ plane
where successful leptogenesis is possible.
We have assumed vanishing (left) or thermal (right) initial sneutrino density,
${\rm Im}A<{\rm TeV}$
and $\sqrt{B m_{N_1}}=100\GeV$ (solid line) and $1\TeV$ (dashed line),
$10\TeV$ (long dashed line).
`Soft leptogenesis' needs a $B$-term much smaller than what suggested
by models of supersymmetry breaking, but not unnaturally small~\cite{softl}.
Soft leptogenesis allows to lower the bounds on $m_{N_1}$.

\section{Leptogenesis with reheating of the universe}\label{reh}
By now there is a wide consensus that the early
universe underwent a primordial
stage of inflation \cite{reviewinf} responsible for the
observed
homogeneity and isotropy of the present universe, as well as
for the
generation of the cosmological perturbations.

The radiation--dominated era of the universe is usually
assumed to be
originated by the decay of the coherent oscillations of a
scalar
field, the inflaton field, whose vacuum energy has driven
inflation
For such a reason the reheating
 process is often associated with the final stage of
inflation.  However, we point out that 
reheating could have been episodic, with several
reheat events after inflation. We will be interested in the
final
reheating  which may just as well
have been the result of the decay of a weakly coupled
scalar field
unrelated to inflation, for instance a modulus. For this
reason the
scalar field $\phi$, whose decay leads to reheating, will
not be
referred to as the inflaton.

The decay of the scalar field $\phi$ into light degrees of
freedom  and 
their subsequent
thermalization, called reheating, leaves the
universe at a temperature $T_{\rm RH}$, which represents
the largest
temperature of the plasma during the subsequent
radiation--dominated
epoch, when temperature is a decreasing function of time.
The onset of
the radiation dominated era is in fact placed at the
temperature
$T_{\rm RH}$, {\em i.e.} at the end of the reheating phase.

Usually $T_{\rm RH}$ is assumed to be very large and this
is the
assumption we have made in the previous sections.  However
the only
information we have on the smallest value of $T_{\rm RH}$
is from
requiring a successful period of primordial
nucleosynthesis,
$T_{\rm RH}\gsim 1$ MeV. Therefore, from a phenomenological
point of view,
$T_{\rm RH}$ is actually a free parameter\footnote{Low
reheating 
scenarios lead as well to a new
perspective on baryogenesis \cite{Davidson:2000dw}, to the
possibility that massive neutrinos may play the role of
warm dark
matter \cite{Giudice:2000dp} or to a change in the
predictions of the relic abundance and resulting model
constraints 
of supersymmetric dark
     matter, axions, massive neutrinos, and other dark
matter candidates
\cite{giudiceetal, fornengo}.}. Any scenario of
baryogenesis
based on the out-of-equilibrium decay of some heavy
particle depends
crucially on the assumption that these particles were
generated during
the reheating process  with abundances sufficiently large
to generate
the observed baryon asymmetry. 
During reheating, particles are generated
through thermal scatterings and quickly thermalize.  Among
them, 
right-handed neutrinos
 may be also produced but their  number depends
strongly on the reheating temperature. If the latter is too
small, the
thermal bath does not give rise to a number of right-handed
neutrinos large
enough to produce the observed baryon asymmetry. This leads
to a
lower limit on $T_{\rm RH}$. Computing this lower bound is
the goal
of this section.

During the reheating epoch, the energy
density of the universe is dominated by the coherent
oscillations of a
scalar field $\phi$. Considering for the moment the case of small
abundance of right-handed
neutrinos\footnote{The case of non-negligible $N_1$ density is
discussed in appendix A.},
as a first step we assume that
the dynamics of reheating is described by the Boltzmann
equations for the energy densities $\rho_{\phi,R}$ of the
two
coupled components: the unstable massive field $\phi$ and
the  radiation
$R$ \cite{book,ckr,giudiceetal}
\begin{eqnarray}
\frac{d \rho_{\phi}}{dt}&=&-3 H
\rho_{\phi}-\Gamma_{\phi}\rho_{\phi} \, ,
\label{eq:rho_phi}\\
\frac{d \rho_{R}}{dt}&=& -4 H \rho_R + \Gamma_{\phi}
\rho_{\phi} 
 \, , \label{eq:rho_R}
\end{eqnarray}
where $H = \dot a/a = \sqrt{8\pi (\rho_\phi  +
\rho_R)/3M_{\rm Pl}^2 }$, and
$M_{\rm Pl}$ is the Planck mass.
Here we have assumed  that the relativistic decay products
of the scalar
field rapidly thermalize and form a relativistic bath of
temperature
$T$ (for a discussion about this point see Ref.
\cite{ckr}).
The key point of our considerations is that reheating is
not an
instantaneous process. On the contrary, the
radiation-dominated phase
follows a prolonged stage of matter domination during which
the energy
density of the universe is dominated by the coherent
oscillations of
the field $\phi$.  The oscillations start at time
$H_I^{-1}$ and end
when the age of the universe becomes of order of the
lifetime
$\Gamma_\phi^{-1}$ of the scalar field. At times $H_I^{-1}
\lsim t
\lsim \Gamma_\phi^{-1}$ the dynamics of the system is quite
involved. During this stage the energy density per comoving
volume of
the $\phi$ field decreases as $\exp(-\Gamma_\phi t)$ and
the light decay products of the scalar field thermalize.
 The temperature $T$ of
this hot plasma, however, does not scale as $T \propto
a^{-1}$ as in
the ordinary radiation-dominated phase (where $a$ is the
Friedmann--Robertson--Walker scale factor) \cite{book,
ckr}, but
reaches a maximum $T_{\rm MAX} \sim (H_I
M_{\rm Pl})^{1/4}T_{\rm RH}^{1/2}$
and then decreases as $T \propto a^{-3/8}$, signalling the
continuous
release of entropy from the decays of the scalar field. 

In fact, until $t\ll  \Gamma_\phi^{-1}$ assuming $\rho_\phi
\gg \rho_R$ the system approximately evolves as
\begin{eqnarray}
\rho_\phi(t)& =& \rho_{\phi}(0) (a_0/a(t))^3
e^{-\Gamma_\phi t}\\
\label{eq:rhoR}
 \rho_R(t)\equiv  \frac{\pi^2g_{*}}{30} T^4
& \approx &  \frac{\sqrt{6/\pi}}{10} \Gamma_\phi M_{\rm Pl}
  \sqrt{\rho_\phi} 
  \left[1 - (a_0/a)^{5/2}\right] .
\end{eqnarray}
This scaling
continues until the time $t \sim \Gamma_\phi^{-1}$, when the
radiation-dominated phase starts with temperature $T\sim
T_{\rm RH}$, defined as
\begin{equation}
T_{\rm RH}= \left[ \frac{45}{4\pi^3 g_*\left(T_{\rm
RH}\right)}\,\Gamma_\phi^2
M_{\rm Pl}^2\right]^{1/4}  .
\end{equation}
The process is described by two extra parameters,
$\rho_{\phi}(0)$ and $\Gamma_\phi$.
It is convenient to replace them with the maximal and
reheating temperatures,  $T_{\rm MAX}$ and $T_{\rm RH}$.
Before reheating is completed, at a given
temperature, the universe expands faster than in the
radiation-dominated phase.  Notice that $T_{\rm RH}$ is not
the maximum
temperature during the reheat process. On the contrary,
$T_{\rm MAX}$
can be much larger than $T_{\rm RH}$.

\medskip

During reheating right-handed neutrinos
may be produced in several
ways. They can be generated directly
through the scalar field  perturbative decay process
\cite{infla} (this
requires that the mass of the $\phi$-field is larger
than $m_{N_1}$) or
 through nonperturbative processes taking place at the
preheating stage \cite{np,np2}. These mechanisms, however,
introduce
new unknown parameters such as the coupling of right-handed
neutrinos
to $\phi$. In this section we take a more conservative
point of view and we limit ourselves to the
case in which right-handed neutrinos are produced 
by thermal scatterings during the reheat stage,
so that in the limit $T_{\rm RH}\gg m_{N_1}$ we obtain the case (0) studied in section~\ref{SM}
(in the opposite case of dominant inflaton decay into $N_1$,
 the $T_{\rm RH}\gg m_{N_1}$ limit  is given by case ($\infty$) of section~\ref{SM}).
For the sake
of concreteness we will focus on the leptogenesis scenario,
but our
findings can be easily generalized to any
out-of-equilibrium scenario. Furthermore, we assume that the mass of the
inflaton field is larger than the reheating temperature
$T_{\rm RH}$; for a discussion of the opposite case, see
ref.~\cite{knr}.

\medskip

We now generalize the Boltzmann equations for thermal
leptogenesis including reheating.
In the standard case it is convenient to write the
 Boltzmann equations that dictate
the time evolution of the number densities $n_X(t)$ of any
species $X$
in terms of $Y_X(z)\equiv n_X/s$ where $z\equiv m_N/T$.
In fact, while particle densities $n_X$ strongly depend on
$t$ because of the expansion of the universe,
their ratios with respect to the entropy density $s$ remain
constant in the absence of interactions.
Since the temperature $T$ is a monotonic decreasing function, one 
usually replaces the time $t$ with $T$.

These two statements no longer hold during reheating.
Nevertheless, we still find convenient to write the
Boltzmann equations for leptogenesis in terms of  $Y_X(z)$.
The first pre-heating phase (when $T$ grows from zero to $T_{\rm MAX}$)
is so fast that it gives
no contribution to leptogenesis: the interesting dynamics
happen during the second reheating  phase,
when the temperature decreases in a non standard
characteristic way, $T\propto a^{-3/8}$.
Therefore {\em corrections to leptogenesis are fully
described by a single parameter, 
the reheating temperature $T_{\rm RH}$}, as long as $T_{\rm MAX}$ is sufficiently
larger than $m_{N_1}$.

Since the temperature is defined in
terms of the 
radiation density 
by eq.~(\ref{eq:rho_R}), we can write
\begin{equation}
\label{eq:d/dt}
\frac{d}{dt} =  - 4 H Z \rho_R  \frac{d}{d\rho_R} = 
HZz  \frac{d}{dz}\qquad
Z \equiv 1 - \frac{\Gamma_\phi \rho_\phi}{4 H \rho_R} =-\frac{a}{T}
\frac{dT}{da}.
\end{equation}
$Z$ vanishes when the maximal temperature $T_{\rm MAX}$ is reached, then
$Z\simeq 3/8$ during reheating, and finally $Z\simeq 1$ in the
standard radiation-dominated phase.
Apart from this ${\cal O}(1)$ correction, reheating affects
leptogenesis in 
two main ways
1) $H$ has a non-standard expression:  $\rho_\phi$ induces
a faster expansion
2) $\phi$ decays create additional matter.
The Boltzmann equations for leptogenesis are explicitly written
in appendix~\ref{Boltz}, eq.~(\ref{sys:Boltz}).

\begin{figure}[t]
$$\includegraphics[width=0.95\textwidth]{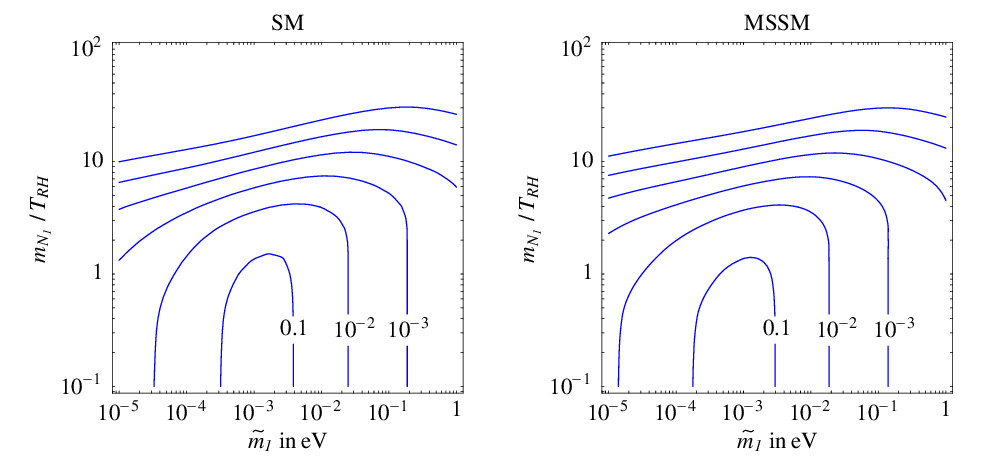}$$
\caption{\em
Isocurves of the efficiency parameter
$\eta=10^{-6,-5,\ldots,-1}$ for leptogenesis with
reheating as function of $(\tilde{m}_1, m_{N_1}/T_{\rm RH}$) in
the SM (left) and MSSM (right)
for $m_{N_1} = 10^{10}\GeV$ (the plot would be only slightly different for any $m_{N_1}\ll 10^{14}\GeV$).
\vspace*{0.5cm}}
\label{fig4}
\end{figure}

\subsubsection{Results}
Figure~\ref{fig4} shows $\eta$ as function of
$\tilde{m}_1$ and of $m_{N_1}/T_{\rm RH}$.
Although it has been obtained for $m_{N_1}=10^{10}\GeV$,
other values of $m_{N_1} < 10^{14}\GeV$ would
essentially give the same result.

We see that the final baryon asymmetry is
strongly suppressed if $T_{\rm RH}\ll m_{N_1}$.
This is due to entropy release from inflaton decays,
which gives a $\sim(m_{N_1}/T_{\rm RH})^5$ suppression of $\eta$.
Furthermore during reheating the universe expands
faster than during the standard thermal phase:
$H/H_{\rm standard} \approx (T/T_{\rm RH})^2$.
This makes both $N_1$ production and washout less efficient,
increasing the value of  $\tilde{m}_1$ at which
leptogenesis is maximally efficient,
as can be seen in fig.~\ref{fig4}.

\begin{figure}[t]
\centerline{\includegraphics[width=0.9\textwidth]{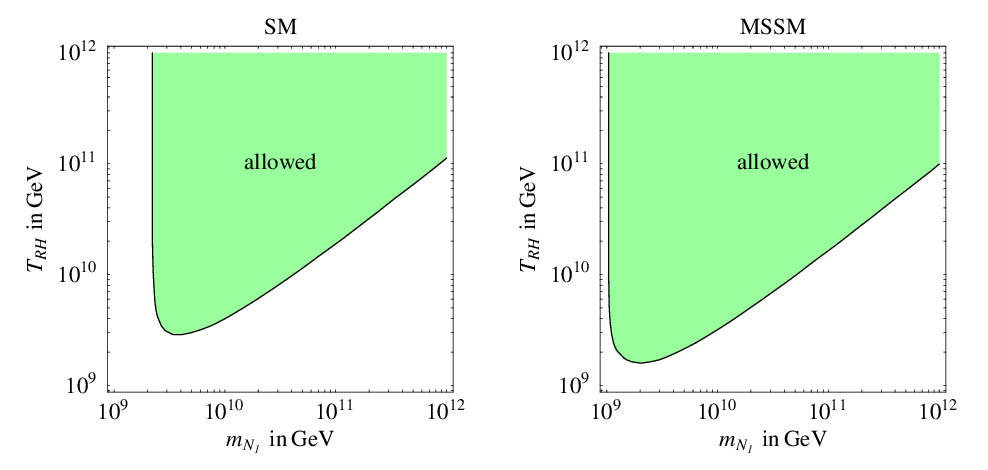}}
\caption{\em 
Lower bounds on $m_{N_1}$ and $T_{\rm RH}$ from
leptogenesis in the SM and MSSM.}
\label{fig5}
\end{figure}

Making use of the bound in \eq{eq:di}~\cite{di2,mBound,Treh}
and marginalizing over $\tilde{m}_1$
we can therefore derive the bound on the reheating temperature
shown in fig.~\ref{fig5}.
This bound holds assuming that thermal leptogenesis
generates the observed baryon asymmetry,
that the 
inflaton decays into SM particles rather than into right-handed neutrinos,
and that right-handed neutrinos are
hierarchical, $m_{N_1}\ll m_{N_{2,3}}$.
In the case of the SM, the lowest reheating temperature allowed
for successful leptogenesis turns out to be 
$2.8\times 10^9$ GeV, while in the case of the MSSM the lowest 
value is $1.6\times 10^9$ GeV. However, 
there are various reasons to suspect that the reheating temperature is
small in
locally supersymmetric theories. 
Indeed, gravitinos
(and other dangerous relics like moduli fields) are produced during reheating.
Unless reheating is delayed, gravitinos will be overproduced, leading to a
large undesired entropy production when they decay after big-bang
nucleosynthesis \cite{ellis}. The limit from gravitino overproduction is
$T_{\rm RH} \lsim 10^{9}$ to $10^{10}$GeV, or even stronger \cite{nucleo}.
This upper bound is at odds with the lower bound we have computed
to achieve successful leptogenesis. 
Fig.~\ref{fig:Mm}b shows that
this conflict can be avoided if $N_1$ and/or $\tilde{N}_1$
decayed while giving a substantial contribution to the total energy density of the universe 
and $\tilde{m}_1\sim\tilde{m}_1^*$.
In such a situation leptogenesis depends on the precise value of the initial
$N_1,\tilde{N}_1$ abundancy, unless it is dominant.
It can be realized if the inflaton decays dominantly into right-handed (s)neutrinos,
or if $\tilde{N}_1$ acquires a large vacuum expectation value, 
as discussed in the next subsection.
An alternative solution to solve the gravitino problem, maintaining a thermal
abundance of $\tilde{N}_1$, is to rely on ``soft leptogenesis'' \cite{softl}. 

\medskip

We can further elaborate on the results presented in figures~\ref{fig4}, \ref{fig5}
by making simple analytical approximations. Since we are interested in the
effects of reheating, we consider the case $m_{N_1}>T_{\rm RH}$, and since we
are studying lower bounds on $T_{\rm RH}$, we restrict ourselves to the most
favorable case in which $T_{\rm MAX}>m_{N_1}$.

The efficiency factor $\eta$
receives three kinds of contributions,
\begin{equation}
\eta=\eta_{\rm ab}\eta_{\rm eq}\eta_{\rm RH}.
\label{etaeq}
\end{equation}
Here $\eta_{\rm ab}$ measures the $N_1$ abundance before decay, relative to
the equilibrium density. In order to estimate it, we first define
\begin{equation}
K(T)=\frac{\Gamma}{H},
\end{equation}
where $\Gamma$ is the $N_1$ decay rate ($\Gamma =(G_F \tilde m_1 m_{N_1}^2)/(2
\sqrt{2} \pi )$ at $T\ll m_{N_1}$ and $\Gamma =(G_F \tilde m_1 m_{N_1}^3)/(2
\sqrt{2} \pi T )$ at $T\gsim m_{N_1}$) and $H$ is the Hubble constant, with
$H=[5\pi^3g_*^2(T)]/[9g_*(T_{\rm RH})]^{1/2} T^4/(T_{\rm RH}^2M_{\rm Pl})$.
Under the assumption\footnote{The expression we are using for $\Gamma$ is
not correct at high temperatures, where the Higgs decay is the relevant 
process. However, for this qualitative discussion, the approximation is
adequate, since 
the right-handed production 
is dominated at temperatures $T\sim  m_{N_1}$, where we can take 
$\Gamma =(G_F \tilde m_1 m_{N_1}^3)/(2
\sqrt{2} \pi T )$.}
that the right-handed neutrino density $n_{N_1}$ is
much smaller than the equilibrium density $n_{N_1}^{\rm eq}$, 
and taking the inverse decay
as the dominant production process, in the relativistic limit
we obtain~\cite{giudiceetal}
\begin{equation}
\frac{d(n_{N_1}/T^8)}{dT}=-\frac{8}{3}K\frac{n_{N_1}^{\rm eq}}{T^9}.
\end{equation}
Using $K\propto T^{-5}$ and $n_{N_1}^{\rm eq}\propto T^3$, we find
\begin{equation}
\frac{n_{N_1}}{n_{N_1}^{\rm eq}}=\frac{4}{15}K .
\label{neqneq}
\end{equation}
Next, we define
\begin{equation}
K_*\equiv K(m_{N_1})=\frac{\tilde m_1}{3\times 10^{-3}~\eV}\left(
\frac{T_{\rm RH}}{m_{N_1}}\right)^2.
\end{equation}
If $K_*\gg 1$, the right-handed neutrinos reach the equilibrium density 
before they become non-relativistic and $\eta_{\rm ab}=1$. If $K_*\ll 1$,
from eq.~(\ref{neqneq}) we obtain $\eta_{\rm ab}=(4/15)K_*$.

The next coefficient in eq.~(\ref{etaeq}) is $\eta_{\rm eq}$, which measures 
the out-of-equilibrium condition at decay. If $K_*\ll 1$, the right-handed
neutrino is decoupled when it becomes non-relativistic, and $\eta_{\rm eq}=1$.
If $K_*\gg 1$, $\eta_{\rm eq}$ can be estimated by computing the $N_1$ density
at the temperature $T_f$ at which the processes that damp the baryon asymmetry
go out of equilibrium,
\begin{equation}
\eta_{\rm eq}= \frac{\sqrt{2\pi}}{3\zeta (3)} 
\left( \frac{m_{N_1}}{T_f}\right)^{3/2}e^{-m_{N_1}/T_f}.
\end{equation}
We assume that the dominant process erasing the asymmetry is the inverse decay,
with $\Gamma_{\rm ID}=(m_{N_1}/T)^{3/2}\exp (-m_{N_1}/T)G_F\tilde m_1 m_{N_1}^2/(8\sqrt{
\pi})$. Then $T_f$ is given by the condition
\begin{equation}
\left. \Gamma_{\rm ID}= H\right|_{T=T_f}.
\label{condit}
\end{equation}
If $T_f>T_{\rm RH}$, eq.~(\ref{condit}) corresponds to $K_*(m_{N_1}/T_f)^\beta
\exp(-m_{N_1}/T_f)\simeq 1$, with $\beta=11/2$ which, in the range $1\ll K_*
<10^4$ is approximately solved by $m_{N_1}/T_f \simeq a (\ln K_*)^b$,
with $a=10$ and $b=0.5$. If $T_f<T_{\rm RH}$, the usual radiation-dominated epoch
determines the dynamics and we find an analogous solution
with $\beta =7/2$, $a\approx 5$, $b\approx 0.5$, and $K_*$ must be computed using the
usual Hubble parameter $H\propto T^2$.

%
%

Finally, $\eta_{\rm RH}$ measures the dilution caused by the expansion during
the reheating phase. Therefore $\eta_{\rm RH}=(T_{\rm RH}/m_{N_1})^5$ for $K_*\ll 1$,
$\eta_{\rm RH}=(T_{\rm RH}/T_f)^5$ for $K_*\gg 1$ and $T_f>T_{\rm RH}$, and
$\eta_{\rm RH}=1$ for $K_*\gg 1$ and $T_f<T_{\rm RH}$.

Putting together the different contributions to $\eta$, we obtain,
for $T_f>T_{\rm RH}$
\begin{eqnsystem}{sys:gian}
\eta\approx 0.1 \left( \frac{\tilde m_1}{10^{-3}~\eV}\right)
\left( \frac{T_{\rm RH}}{m_{N_1}}\right)^7& ~~~&{\rm for}~~
\tilde m_1 \ll 3\times 10^{-3}~\eV \left( \frac{m_{N_1}}{T_{\rm RH}}\right)^2\\
\eta\approx 20 \left( \frac{10^{-3}~\eV}{\tilde m_1}\right)
\left( \frac{T_{\rm RH}}{m_{N_1}}\right)^3\left( \ln K_* \right)^{0.5}
 &~~~&{\rm for}~~
\tilde m_1 \gg 3\times 10^{-3}~\eV \left( \frac{m_{N_1}}{T_{\rm RH}}\right)^2.
\end{eqnsystem}

\subsubsection{Leptogenesis from inflaton sneutrino decays}\label{snuCondensate}

In supersymmetric seesaw models there is a distinctive 
possibility that inflaton itself is a scalar superpartner of the lightest 
heavy neutrino~\cite{sn1,sn2}. This is an interesting scenario
because both the reheating of the Universe and the thermal leptogenesis 
efficiency depend on a single neutrino parameter $\tilde m_1.$  Therefore
this is a predictive example of a realistic scenario of the early universe.
In this case there is an additional source of the 
lepton asymmetry from the direct decays of the inflaton sneutrino.

We do not study in detail how the sneutrino condensate decays,
and assume a decay width $\Gamma_\phi=\Gamma_{{\tilde N}_1}(T=0)$ with
 CP asymmetry $\epsilon_1=\epsilon_{\tilde N_1}(T=0)$.
 This is not correct if $T_{\rm RH}\circa{>} m_{N_1}$~\cite{knr},
 that, in our case, happens for $\tilde{m}_1\circa{>} 10^{-3}\eV$.
 However if $\tilde{m}_1\circa{>} 10^{-2}\eV$
 thermalization is so efficient that details of the past thermal history
 negligibly affect our final result.

In order to study this scenario we solved the Boltzmann equations for
the $\tilde N_\pm \equiv \tilde N_1 \pm \tilde N_1^\dagger$ and $N_1$
abundancies, and for the $\tilde{N}_1$ condensate.
In our calculation we take into account 
the temperature dependent interaction rates and
CP asymmetries, and the on-shell resonances are correctly subtracted. 
The Hubble constant $H$ and the parameter $Z$ are the obvious
supersymmetric extensions of eq.s~(\ref{sys:Boltz}--\ref{eq:Z}) including the reheating effects of
thermal (s)neutrinos.

\begin{figure}[t]
\centerline{\includegraphics[width=0.45\textwidth]{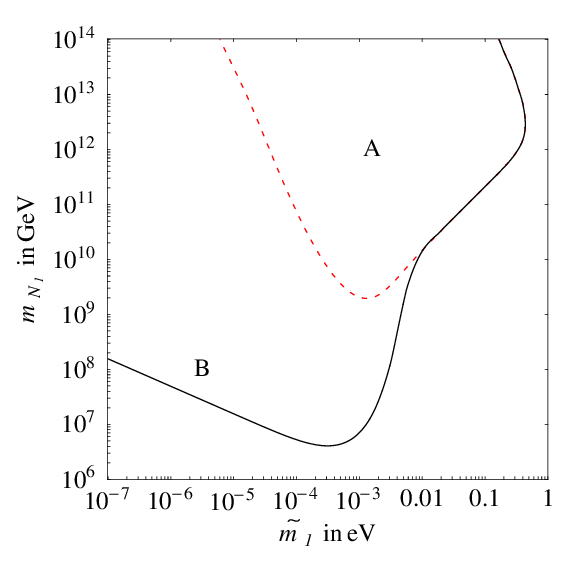}
\hfill
\includegraphics[width=0.45\textwidth]{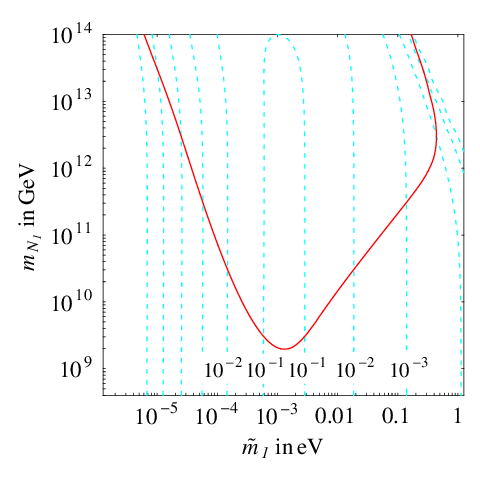}}
\caption{\em 
Lower bounds on $m_{N_1}$ from sneutrino inflaton 
leptogenesis (left) and the efficiency of thermal leptogenesis 
in this scenario (right).}
\label{figsnreh}
\end{figure}

Assuming as before hierarchical light neutrinos and using the maximal CP asymmetry of \eq{eq:di} 
for supersymmetric case, the solutions to the Boltzmann 
equations are presented in fig.~\ref{figsnreh} where we plot
the lower bound on $m_{N_1}$ as a function of $\tilde m_1$ from the observed
baryon asymmetry of the Universe. This parameter space has two distinct 
regions. The one denoted by {\bf A} is the region of purely thermal leptogenesis.
The corresponding curve in \cite{ery} was obtained with wrong subtraction
of on-shell resonances and with a constant CP asymmetry.
For the region {\bf A} we plot also the isocurves of the leptogenesis 
efficiency $\eta$ (right plot) which decreases very fast for small $\tilde m_1$.
This is because of  the suppression of $T_{RH}$ in that region of the
parameter space.

The region denoted by  {\bf B} is the one of direct leptogenesis from
the inflaton decay~\cite{sn2}.

Between those two regions leptogenesis is
a mixture of the two scenarios. Thus the inflaton sneutrino 
 scenario allows to lower the lower
 bound on $m_{N_1}$ and on $T_{RH}$ from successful leptogenesis over a large
region of $\tilde m_1.$ This is very desirable from the point of view of the
gravitino problem, as discussed in the previous subsection.

\section{Conclusions}\label{secconc}
We have performed a
thorough study of thermal leptogenesis which, at present, is one
of the most attractive
mechanism to account for the baryon asymmetry of the universe.
The final prediction of leptogenesis for the baryon asymmetry can be written
in terms of the CP-asymmetry at zero temperature, $\epsilon_{N_1}$, and of
the efficiency $\eta$ of leptogenesis as
$n_{\cal B}/s = -1.37~10^{-3}\epsilon_{N_1}\eta$ in the SM,
and as in \eq{uums} for the MSSM.
Figures~\ref{figSM} (SM) and \ref{figMSSM} (MSSM) show $\eta$
as function of the relevant unknown high energy parameters, {\it i.e.}
the mass $m_{N_1}$ of the lightest right-handed neutrino $N_1$
and $\tilde{m}_1$, its contribution to light neutrino masses.
All the new effects discussed  in this paper
have been added: cumulatively the final baryon asymmetry 
gets typically corrected by a order unity factor
 with respect to previous studies. 
For example, $n_{\cal B}$ gets roughly doubled if $\tilde{m}_1\sim(\Delta m^2_{\rm atm,sun})^{1/2}$.
Since individual terms give larger corrections to the final 
result, in general, it is necessary to include these corrections 
to obtain a trustworthy result. The most important sources of corrections
are summarized in section~\ref{sappr}.

Although thermal leptogenesis allows to compute the baryon asymmetry in terms of particle physics,
a few relevant parameters are presently unknown.
Improving on this issue is as crucial as hard.
In the meantime, by making some assumptions on the high-energy parameters
(the most relevant  one being that right-handed neutrinos are hierarchical)
one can get interesting constraints~\cite{di2,mBound}.
Including all the new effects discussed in this paper
and combining uncertainties at $3\sigma$,
we have found
that successful leptogenesis needs 
$$
m_{N_1} >  \left\{\begin{array}{rl}
2.4\times 10^{9} \GeV    &   \hbox{if $N_1$ has zero} \\
 4.9\times 10^{8}\GeV     &\hbox{if $N_1$ has thermal} \\  
1.7\times 10^{7}\GeV     &\hbox{if $N_1$ has dominant}
\end{array}\right.
\hbox{initial abundancy at $T\gg m_{N_1}$}
$$
and  $m_3 < 0.15\eV$,
where $m_3$ is the heaviest left-handed neutrino mass.
In the MSSM we get similar values.

Furthermore, in inflationary cosmologies, 
we obtained a 
lower  bound on the reheating temperature,
$T_{\rm RH} > 2.8\times 10^9\GeV$ 
assuming that inflaton reheats  SM particles but not directly
right-handed neutrinos. 
Within the MSSM the bound is  $T_{\rm RH} > 1.6\times 10^9\GeV$, which is at odds with the
lower bound from gravitino overproduction.
This seems to suggest that one has to rely on 
alternative  (non-thermal)  mechanisms
to generate right-handed (s)neutrinos after inflation
(like the sneutrino condensate studied at page~\pageref{snuCondensate}), or to invoke
leptogenesis with degenerate right-handed neutrinos~\cite{pilaf1,pilaf} or
``soft leptogenesis''~\cite{softl} (that we study at page~\pageref{soft}).

\medskip

We stress that all these constraints are based on untested assumptions
and therefore cannot be considered as absolute bounds.

$$ *\qquad*\qquad *\qquad$$

\medskip

Finally, 
we would like to emphasize some weak points and possible 
refinements of our analysis.
At $\tilde{m}_1\gg 10^{-3}\eV$ the relevant abundances are
close to thermal equilibrium, suppressing the dependence
on initial conditions.
In this region we are not aware of any missing effect larger than $\sim 10\%$.
Our inclusion of thermal effects 
focussed on thermal corrections to kinematics:
by resumming corrections to propagators
we could study effects which become large at $T\gsim m_{N_1}$.
We approximately included corrections to couplings  
by renormalizing them at $\sim 2\pi T$.
Although this is a significant improvement
with respect to previous computations which
used couplings renormalized at the weak scale,
a somewhat  different approach could give a more precise 
result valid for $\tilde{m}_1\gg 10^{-3}\eV$.
As explained in the text, one should concentrate on computing
corrections to the $N_1\to HL$ decay rate at 
relatively small temperature, $T\lsim m_{N_1}$,
without making our simplifying approximations,
with the goal of including all few $\%$ 
corrections of relative order $\sim g^2/\pi^2$ and $\lambda_t^2/\pi^2$:
$\Delta L=1$ scatterings and their CP-asymmetry,
 three-body decays $N_1\to LQ_3U_3,LHA$ and radiative corrections 
to $N_1\to LH $ decay and its CP-asymmetry.

At $\tilde{m}_1 \lsim 10^{-3}\eV$ the final result depends on initial conditions.
Starting with zero initial abundancy, 
the final baryon asymemtry also depends strongly on
 thermal corrections to the CP-asymmetry $\epsilon_{N_1}$.
Unfortunately we found that, as the temperature rises,
thermal corrections first reduce, then enhance, 
reduce and finally enhance $\epsilon_{N_1}$.
Since the correction does not  go in a clear direction,
a more accurate computation might be welcome.

  \paragraph{Acknowdlegements}
      We thank R. Barbieri, S. Catani, M. Ciafaloni, S. Davidson,
      T. Hambye,
      M. Laine,
      M. Mangano, M. Moretti, M. Papucci, M. Passera,
     M. Pl\"umacher, R. Rattazzi, F. Vissani.

\appendix

\section{Boltzmann equations}\label{Boltz}
Elastic scatterings keep the SM particles in {\em kinetic}
equilibrium,
so that their
energy distribution approximatively follows the
  Maxwell-Boltzmann distribution $f=e^{-E/T}$
(unless otherwise specified we neglect the slightly different energy distributions of
bosons and fermions. This is a good approximation because
$f\circa{<}0.05$ at the average energy $\langle E\rangle\sim 3T$).
Each particle species is simply characterized by its total
abundancy,
that can be varied only by  inelastic processes.
When they are sufficiently fast to maintain also {\em
chemical} 
equilibrium,
the total number $n^{\rm eq}$ and energy density $\rho^{\rm
eq}$ of
ultra-relativistic particles at temperature $T$ are
$$n^{\rm eq} = g\int {d^3p\over(2\pi)^3}~f ^{\rm eq}=
\frac{gT^3}{\pi^2},\qquad \rho^{\rm eq} =
g\int {d^3p\over(2\pi)^3}~E f^{\rm eq}  =
\frac{3gT^4}{\pi^2}
$$
where $g$ is the number of spin degrees of freedom.
All SM particles have $g_{\rm SM} = 118$, and
a right handed neutrino has $g_N = 2$.

In kinetic equilibrium, the phase space density is
$f{n/ n^{\rm eq}}$.
The Boltzmann equation describing the evolution of
the total abundancy $n_N$ of a particle $N$ is
\begin{equation}
\dot n_N + 3H n_N = -
     \sum_{a,i,j,\ldots}[Na\cdots\leftrightarrow ij\cdots] ,
     \label{eq:Boltzmann}
\end{equation}
where $H = {\dot a}/{a} = \sqrt{{8\pi \rho}/{3
M_{\rm
Pl}^2}}$,
$$[Na\cdots\leftrightarrow ij\cdots]=
{n_N n_a\ldots\over
      n_N^{\rm eq}n_a^{\rm eq}\ldots}\,\gamma^{\rm
eq}(Na\cdots\to
      ij\cdots)-
{n_in_j\cdots\over      n_i^{\rm eq}n_j^{\rm eq}\cdots}
      \gamma^{\rm eq}\left(ij\cdots\to
Na\cdots\right).$$
A symmetry factor should be added when there are identical particles
in the final state or in the initial state.
$\gamma^{\rm eq}$
    is the space-time density of scatterings in thermal
equilibrium
of the various interactions in which a $N$ particle takes part:
\beq
\label{gammageneral}
\gamma^{\rm eq}(N a\to ij) = 
\int d\vec{p}_N  \,d\vec{p}_a ~f_N f_a    \int  d\vec{p}_id\vec{p}_j  
\, (2\pi)^4\delta^4(P_N+P_a-P_i-P_j) |A|^2 , 
\eeq
where $d\vec{p}_X = d^3p_X/2E_X(2\pi)^3$ and $|A|^2$ is the squared 
transition amplitude
summed over initial and final spins.
Neglecting CP-violating effects, the inverse processes 
have the same
reaction densities.

Having neglected Pauli-blocking and stimulated emission factors,
and assuming that $|A|^2$ does not depend on the relative motion of particles with respect
 to the plasma, the expression for the scattering rates $\gamma^{\rm eq}$ can
 be conveniently simplified.

For a decay, \eq{gammageneral} reduces to
 \beq\label{eq:gammaDgeneral}
\gamma^{\rm eq}(N\to ij\cdots)=\gamma^{\rm eq}(ij\cdots\to
N)=
      n^{\rm eq}_N
{\mbox{K}_1(z)\over\mbox{K}_2(z)}\,\Gamma_N ,
\eeq
  where $z=m_N/T$ and
$\Gamma_N$ is the decay width in the rest system of
     the decaying particle.
The Bessel $K$ functions arise from the
thermal average of the time dilatation factor $m_N/E$.

The thermal rate $\gamma^{\rm eq}$ of a two-body scattering
can be conveniently rewritten by multiplying \eq{gammageneral} times
$1=\int d^4P~ \int \delta^4 (P-P_N-P_a) $
\beq\label{eq:gammascattgeneral}
\gamma^{\rm eq}(Na\leftrightarrow ij\cdots)= \frac{1}{8\pi} \int \frac{d^4P}{(2\pi)^4}\
\hat\sigma(Na\to ij\ldots)\ e^{-P_0/T}. \eeq
Here 
$\hat\sigma\equiv 2s\ \lambda[1,{m_N^2/ s},{m_a^2/ s}]  \,\sigma$
is the `reduced cross section', $\sigma$ is the total cross section  summed
over final {\em and initial} quantum numbers (spin, gauge multiplicity, 
\ldots), 
$\lambda[a,b,c]\equiv (a-b-c)^2 - 4 b c$.
If the total cross section $\sigma$ depends only on $s$ (and not on the thermal motion with respect to the plasma)
we can use $d^4P = 2\pi \sqrt{P_0^2-s} ~ds~dP_0$ to get
\beq\label{gammascatt}
\gamma^{\rm eq}(Na\leftrightarrow ij\cdots)=
{T\over 64\pi^4}\int_{s_{\rm min}}^{\infty}
ds~s^{1/2}\, {\hat\sigma}(s)\  
\mbox{K}_1\!\left(\frac{\sqrt{s}}{T}\right) ,
\eeq
where $s$ is the squared center of mass energy,
$s_{\rm min} = \max[(m_N+m_a)^2,(m_i+m_j+\cdots)^2]$.

\bigskip

The $3H$ term in the Boltzmann equations, \eq{eq:Boltzmann}, accounts for the dilution due to the overall
expansion of the universe.
It is convenient to reabsorb it  by normalizing the number density $n$ to
the entropy density $s$.
Therefore we study the evolution of  $Y = n/s$
as function of $z=m_N/T$
in place of time $t$
($H\, dt = d\ln R = d\ln z$ since during adiabatic
expansion $sR^3$ is 
constant, i.e.\ $R\propto 1/T$).
The Boltzmann equations become
\begin{equation}
zHs \frac{dY_{N}}{dz}  = -\sum_{a,i,j,\ldots} [Na\leftrightarrow ij] .
\label{eq:BoltzmannY}
\end{equation}

\subsubsection{Leptogenesis}
We now specialize to leptogenesis.
Neglecting sphalerons,
the scattering processes relevant  for leptogenesis are:\footnote{We
add $\Delta L=1$ scatterings involving gauge bosons.
We neglect three-body decays $N_1\to LQ_3U_3,LHA$ and radiative corrections 
to $N_1\to LH $ decay and its CP-asymmetry,
although in most of the parameter space they
enter at the same order as $\Delta L=1$ scatterings, giving
 $\sim g^2/\pi^2\sim \hbox{few}~\%$ corrections.
 Ref.~\cite{pilaf} suggests that $\Delta L=1$ scatterings
 have, at $T=0$, the same CP-asymmetry as decays.
 If true this gives $\%$ corrections, that we prefer not to include.}
\beq\begin{array}{lll}
\Delta L = 1:\qquad& D = [N_1\leftrightarrow LH] &
S_s = H_s + A_s \qquad
S_t= H_t + A_t \\[3mm]
\Delta L = 2: & N_s=[LH \leftrightarrow\bar{L}\bar{H}]&
N_t=[LL \leftrightarrow\bar{H}\bar{H}]\end{array}\eeq
where
\beq\begin{array}{ ll}
H_s = [LN_1 \leftrightarrow Q_3U_3],      &  
2H_t =[N_1\bar U_3 \leftrightarrow Q_3\bar L]+  [N_1\bar Q_3\leftrightarrow U_3\bar L],  \\
A_s=[LN_1\leftrightarrow \bar H A],      &   
2A_t=  [N_1 H \leftrightarrow A {\bar L}]+  [N_1 A\leftrightarrow {\bar H}
{\bar L}].
\end{array}\eeq
We separated $\Delta L = 1$ scatterings $S_{s,t}$ into top ($H_{s,t}$) and
gauge contributions ($A_{s,t}$).
The relative Feynman diagrams are plotted in fig.\fig{leptogDiags}
and computed, including finite temperature effects, in appendices~\ref{gamma} and~\ref{AppEpsilon}.

We assume that $N_1$, $L$ and $\bar L$ can be out
of thermal equilibrium, while $Y_X = Y_X^{\rm eq}$ for $X=\{H,Q_3,U_3\}$.
The Boltzmann equations are
\begin{eqnarray*}
zHs Y'_{N_1} &=& -D-\bar{D}-S_s-\overline{S_s}-S_t-\overline{S_t}\\\
zHs Y'_L &=&D- N_s-N_t- S_s+\overline{S_t} \\
zHs Y'_{\bar{L}} &=& \bar D+ N_s  -\overline{N_t} -\overline{S_s}+S_{t} .
\end{eqnarray*}
We write the decay rates in terms of the CP-conserving total 
decay rate $\gamma_D$ and of the CP-asymmetry $\epsilon_{N_1}\ll1$: 
\begin{equation}\label{eq:gammaD}
\begin{array}{l}
\gamma^{\rm eq}(N_1\to LH)=\gamma^{\rm eq}(\bar{L}\bar{H}\to N_1)=
(1+\epsilon_{N_1}){\gamma_D}/{2},\\
\gamma^{\rm eq}(N_1\to \bar{L}\bar{H})=\gamma^{\rm eq}(LH\to N_1)=
(1-\epsilon_{N_1}){\gamma_D}/{2} 
\end{array}\end{equation}
so that
$$D = \frac{\gamma_D}{2}\bigg[\frac{Y_{N_1}}{Y_{N_1}^{\rm eq}} 
(1+\epsilon_{N_1})-\frac{Y_L}{Y_L^{\rm eq}} (1-\epsilon_{N_1})\bigg]\ ,\qquad
\bar D = \frac{\gamma_D}{2}\bigg[\frac{Y_{N_1}}{Y_{N_1}^{\rm eq}} 
(1-\epsilon_{N_1})-\frac{Y_{\bar L}}{Y_{\bar L}^{\rm eq}} (1+\epsilon_{N_1})
\bigg].$$
Here $Y_{N_1}^{\rm eq}$, $Y_L^{\rm eq}$ and $Y_{\bar L}^{\rm eq}$ are
fermionic equilibrium densities each with 2 degrees of freedom, and therefore
are all equal to high temperature.
Keeping only decays and inverse decays, a baryon asymmetry 
would be generated even in thermal
equilibrium, since CPT invariance implies
that if $N_1$ decays preferentially produce $L$, 
than inverse decays preferentially destroy $\bar L$ (or, in
formul\ae{}, $D-\bar D\ne 0$)~\cite{DolgovZeldovich}.
In order to obtain Boltzmann equations with the correct behavior
one needs to correctly address some subtlety, as discussed in
ref.~\cite{Wolfram}
in the context of baryogenesis. 

\subsubsection{Subtractions of on-shell propagators}
At leading order in the couplings,
$2\leftrightarrow 2$ scatterings must be computed at tree level and
are consequently CP-conserving.
However $N_s$, the $LH\leftrightarrow \bar L\bar H$ 
scattering rate mediated by $s$-channel exchange of $N_1$
shown in fig.\fig{leptogDiags}d,
must be computed by subtracting the CP-violating contribution 
due to on-shell $N_1$ exchange,
because in the Boltzmann equations this effect is already taken 
into account by successive decays,
$LH\leftrightarrow N\leftrightarrow\bar L\bar H$.
Since the on-shell contribution is $\gamma_{Ns}^{\rm on-shell}(
LH\to\bar{L}\bar{H}) =\gamma^{\rm eq}(LH\to N_1) \hbox{BR}(N_1\to \bar{L}\bar{H})$,
where $BR(N_1\to \bar{L}\bar{H})=(1-\epsilon_{N_1})/2$, we obtain 
\begin{eqnarray}
\gamma^{\rm eq}(LH\to\bar{L}\bar{H})&=&\gamma_{Ns}-(1-\epsilon_{N_1})^2
{\gamma_D}/{4},\\
\gamma^{\rm eq}(\bar{L}\bar{H}\to LH)&=&\gamma_{Ns}-(1+\epsilon_{N_1})^2
{\gamma_D}/{4},
\end{eqnarray}
so that
\begin{eqnarray}
N_s &=& \frac{Y_L}{Y_L^{\rm eq}}\gamma^{\rm eq}(LH\to\bar{L}\bar{H}) - 
\frac{Y_{\bar L}}{Y_{\bar L}^{\rm eq}}\gamma^{\rm eq}(\bar{L}\bar{H}\to LH)
\\ &=&
\frac{Y_{\cal L}}{Y_{L}^{\rm eq}}\left( \gamma_{Ns}-\frac{\gamma_D}{4}\right)
+\epsilon_{N_1} \gamma_D+
{\cal O}(\epsilon_{N_1}^2) ,
\end{eqnarray}
having defined the lepton number asymmetry 
$Y_{\cal L} = Y_L-{Y}_{\bar L}$ and used
$Y_L+ Y_{\bar L}= 2 Y_{L}^{\rm eq}+{\cal O}(\epsilon_{N_1})$.
Expanding at leading order in $\epsilon_{N_1}$ gives the  Boltzmann equations
\begin{eqnarray}
zHs Y'_{N_1} &=& -\bigg(\frac{Y_{N_1}}{Y_{N_1}^{\rm eq}}-1\bigg)
(\gamma_D + 2 \gamma_{Ss} + 4\gamma_{St})\\
zHs Y'_{\cal L} &=&\gamma_D \bigg[\epsilon_{N_1} 
\bigg(\frac{Y_{N_1}}{Y_{N_1}^{\rm eq}}-1\bigg) - 
\frac{Y_{\cal L}}{2Y_{L}^{\rm eq}}\bigg]\label{eq:B2}
 -\frac{Y_{\cal L}}{Y_{L}^{\rm eq}}\bigg(2\gamma_{N}^{\rm sub}+2\gamma_{St} 
 + \gamma_{Ss}\frac{Y_{N_1}}{Y_{N_1}^{\rm eq}}\bigg) ,
\end{eqnarray}
where $\gamma_{N}^{\rm sub} = \gamma_{Nt}+\gamma_{Ns} - \gamma_D/4$.
 
Alternatively, one can simply not
include the decay contribution to washout of $Y_{\cal L}$
 because it is already accounted by resonant decays.
 Then one gets
\beq 
zHs Y'_{\cal L} =\gamma_D\epsilon_{N_1} \left(\frac{Y_{N_1}}
{Y_{N_1}^{\rm eq}}-1\right) -\frac{Y_{\cal L}}{Y_{L}^{\rm eq}}\left(
2\gamma_{N}+2\gamma_{St} + \gamma_{Ss}\frac{Y_{N_1}}{Y_{N_1}^{\rm
eq}}\right) , 
\eeq
 which is equivalent to \eq{eq:B2}.
 Therefore it is not necessary to compute subtracted rates.
 For our purposes it is convenient to compute $\gamma_N$ as 
$\gamma_N^{\rm sub}+\gamma_D/4 $.
 Since $\gamma_D$ is a simple and important quantity, we will compute
 it accurately going beyond the Boltzmann approximation.
 We now discuss how to directly compute $\gamma_N^{\rm sub}$, 
 and show that it is subdominant,
 unless the $N_1$ Yukawa couplings are large.

We can directly compute $\gamma_{Ns}^{\rm sub}=\gamma_{Ns}-\gamma_D/4$ by 
subtracting to the intermediate $N_1$ propagator the resonant part
in the narrow-width approximation\footnote{In numerical computations one employs
any representation of $\delta(x)$ that, like 
$\delta(x) = (2\epsilon^3/\pi)/(x^2+\epsilon^2)^2$ and unlike 
$\delta(x)=(\epsilon/\pi) /(x^2+\epsilon^2)$,
goes to zero faster than the propagator away from the peak.
The value of $\epsilon$ can be conveniently chosen
to be somewhat smaller than the width of $N_1$, although
this is not necessary if it is narrow.
In this limit, which covers almost all the parameter space where thermal
leptogenesis can generate the observed baryon asymmetry,
one can simply set $\Gamma_{N_1}/m_{N_1}=\epsilon$,
getting the subtracted propagator of \eq{DN},
and assign to the width any sufficiently small value.}
\beq
|D_{N_1}|^2\to |D_{N_1}^{\rm sub}|^2 =
|D_{N_1}|^2 - \frac{\pi}{m_{N_1} \Gamma_{N_1}}\delta 
(s-m_{N_1}^2) ~~~~~
 D_{N_1} \equiv \frac{1}{s-m_{N_1}^2 + i m_{N_1} 
\Gamma_{N_1} }
\eeq
as discussed, in a different context in ref.~\cite{Olive49}.
Using eqs.~(\ref{eq:gammaD}) and (\ref{gammascatt}), and recalling that,
for $s$-channel exchange $\sigma(LH \to \bar L \bar H)=8\pi \Gamma_{N_1}^2
|D_{N_1}|^2$, we obtain
\beq
\gamma_{Ns}^{\rm sub}=
{T\over 64\pi^4}\int_{s_{\rm min}}^{\infty}
ds~s^{1/2}\, {\hat\sigma}(s)\  
\mbox{K}_1\!\big(\frac{\sqrt{s}}{T}\big)
\left[ 1-\frac{\pi}{\Gamma_{N_1} m_{N_1}|D_{N_1}|^2}\delta(s-m_{N_1}^2)
\right] = \gamma_{Ns}-\frac{\gamma_D}{4}.
\eeq
This corresponds to the result previously obtained.

The subtraction employed in ref.s~\cite{k-sm,mBound,pilaf} is
$$|D_{N_1}|^2\to |\hbox{Re} D_{N_1}|^2 = |D_{N_1}|^2 - ( {\rm Im}\,{D_{N_1}})^2
\stackrel{\Gamma_{N_1}\ll m_{N_1}}{\simeq}
 |D_{N_1}|^2- \frac{1}{2}\frac{\pi  }{m_{N_1} \Gamma_{N_1}}\delta 
(s-m_{N_1}^2)
$$
which leads to
$\gamma_{Ns}^{\rm sub}\simeq \gamma_{Ns}-\gamma_D /8$.
This corresponds to subtracting only
1/2 of the on-shell contribution, thereby over-estimating
washout by 3/2 when $\gamma_D$ is the dominant process
(i.e.\ when the neutrino Yukawa coupling is small). 
As shown in fig.\fig{fig0},
the properly subtracted rate has no spurious peaks around the resonance
region, in contrast with the result of refs.~\cite{k-sm,mBound}. 

\subsubsection{Sphalerons}
Finally, we have to include sphaleronic scatterings, 
that convert the lepton asymmetry into a baryon asymmetry.
This is conveniently done by converting the Boltzmann equation for $Y_{\cal L}$ into 
a Boltzmann equation for $Y_{{\cal B} - {\cal L}}$:
since $Y_{{\cal B} - {\cal L}}$ is not affected by sphalerons
we only need to find the relation between $Y_{{\cal B} - {\cal L}}$ and $Y_{\cal L}$.
At temperatures larger than $10^{10}\GeV$
sphaleronic scatterings are expected to be negligibly slow with respect to
the expansion rate of the universe, 
so that $Y_{{\cal L}} = - Y_{{\cal B}-{\cal L}}$.
At lower temperatures sphaleronic processes keep thermal equilibrium
and the relation would become
 $Y_{\cal L} = -(63/79) Y_{{\cal B}-{\cal L}}$ ($Y_{{\cal L}} = -(9/8) Y_{{\cal B}-{\cal L}}$)
 if all SM Yukawa couplings were large (negligible).
In reality some couplings mediate equilibrium reactions ($y_t, \ldots$)
and some others are negligible ($y_e,\ldots$) 
so that without making approximations
one cannot ignore flavour and must proceed as in ref.~\cite{BRS}.
In particular we stress that in order to study
how the three generations share the lepton asymmetry 
one must consider the evolution of the full flavour $3\times3$ density matrix.
Within $10\%$ accuracy we may approximate $Y_{{\cal L}} \approx - Y_{{\cal B}-{\cal L}}$
and solve the Boltzmann equation of \eq{eq:BoltzB-L}.
After computing $Y_{{\cal B}-{\cal L}}$ the baryon asymmetry $Y_{\cal B}$ is 
obtained by means of \eq{eq:YB}.

 \subsubsection{Inflaton and $N_1$ reheating}
  We now add one refinement.
 We described in section~\ref{reh} how the Boltzmann equations are modified 
 by the presence of a field $\phi$, whose decays into SM particles reheat the universe.
 Proceeding along the same lines 
 we also take into account that reheating due to $N_1$
 decays might be non-negligible.
 Terms of relative order $\rho_{N_1}/\rho_R$ are neglected in the `standard' Boltzmann equations: 
 in thermal equilibrium this factor is small, $\rho_{N_1}/\rho_R\circa{<} g_{N_1}/g_\star\sim 0.02$ since
 $N_1$ is one out of many more SM particles.
 Away from thermal equilibrium it can be sizable.
 Including these effects, the Boltzmann equations become
  \begin{eqnsystem}{sys:Boltz}
  HZz\frac{d\rho_\phi}{dz}&=& -
{3H\rho_\phi}-{\Gamma_\phi\rho_\phi}\,
,\\
sHZz \frac{dY_{N_1}}{dz} &=& 3sH(Z-1)Y_{N_1}
 -\bigg(\frac{Y_{N_1}}{Y_{N_1}^{\rm eq}}-1\bigg)(\gamma_D + 2 \gamma_{Ss} + 4\gamma_{St})\\
sHZz \frac{dY_{{\cal B} - {\cal L}}}{dz} &=& 
3sH(Z-1)Y_{{\cal B} - {\cal L}}
-\gamma_D \epsilon_{N_1} \bigg(\frac{Y_{N_1}}{Y_{N_1}^{\rm eq}}-1\bigg) +\nonumber
\\&&
-\frac{Y_{{\cal B} - {\cal L}}}{Y_{L}^{\rm eq}}\bigg(\frac{\gamma_D}{2}+
2\gamma_{N}^{\rm sub}+2\gamma_{St}  + \gamma_{Ss}\frac{Y_{N_1}}{Y_{N_1}^{\rm eq}}\bigg)
 \label{eq:BoltzB-L}
\end{eqnsystem}
 where $s$ and $\rho_R$ are the entropy and energy density of SM particles,
\beq H= \sqrt{\frac{8\pi}{3M_{\rm Pl}^2} (\rho_R + \rho_{N_1} + \rho_\phi)}\eeq
is the Hubble constant at temperature $T$ and
\beq \label{eq:Z}
Z=1-\frac{\Gamma_\phi\rho_\phi}{4H\rho_R}-\frac{\gamma_D+2\gamma_{Ss}+4\gamma_{St}}{4H\rho_R}
 \frac{ \rho_{N_1}^{\rm eq}}{n_{N_1}^{\rm eq}}\bigg(\frac{Y_{N_1}}{Y_{N_1}^{\rm eq}}-1\bigg)
 \eeq
 takes into account reheating.
  Since interactions with SM field can keep $N_1$ close to thermal equilibrium,
 its contribution to $Z$ is not just given by
 $\Gamma_{N_1}\rho_{N_1}/4H\rho_R=( \gamma_D/4H\rho_{\rm SM})
 ( \rho_{N_1}^{\rm eq} /n_{N_1}^{\rm eq})(Y_{N_1}/Y_{N_1}^{\rm eq})$.

\medskip

 In the next appendices we will see how  
 thermal effects can be included by modifying the scattering rates and the CP-asymmetries,
 but not the form of the equation themselves.

\section{Thermal corrections to decays and scatterings}\label{gamma}
We present the temperature-dependent decay rates and
cross sections that generate and washout the lepton
asymmetry.
Since we consider temperatures $T\sim m_{N_1}$
much above the electroweak scale,  the thermal  masses of the
Higgs doublet, lepton doublet, top quarks and
electroweak gauge bosons
are given by~\cite{thermalmasses}
\begin{eqnsystem}{sys:SMm}
\frac{m_H^2}{T^2}&=& \frac{3}{16} g^2_2 + \frac{1}{16}
g^2_Y + \frac{1}{4} y^2_t + \frac12 \lambda, \\
\frac{m_L^2}{T^2}&=& \frac{3}{32} g^2_2 + \frac{1}{32}
g^2_Y, \\
\frac{m_{Q_3}^2}{T^2}&=& \frac{1}{6} g^2_3 +\frac{3}{32}
g^2_2 + 
\frac{1}{288} g^2_Y + \frac{1}{16}  y^2_t, \\
\frac{m_{U_3}^2}{T^2}&=& \frac{1}{6} g^2_3 + \frac{1}{18}
g^2_Y +
\frac{1}{8} y^2_t, \\
\frac{m_W^2}{T^2} &=& \frac{11}{12}g_2^2,\qquad
\frac{m_B^2}{T^2} =\frac{11}{12}g_Y^2 ,
\label{Mqz}
\end{eqnsystem}
where all couplings are renormalized at the RGE scale $\mu=2\pi T$.
The Higgs boson, lepton and quark masses are
functions of $T$, even if not explicitly denoted.
We neglected their zero-temperature values.
At leading order the quartic Higgs coupling $\lambda$ is given in terms
of the zero-temperature Higgs mass $m_h$ as $\lambda(\mu=v) = (m_h/2v)^2$
where $v=174\GeV$.
The top Yukawa couplings is similarly given by $y_t(\mu=v) = m_t/v$.
Their RGE running is known up to next-to-leading order~\cite{RGESM}.

In the MSSM the relevant thermal masses are~\cite{thermalmasses}
\beq\begin{array}{rcccl}
\displaystyle\frac{m_{H_{\rm
u}}^2}{T^2}&=&2\displaystyle\frac{m_{\tilde{H}_{\rm
u}}^2}{T^2} &=&\displaystyle 
\frac{3}{8} g^2_2 + \frac{1}{8} g^2_Y+ \frac34 \lambda_t^2,

\\
\displaystyle\frac{m_{\tilde{L}}^2}{T^2} &=&
2\displaystyle\frac{m_{L}^2}{T^2}  &=&\displaystyle 
\frac{3}{8} g^2_2 + \frac{1}{8} g^2_Y ,
\\
\displaystyle\frac{m_{\tilde{Q}_3}^2}{T^2}&=&2\displaystyle\frac{m_{Q_3}^2}{T^2}&=&
\displaystyle 
\frac23 g_3^2 +  \frac{3}{8} g^2_2 + \frac{1}{72} g^2_Y + 
\frac14 \lambda_t^2,
\\
\displaystyle\frac{m_{\tilde{U}_3}^2}{T^2}&=&
2\displaystyle\frac{m_{U_3}^2}{T^2}&=&
\displaystyle\frac23 g_3^2 +  \frac{2}{9} g^2_Y + \frac12
\lambda_t^2 .
\end{array}\eeq
Here and in the following
 appendix the neutrino couplings $Y_\nu$ are renormalized
at
the high-scale.
The one-loop RGE equations for the Majorana neutrino mass
matrix $m_{ij}$
valid from the Fermi scale (below which it is not affected
by
quantum corrections)
up to $m_{N_1}$ are~\cite{SMRGE}
\begin{equation}
\label{eq:RGESM}
(4\pi)^2 \frac{dm}{d\ln\mu}= m (\lambda  - 3
g_2^2+6\lambda_t^2) 
-\frac{3}{2}\bigg[m \cdot (Y_E^\dagger \cdot Y_E)^T+
 (Y_E^\dagger  \cdot Y_E)\cdot m\bigg]
\end{equation}
 in the SM and
\begin{equation}
\label{eq:RGEMSSM}{
(4\pi)^2 \frac{dm}{d\ln\mu}= m (-2g_Y^2 - 6
g_2^2+6\lambda_t^2) 
+m \cdot (Y_E^\dagger \cdot Y_E)^T+
 (Y_E^\dagger  \cdot Y_E)\cdot m } 
\end{equation}
in the MSSM.
Higher order effects (two-loop RGE, thresholds) are partially included in our codes.
Here $Y_E$ and $\lambda$ are the charged lepton and Higgs
couplings.
The solution of these RGEs is described in
section~\ref{couplings}.
The Yukawa couplings of right-handed neutrinos $N_1$ give
extra RGE effects at scales above $m_{N_1}$.
We neglect these effects, as large Yukawa couplings anyhow
lead to
an exponentially small efficiency for leptogenesis.

The $N_1$ mass does not receive thermal corrections, as
long as we 
neglect the relevant neutrino Yukawa couplings, which are
indeed small
in most of the interesting parameter region, since
$|Y_{\nu 1i}|^2<3\times 10^{-7}(m_{N_1}/10^{10}~{\rm GeV})
(\tilde{m}_1/10^{-3}~{\rm eV})$.   
We define 
\begin{equation}
z=\frac{m_{N_1}}{T}  , \qquad
x=\frac{s}{m_{N_1}^2},\qquad
a_{H,L,Q,U,W,B}= \frac{m_{H,L,Q_3,U_3,W,B}^2(z)}{m_{N_1}^2}, \qquad
a_\Gamma= \frac{\Gamma_{N_1}^2}{m_{N_1}^2},
\end{equation}
\begin{equation}\label{DN}
D_{N_1}= \frac{1}{x -1  + i a^{1/2}_\Gamma},
\qquad
|D_{N_1}^2|^{\rm sub} = \frac{(x-1)^2 - a_\Gamma}{[(x-1)^2+a_\Gamma]^2}
\end{equation}
and $\lambda[a,b,c]=(a-b-c)^2 - 4 b c$.

\subsubsection{Decays}
At low temperature $m_{N_1}>m_H+m_L$ so that one has the
usual $N_1$ decay  with total width
\beq\label{GammaN1}
\Gamma_{N_1}  =
\frac{1}{8\pi} (Y_\nu Y_\nu^\dagger)_{11} 
\lambda^\frac{1}{2}[1,a_H,a_L]
(1-a_H+a_L)\; m_{N_1}.
\eeq
Analytical approximate solutions of the Boltzmann equations~\cite{bcst} show that
in the most interesting region ($\tilde{m}_1\gg 10^{-3}\eV$ and small $N_1$ Yukawa couplings),
the efficiency of thermal leptogenesis is mainly controlled by
$\gamma_{N} = \gamma_{N}^{\rm sub} +\gamma_D/4 \simeq \gamma_D/4$.
Therefore we compute  the thermally averaged $N_1\to LH$ decay rate
$\gamma_D$ accurately.
Instead of using the Boltzmann approximation of \eq{eq:gammaDgeneral},
 we compute $\gamma_D$
using the full Bose-Einstein $f_H$ and Fermi-Dirac $f_L,f_N$ distributions,
including stimulated emission and Pauli blocking factors.
The relation $\gamma_N\simeq\gamma_D/4$ remains unaltered where now
\begin{eqnarray}\label{gammaDBEFD}
 \gamma_D &=&2  \int d\vec{p}_N d\vec{p}_L d\vec{p}_H  \, f_L f_H  f_N e^{E_N/T}
(2\pi)^4 \delta^4(P_N- P_L-P_H)|A|^2\\
&=&\frac{ \Gamma_{N_1}}{{\pi^2}} \int_{m_{N_1}}^\infty dE_N\,{m_{N_1} \sqrt{E_N^2 - m_{N_1}^2}} f_N e^{E_N/T}
\frac{1}{2}\int_{-1}^1 d\cos\theta_0    f_L(E_L) f_H (E_H)\nonumber
\end{eqnarray}
where $d\vec{p}_X = d^3p_X/2E_X(2\pi)^3$ is the relativistic phase space,
$\theta_0, E_L^0,p_L^0$ are the decay angle, the $L$ energy,
the $L$ momentum in the $N_1$ rest frame
and $E_L = \gamma  E_L^0+\sqrt{\gamma^2-1} p_L^0\cos\theta_0$
($\gamma = E_N/m_{N_1}$).
The integral in $d\cos\theta_0$ can be done analytically.

As expected, the Boltzmann approximation is accurate at 
low $T\circa{<}0.1\,m_{N_1}$ and differs from the full result by 
a few $10\%$ at $T\sim m_{N_1}$.
Actually, for the specific SM values of the thermal masses,
 the Boltzmann approximation is accurate within $10\%$.

\medskip

We stress that our inclusion of thermal effects is only approximate.
In the present work we focussed on those corrections which become large
at $T\circa{>} m_{N_1}$.

In particular, at sufficiently high temperature
the $N_1$ decay becomes kinematically forbidden at higher
temperature by the $H$ thermal mass.
When $m_H>m_{N_1}+m_L$, the $H\to LN_1$ decay is allowed. Its
width is
\begin{eqnarray}
\Gamma_H  =
\frac{1}{16\pi} (Y_\nu Y_\nu^\dagger)_{11} 
\lambda^\frac{1}{2}[a_H,1,a_L]
\frac{a_H-1-a_L}{a_H^2}\; m_{H}.
\end{eqnarray}
Despite the change in the decay process the Boltzmann equations keep the same form,
with the $N_1$ decay rate replaced by the $H$ decay rate.

Even including thermal effects,
at intermediate temperatures all $1\leftrightarrow 2$ and $3\leftrightarrow 0$
processes are kinematically forbidden, so that $\gamma_D=0$.

\subsubsection{$\Delta L=2$ scatterings}
We now consider lepton-number violating scatterings:
the $\Delta L = 2$ processes
$LH^\dagger \leftrightarrow \bar LH$  and
$LL\leftrightarrow H^\dagger H$ (middle row of fig.\fig{leptogDiags}).

As explained in appendix~\ref{Boltz} we compute
the $LH \leftrightarrow \bar LH^\dagger$
subtracting the contribution
from on-shell intermediate $N_1$,
to avoid double counting with two-body decays.
At low temperatures the computation is similar to the $T=0$ case:
$N_1$ can be on-shell in the $s$-channel diagram of  
fig.\fig{leptogDiags}d
so that one must replace the $s$-channel propagator $D_{N_1} =  
1/(s-m_{N_1}^2 + i m_{N_1} \Gamma_{N_1})$
with a subtracted propagator $D^{\rm sub}$.

At high temperatures, when $H\to N_1L$ replace $N_1\to HL$ decays,
the situation becomes more tricky.
$N_1$ can be on-shell in the $u$-channel diagram of  
fig.\fig{leptogDiags}e.
As this corresponds to on-shell $N_1$ in the decay $H\to N_1L$,
it has to be subtracted similarly to the $s$-channel resonance.
The imaginary part of the $N_1$ propagator
which renders finite
the $LH \leftrightarrow \bar LH^\dagger$ rate
is no longer given by $N_1$ decay, but by
thermal absorption of $N_1$, given by~\cite{Gamma(T)}
\begin{eqnarray}
-\hbox{Im}\,\Pi_{N_1}(E_N) &=& E_N \Gamma_{N_1}(E_N) = \nonumber
\frac{1}{2} \int d\vec{p}_L\,d\vec{p}_H\,(2\pi)^4 \delta^4(P_N+P_L-P_H)  
|A|^2 [f_L+f_H]
\end{eqnarray}
Unlike a decay at $T=0$, for which $\Gamma_{N_1}(E_N) =  
\Gamma_{N_1}(m_{N_1}) m_{N_1}/E_N$,
the $N_1$ width at finite temperature depends on thermal motion of  
$N_1$ with respect to the plasma,
giving rise to lengthy expressions.
For simplicity we give the expression corresponding to
the narrow-resonance limit $\Gamma_{N_1}\to 0$,
which is always valid.
In fact, if the $N_1$ Yukawa coupling is large, $N_1$ gets a thermal  
mass
avoiding $H\to N_1 L$ decays.

The reduced $LH \leftrightarrow \bar LH^\dagger$ cross section is given  
by\footnote{We thank H. McKenna, M. Gorbahn and J. Smirnov  for noticing a typo. It was not present in the code, so results remain unchanged.}
\begin{eqnarray}
\hat\sigma_{Ns}^{\rm sub}&=&\nonumber
\frac{(Y_\nu Y_\nu^\dagger)_{11}^2}{4\pi x} \bigg[
(1+(1-2a_H+x)(\hbox{Re}D_{N_1}-3\xi ))\ln\bigg(\frac{R_2^2/x^2+\epsilon}
{R_1^2+\epsilon}\bigg)+\\
&&+\label{sigNs}
\frac{1}{x^2}(a_H^2+(a_L-x)^2-2a_H(a_L+x))
\bigg(|D_{N_1}^2|^{\rm sub} (x-a_H+a_L)^2+
\\&& \nonumber
+2x\hbox{Re}D_{N_1}+\frac{2(1+x-2a_H) R_2 R_1}
{[R_2^2/x^2+\epsilon ][R_1^2+\epsilon]}
-3\xi (2x+(x-a_H+a_L)^2(\hbox{Re}D_{N_1}-\xi )
 \bigg)\bigg]
\end{eqnarray}
where $R_1=1-2a_H-2a_L+x$ and $R_2= x-(a_H-a_L)^2$
and $\epsilon$ is any small number, $\epsilon\ll a_L^2$.

For the $\Delta L = 2$ scattering $LL\leftrightarrow HH$ we
obtain
\begin{equation}
\hat\sigma_{Nt}  =
\frac{(Y_\nu Y_\nu^\dagger)_{11}^2}{2\pi}  \frac{x-2a_L
}{x}
\bigg[\frac{r}{R_1+(a_H-a_L)^2}+\frac{3}{2}r\xi ^2
+(\frac{1}{R_1+1}+3\xi )\ln\frac{R_1+1+r}{R_1+1-r}\bigg]\label{sigNt}
\end{equation}
where $r\equiv  \sqrt{(x-4a_H)(x-4 a_L)}$.
The parameter $\xi $ takes into account scatterings mediated by heavier right-handed neutrinos
$m_{N_{2,3}}\gg m_{N_1}$, as explained in section~\ref{SM}.

\subsubsection{$\Delta L=1$ scatterings}
Let us now consider the $\Delta L = 1$
processes $L N\leftrightarrow Q_3 U_3$ and
$\bar U_3 N\leftrightarrow Q_3 \bar L$ (bottom row of  
fig.\fig{leptogDiags}).
We can neglect the small difference between the thermal masses of
$Q_3$ and $U_3$ (see fig.\fig{mT}a), setting $a_U\approx a_Q$.
The $L N\leftrightarrow Q_3 U_3$ reduced cross section  is
\begin{equation}
\hat\sigma_{Hs} =
\frac{3}{4\pi} (Y_\nu Y_\nu^\dagger)_{11} y_t^2
\frac{(x-1-a_L)(x-2a_Q)}{x(x-a_H)^2}\sqrt{[(1+a_L-x)^2-4a_L][1-4a_Q/x]}.
\end{equation}
The  $\bar U_3 N\leftrightarrow Q_3 \bar L$ and the
$\bar Q_3 N\leftrightarrow U_3 \bar L$ cross sections are equal and given by
\begin{eqnarray}
\hat\sigma_{Ht}  &=&
\frac{3}{4\pi} (Y_\nu Y_\nu^\dagger)_{11} y_t^2 
\frac{1}{x} \bigg[t_+ - t_- -
 (1-a_H+a_L)(a_H-2a_Q)\bigg(\frac{1}{a_H-t_+}-\frac{1}{a_H-t_-}\bigg)+\nn
&&-
(1-2a_H+a_L+2a_Q)\ln\frac{t_+-a_H}{t_--a_H}\bigg]
\end{eqnarray}
where
\begin{eqnarray} 
t_\pm&\equiv& \frac{1}{2x}\bigg[a_Q + x -
(a_Q-x)^2+a_L(x+a_Q-1)+\nn
&&\pm \sqrt{[a_Q^2+(x-1)^2-2a_Q(1+x)][a_L^2 +
(x-a_Q)^2-2a_L(x+a_Q)]}\bigg]
\end{eqnarray}

%
%

Neglecting thermal masses, all these interaction rates agree with those used
in the literature (up to typos present in older papers).
As explained in the text, we must include 
$\Delta L = 1$ scatterings involving gauge bosons.
We do not compute the full thermally corrected rates, since
we should take into account thermal motion of $A$ and $L$ with respect to
the plasma, which gives a complicated result.
We cannot fully neglect thermal effects, since exchange of
massless  $H$ and $L$ would give an IR divergent result.
Therefore we keep thermal masses of $A,L,H$ only when they
act as regulators of IR enhanced contributions.
This approximation is accurate at $T\ll m_{N_1}$.
At larger $T$ it neglects terms suppressed by higher powers of
$g^2 \sim 1/3$.
The result is
\begin{eqnarray}\label{eq:As}
\hat\sigma(N_1 \bar{L}\to HA)&=&
 \frac{3 g_2^2 (Y_\nu Y_\nu^\dagger)_{11}}{16\pi x^2}\bigg[
2t(x-2)+(2-2x+x^2)\ln[(a_L-t)^2+\epsilon]+
\\&&\hspace{15ex} + 2\frac{x(a_L-t)(a_L +a_L x - a_W)+\epsilon(2-2x+x^2)}{(a_L-t)^2+\epsilon}\nonumber
\bigg]_{t_-}^{t_+}\\
\hat\sigma(LH\to N_1A)&=& 
\frac{3 g_2^2 (Y_\nu Y_\nu^\dagger)_{11}}{8\pi x(1-x)}\bigg[2x\ln(t-a_H)-(1+x^2)\ln(t+x-1-a_W-a_H)
\bigg]_{t_-}^{t_+}\nonumber\\
\hat\sigma(\bar LA\to N_1H)&=&
\frac{3 g_2^2 (Y_\nu Y_\nu^\dagger)_{11}}{16\pi x^2}\bigg[
t^2+2t(x-2)-4(x-1)\ln(t-a_H)+x\frac{a_W-4a_H}{a_H-t}
\bigg]_{t_-}^{t_+}\nonumber
\end{eqnarray}
We wrote only the ${\rm SU}(2)_L$ contribution.
One must add the ${\rm U}(1)_Y$ contribution,
obtained by substituting $a_W\to a_Y$ and
$\frac{3}{2}g_2^2 \to \frac{1}{4} g_Y^2$.\footnote{Gauge scatterings have
been estimated in a recent paper~\cite{pilaf},
by introducing some infra-red cutoff,
which should give a qualitatively correct result at low temperature.
We can only compare the ratio between ${\rm SU}(2)_L$ and ${\rm U}(1)_Y$
contributions, which is different from our value.
We do not use simplified expressions for $t_\pm$,
valid when $m_{L,W,H}^2\ll s,m_{N_1}^2$, because
they not even accurate at low temperature, where small
$s- m_{N_1}^2\simeq m_{L,W,H}^2$ is a relevant kinematical range.
Due to the $1-x$ factor, at $T\ll m_{N_1}$ the $LH\to N_1A$ 
interaction rate is $\gamma_{At}\sim (g/\pi)^2\gamma_D$,
of the same order as one loop corrections to the decay rate
(that we do not include).}
The expression $[f(t)]^{t_+}_{t_-}$ denotes $f(t_+/m_{N_1}^2) - f(t_-/m_{N_1}^2)$ where
 $t_\pm$
are the usual kinematical ranges for $t = (P_1 - P_3)^2$ in the various 
$12\to 34$ scatterings:
$$t_\pm = \frac{(m_1^2-m_2^2-m_3^2+m_4^2)^2}{4s}-
\bigg(\sqrt{\frac{(s+m_1^2-m_2^2)^2}{4s}-m_1^2}\pm 
\sqrt{\frac{(s+m_3^2-m_4^2)^2}{4s}-m_3^2}\bigg)^2$$
In some parameter range the process $N_1 \bar{L}\to HA$ can have on-shell $L$ in the $t$-channel,
that we have subtracted:
in such a case one should  use any finite value $\epsilon\ll a_L^2$.
Otherwise one can set $\epsilon=0$.

The rates that enter in Boltzmann equations are given by
$$\hat{\sigma}_{As} = \hat\sigma(L N_1 \to {\bar H}A),\qquad
\hat{\sigma}_{At} = \frac12 \left[
\hat\sigma(A{\bar L}\to N_1H)+\hat\sigma(\bar H \bar L\to N_1A)\right] $$

\subsubsection{Resonances in $s$ and $u$ channels}

We here explicitly verify that the relation between decay and resonant scatterings,
$\gamma_{N}^{\rm on-shell}  = \gamma_D/4$
(up to CP-violating corrections),
remains valid without approximating thermal distributions with Boltzmann statistics,
and that it applies to both $s$-channel as well as $u$-channel resonances.
This last issue is non-trivial, as computing a
cross section mediated by an unstable particle which can be on-shell
is a difficult problem even 
  at zero temperature and at tree level. In that case, the beam size and
  the width of the external unstable particle act
as a regulator of the divergence~\cite{russi}.

\medskip

We first  consider the $s$-channel case.
The $N_1\leftrightarrow L H$ rates are
{\small \begin{eqnarray*}
\label{gammad+i}
\gamma^{\rm eq}(N_1\to L H) &=&
\int d\vec{p}_{N_1} d\vec{p}_L d\vec{p}_H  ~f_{N_1} (1-f_L) (1+f_H)
\, (2\pi)^4\delta^4(P_{N_1}-P_L-P_H) |A( N_1 \to LH )|^2  \\
\gamma^{\rm eq}(LH\to N_1) &=& \int  d\vec{p}_{N_1}  d\vec{p}_L d\vec{p}_H ~f_L f_H (1-f_{N_1})
\, (2\pi)^4\delta^4(P_{N_1}-P_L-P_H) |A( LH \to N_1 )|^2.
\end{eqnarray*}}
Using $E_{N_1}=E_H+E_L$, we obtain 
$f_{N_1} (1-f_L) (1+f_H)=  f_L f_H (1-f_{N_1})=f_L f_{N_1} f_H e^{E_{N_1}/T}$.
The on-shell contribution to the $LH \rightarrow \bar{L} \bar{H} $ equilibrium interaction rate is
\begin{eqnarray*}
\label{gammaonshellExact}
\gamma_{\rm on-shell}^{\rm eq}(L H \to \bar{L} \bar{H}) &=&
 \int d\vec{p}_H d\vec{p}_L d\vec{p}_{\bar{L}} d\vec{p}_{\bar{H}} ~f_L f_H (1- f_{\bar{L}}) (1+f_{\bar{H}})
|A( LH \to N_1 )|^2  \\&& \left( \frac{\pi \delta(s-m_{N_1}^2)}{m_{N_1} \Gamma^{\rm th}_{N_1}} \right)
|A( N_1 \to \bar{L}{\bar{H}} )|^2
 (2\pi)^4\delta^4(P_L+P_H-P_{\bar{L}}- P_{\bar{H}})
\end{eqnarray*}
where the width that cutoffs the resonance is $\Gamma_{N_1}^{\rm th}$ \cite{Gamma(T)}, the damping rate at finite temperature.
We can rewrite the product of thermal distributions as:
\beq
(1- f_{\bar{L}}) (1+f_{\bar{H}})=f_{N_1} e^{E_{N_1}/T} \left[
(1-f_{\bar{L}})(1+f_{\bar{H}})+f_{\bar{L}} f_{\bar{H}}\right]
\eeq
and insert $1=\int d^4 P_{N_1} \delta^4(P_{N_1}-P_L-P_H)$, obtaining:
\beq
   \label{gammaonshellExact2}
  \begin{split}
&\gamma_{\rm on-shell}^{\rm eq} (L H \rightarrow \bar{L} \bar{H}) =
 \int d\vec{p}_H d\vec{p}_L \left( \frac{d^4 P_{N_1}}{(2 \pi)^4}
 2\pi \delta(P_{N_1}^2-m_{N_1}^2)\right)\\
 & \qquad f_L f_H f_{N_1} e^{E_{N_1}/T} |A( LH \to {N_1} )|^2
 (2\pi)^4\delta^4(P_L+P_H-P_{N_1})
\left(\frac{1}{ 2 m_{N_1} \Gamma^{\rm th}_{N_1} } \right) \\
& \qquad \int  d\vec{p}_{\bar{L}} d\vec{p}_{\bar{H}}
\left[ (1-f_{\bar{L}})(1+f_{\bar{H}})+f_{\bar{L}} f_{\bar{H}}\right]
|A( {N_1} \to \bar{L} \bar{H} )|^2
 (2\pi)^4\delta^4(P_{N_1}-P_{\bar{L}}-P_{\bar{H}})
 \end{split}
\eeq
The integrals over final-state particles reconstruct the thermal width 
of $N_1$~\cite{Gamma(T)}
\beq
\Gamma_{N_1}^{\rm th}
=\frac{1}{m_{N_1}}  \int  d\vec{p}_{\bar{L}} d\vec{p}_{\bar{H}}
\left[ (1-f_L)(1+f_H)+f_L f_H\right]
|A_{\rm tree}|^2
(2\pi)^4\delta^4(P_{N_1}-P_{\bar{L}}-P_{\bar{H}})
\eeq
In Boltzmann approximation the term in square brackets can be approximated with 1,
and $\Gamma^{\rm th}_{N_1}$ reduces to the standard expression, \eq{GammaN1}.
The integrals over initial-state particles reconstruct $\gamma_D$,
giving the desired relation $\gamma_{Ns}^{\rm on-shell} = \gamma_D/4$.

\medskip

In an analogous way we can deal with $u$-channel
resonance, present in $L H\to \bar{L} \bar{H}$ scatterings
at high  temperatures when $H$ decays to $N_1 \bar{L}$ while $N_1$ is stable.
 At finite $T$ the propagator of a particle involved in a $1\leftrightarrow 2$ process gets an
 imaginary part, even if it is not the decaying particle.
This thermal $\Gamma_{N_1}$ cutoffs the $u$-channel
 resonance and gives consistent Boltzmann equations
 (no ${\cal L}$ asymmetry generated in thermal equilibrium).
To show this fact we follow the same procedure.
The interaction rates for $\bar{H}\leftrightarrow N_1 L$ are
{\small\begin{eqnarray*}
\label{gammaH}
\gamma^{\rm eq}(\bar{H}\to {N_1} L ) &=&
\int d\vec{p}_{\bar{H}} d\vec{p}_L d\vec{p}_{N_1} ~f_{\bar{H}} (1-f_{N_1}) (1-f_L)
\, (2\pi)^4\delta^4(P_{\bar{H}}-P_L-P_{N_1}) |A( \bar{H} \to L {N_1} )|^2  \\
\gamma^{\rm eq}({N_1} L\to \bar{H}) &=&\int d\vec{p}_{N_1}  d\vec{p}_L d\vec{p}_{\bar{H}} ~f_L f_{N_1} (1+f_{\bar{H}})
\, (2\pi)^4\delta^4(P_{\bar{H}}-P_L-P_{N_1}) |A( {N_1} L \to \bar{H} )|^2.
\end{eqnarray*}}
Since $E_{\bar{H}}=E_{N_1} + E_L$ one can show that
$f_{\bar{H}} (1-f_{N_1}) (1-f_L)=  f_L f_{N_1} (1+f_{\bar{H}})$, so that, again
\begin{eqnarray}
\gamma^{\rm eq} (\bar{H} \rightarrow {N_1} L)=\gamma^{\rm eq} (N_1 \bar{L} \rightarrow H)=\frac{\gamma_D}{2}(1+\epsilon_{H})  \\
\gamma^{\rm eq}(N_1 L \rightarrow \bar{H})  =\gamma^{\rm eq} (H \rightarrow N_1 L )=\frac{\gamma_D}{2}(1-\epsilon_{H})
\end{eqnarray}
where now
\beq
\label{gammaD}
\gamma_D = 2
\int d\vec{p}_{\bar{H}} d\vec{p}_L d\vec{p}_{N_1} ~f_{\bar{H}} (1-f_{N_1})(1-f_L)
\, (2\pi)^4\delta^4(P_{\bar{H}}-P_L-P_{N_1}) |A_{\rm tree}(\bar{H} \to N_1 L)|^2.
\eeq
The $u$-channel on-shell contribution to
$L H \to \bar{L} \bar{H}$ is
\beq
\begin{split}
\label{gammaonshellExactH}
&\gamma_{\rm on-shell}^{\rm eq} (L H \rightarrow \bar{L} \bar{H})=
 \int d\vec{p}_H d\vec{p}_L  d\vec{p}_{\bar{L}} d\vec{p}_{\bar{H}} ~f_L f_H (1- f_{\bar{L}}) (1+f_{\bar{H}})  \\
&
|A( H \to {N_1} \bar{L})|^2 \left( \frac{\pi \delta(u-m_{N_1}^2)}{E_{N_1} \Gamma^{\rm th}_{N_1}(E_{N_1})} \right)|A( {N_1} L \to \bar{H})|^2
 (2\pi)^4\delta^4(P_L+P_H-P_{\bar{L}}- P_{\bar{H}}) .
\end{split}
\eeq
The product of the distributions can be rewritten, making use of  $E_{\bar{H}}=E_{N_1} + E_L$ as
\beq
f_L (1+f_{\bar{H}})=(1-f_{N_1})\left[ (f_L)(1+f_{\bar{H}})+f_{\bar{H}} (1-f_L)
\right]
\eeq
Inserting $1=\int d^4 P_{N_1} \delta^4(P_H-P_{\bar{L}}-P_{N_1})$ we get
\beq
   \label{gammaonshellExact2H}
  \begin{split}
&\gamma^{\rm eq}_{\rm on-shell} (L H\rightarrow \bar{L} \bar{H}) =
 \int d\vec{p}_H d\vec{p}_{\bar{L}} \left( \frac{d^4 P_{N_1}}{(2 \pi)^4}
 2\pi \delta(P_{N_1}^2-m_{N_1}^2)\right)\\
 & ~f_H (1-f_{N_1})(1-f_{\bar{L}}) |A( H \to {N_1} \bar{L})|^2
 (2\pi)^4\delta^4(P_H-P_{\bar{L}}-P_{N_1})
\left(\frac{1}{ 2 E_{N_1} \Gamma^{\rm th}_{N_1}(E_{N_1})} \right)\\
& \int  d\vec{p}_{L} d\vec{p}_{\bar{H}}[(f_L)(1+f_{\bar{H}})+f_{\bar{H}} (1-f_{L})]
|A( N_1 L \to \bar{H} )|^2
 (2\pi)^4\delta^4(P_{N_1}+P_{L}-P_{\bar{H}})
 \end{split}
\eeq
The last terms reconstructs the
thermal damping rate of $N_1$~\cite{Gamma(T)} 
(when the decaying particle is not $N_1$, but $H$),
due to interactions of $N_1$ with the $L$ present in the plasma 
producing a $\bar{H}$ (and the inverse process),
\beq
\begin{split}
\Gamma^{\rm th}_{N_1}(E_{N_1})=\frac{1}{2 E_{N_1}} \cdot 2 \int  d\vec{p}_{L} d\vec{p}_{\bar{H}}[(f_L)(1+f_{\bar{H}})+f_{\bar{H}} (1-f_L)] \\
|A_{\rm tree}(\bar{H} \to L {N_1})|^2
(2\pi)^4\delta^4(P_{N_1}+P_{L}-P_{\bar{H}}) .
\end{split}
\eeq

\section{Thermal correction to SM CP-asymmetries\label{AppEpsilon}}
In this appendix we show explicitly the calculation of
the CP asymmetry parameter $\epsilon$ for both
 the $N_1$ decay  and  the $H$ decay, in the SM.
 Since the final result for leptogenesis turns out to be
very weakly dependent on $\epsilon_H$,
we compute $\epsilon_H$ neglecting the motion of $H$ with
respect to the plasma.
On the contrary we make the analogous approximation for
$\epsilon_{N_1}$
only in the analytic expression presented in the main text,
eq.~(\ref{eps0}),
and present here the full result, averaged over
thermal $N_1$ motion.

\begin{figure}[t]
\centerline{\includegraphics[width=15cm]{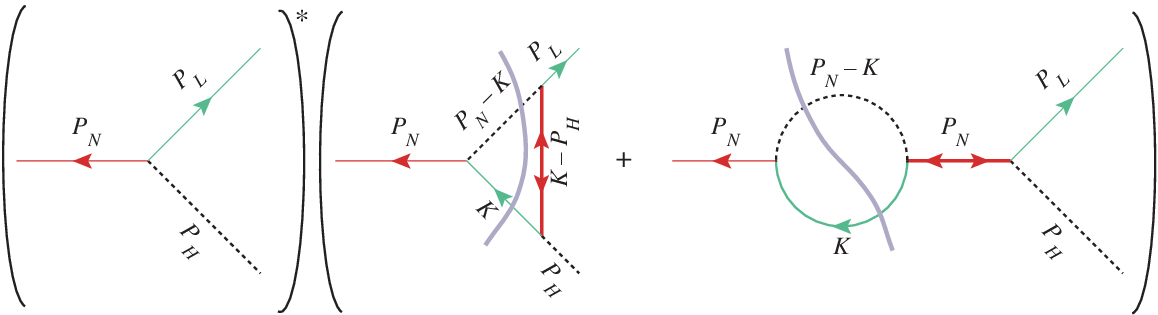} }
\caption{\em Im$[I_0^* (I_1^V+I_1^W)]$ in $N_1\to LH$
decay.
The momenta flow in the direction in which the labels are
written. }
\label{Ndecay}
\end{figure}

\begin{figure}[t]
\centerline{\includegraphics[width=15cm]{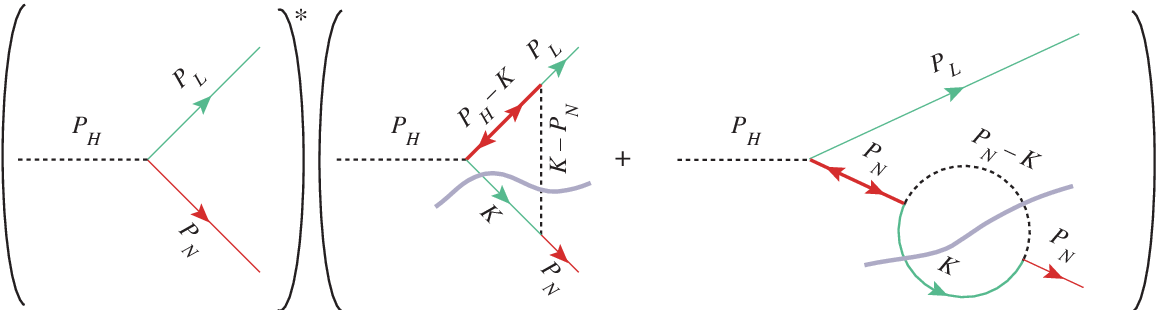} }
\caption{\label{Hdecay}\em Im$[I_0^* I_1]$ in $H\to LN$ decay.}
\end{figure}

We choose to work in the rest system of the plasma.
The CP-asymmetry in $i\to f$ decays
which enters into the Boltzmann equations is (see
\eq{eq:gammaD})
\begin{equation}
\epsilon_i=\frac{\gamma^{\rm eq}(i\rightarrow
f)-\gamma^{\rm eq}(\bar{\hbox{\em \i}}
\rightarrow\bar{f})}{\gamma^{\rm eq}(i \rightarrow
f)+\gamma^{\rm eq}(\bar{\hbox{\em \i}}\rightarrow
\bar{f})} = \frac{\int d^3\!p_i \,
f(E_i)\,\int d\phi\,\epsilon_A |A|^2 }
{\int d^3\!p_i  \, f(E_i)\,\int d\phi\, |A|^2} + {\cal
O}(\epsilon_i^2)
\end{equation}
where $d^3p_i~f(E_i)$ is the thermal distribution of the
initial state,
$d\phi$ is the standard relativistic phase-space for $i\to
f$ decay,
\begin{equation}
\epsilon_A=
\frac{|A(i\rightarrow f)|^2-
|A(\bar{\hbox{\em \i}} \rightarrow\bar{f})|^2}{|A(i
\rightarrow
f)|^2+|A(\bar{\hbox{\em \i}}
\rightarrow \bar{f})|^2}\, ,
\end{equation}
and $A$ is the amplitude for the process computed
at given quadri-momenta $P_{i,f}$. We define
$ A_{0}(i\rightarrow f)$ as the amplitude of  the
tree level
process  and parametrize it as $A_{0}=\lambda_0
I_0$, where
$\lambda_0$ contains the coupling constants. In the
same way we define the amplitude up to the one-loop
level as
 $A=\lambda_0 I_0+\lambda_1 I_1$.
With this notations  $\epsilon_A$ at one-loop is expressed
as
\begin{equation}
\epsilon_A=\frac{|\lambda_0 I_0+\lambda_1
I_1|^2-|\lambda_0^*
I_0+\lambda_1^* I_1|^2}{|\lambda_0 I_0+\lambda_1
I_1|^2+|\lambda_0^* I_0+\lambda_1^* I_1|^2} \simeq -2
\frac{\textrm{Im}[\lambda_0^* \lambda_1]  }{|\lambda_0 |^2}
\frac{\textrm{Im}[I_0^*I_1]}{|  I_0|^2}\, .
\end{equation}
In our case $I_1$ is the sum of two diagrams: a ``vertex''
correction and a ``wave'' correction (fig.s \ref{Ndecay},
\ref{Hdecay}).
For both diagrams the couplings and the
$|A_{0}(i\rightarrow f)|^2$
factor are the same:
\begin{equation}
\epsilon_A=-2 \frac{\textrm{Im}[(Y^{\dagger}
Y)_{1j}^2]}{(Y^{\dagger} Y)_{11}}
\frac{\textrm{Im}[I_0^* I_1]}{2 P_N \cdot P_L}
\label{genericepsilon}
\end{equation}
where the quadri-momenta $P_N$ and $P_L$ are defined in
fig.~\ref{Ndecay} for
$N_1$ decay
and in fig.~\ref{Hdecay} for $H$ decay.
We have to compute  $\textrm{Im}[I_0^* I_1]$.
Using the cutting rules, we obtain
\begin{equation}
\textrm{Im}[I_0^* I_1]=\frac{1}{2 i}I_0 \sum_{\rm cuttings}
I_1
\, .
\label{cuttings}
\end{equation}
$I_1$ is the sum of two diagrams: the ``vertex'' and the
``wave'' one:
 $I_1=I_1^V+I_1^W$.

 \subsubsection{Computation of $\epsilon_{N_1}$}
We consider here the case
of $N_1\rightarrow L H$.
In principle there are
three possible cuttings (or six circlings in the notation
of ref.~\cite{diagrammar,Kobes}) for $I_1^V$, and one
(or two circlings) for $I_1^W$, and none of them
 is forbidden at finite $T$. In fact, at finite $T$, energy
is
 no more forced to flow from uncircled to circled vertices
(see ref.~\cite{Kobes}). The reason is that, while the cut
propagators at
zero $T$ are proportional to $\theta(E)$, at finite $T$
they have a
new contribution, see eqs. (\ref{bosonpropagator})--(\ref{fermionpropagator}),
proportional to $f_F$ (or $f_B$)  if the cut particle is  a
fermion (or a
boson). This accounts for  particles in the thermal bath
which do not have  the $\theta (E)$ function, since 
they can be emitted by the bath (positive energy)
or absorbed from the bath (negative energy).
However, recalling that we are working under
the assumption that the $N_j$ (with $j\neq 1$) are very
heavy,
the cuttings which involve the $N_j$ can
 be neglected since they are exponentially suppressed
 by $m_{N_j}/T$.
Moreover the graph with the $N_1$ circulating in the loop
does not
contribute to the CP asymmetry since its
Yukawa
couplings are real (see eq.~(\ref{genericepsilon}) with
$j=1$).
So, the only relevant cutting in $I_1^V$ is the one with
lepton
line and Higgs line cut (fig.~\ref{Ndecay}), as in the 
zero-temperature case.

We compute here the cutting in
the vertex part
\begin{equation}
\textrm{Im}[I_0^* I_1^V] =\frac{1}{2i} \int
\frac{d^4\!K}{(2 \pi)^4} D_N(P_L-P_N+K)\left[ D_H^+
(P_N-K) D_L^+(K)+D_H^-(P_N-K) D^-_L(K)\right] T(K) ,
 \label{integrale}
 \end{equation}
which is the first contribution shown in fig.~\ref{Ndecay}
(the single cutting in the figure stands for two possible
circlings
of the vertices).
Here  $T(K)$ is the result of the traces over the spinor
indices,
$D$ are the propagators (without numerators): $D_N$ is the
propagator of the $N_j$
(which we choose as the zero-temperature  one,
since we neglect $N_j$ interactions with the plasma),
$D_H^{\pm}$ and $D_L^{\pm}$  are the finite-$T$
cut propagators of the Higgs and of the lepton respectively,
see eq.~(\ref{bosonpropagator}) and
(\ref{fermionpropagator}),
\begin{eqnarray}
 D_N(P_L-P_N+K)&=&\frac{1}{(P_L-P_N+K)^2-m_{N_j}^2} 
\nonumber \\
 D_H^{\pm}(P_N-K)&=&2 \pi
\delta\big((P_N-K)^2-m_H^2\big)\left[
 \theta(\pm(E_N-\omega)) +f_B(|E_N-\omega|) \right]\, ,
 \\
D_L^{\pm}(K)&=& 2 \pi
\delta\big(\delta_L(\omega,k)\big) \left[ \theta(\pm
\omega) -f_F(|\omega|)
\right] \, ,
\nonumber
\end{eqnarray}
where
\begin{equation}
\delta_L= [(1+a)\omega + b]^2 -(1+a)^2k^2
   \label{deltaL}
\end{equation}
and $a,b$ are defined in \eq{aaa}.
Keeping only relative angles relevant for our computation,
we can conveniently parameterize the quadri-vectors as
$$
U = (1,0,0,0),\qquad
P_N = (E_N, p_N,0,0), \qquad
P_H = P_N - P_L = (E_H,\vec{p}_H)$$
$$
P_L =((E_L^0 E_N+ p_0 p_N\cos\theta_0)/m_{N_1},(p_0
E_N\cos\theta_0 + E_L^0 p_N)/m_{N_1},
p_0\sin\theta_0,0)=(E_L,\vec{p}_L)$$
$$K = (\omega,
k\cos\theta,k\sin\theta\cos\varphi,k\sin\theta\sin\varphi)=(\omega,\vec{k})$$
where $E_L^0 = (m_{N_1}^2+m_L^2-m_H^2)/2m_{N_1}$ and
$p_0 = \sqrt{(E_L^0)^2-m_L^2}$ are the $L$ energy and
momentum
with respect to the $N$ rest frame and $\theta_0$ is a
decay angle
in the same frame.

Now we compute the trace part.
Using four-component Majorana spinors\footnote{The direction of 
the arrows for Majorana spinors in fig.~\ref{Ndecay} is arbitrary,
and one is free
to choose it as a matter of convenience;
the particular choice made dictates which Feynman rules are
used.},
we get
$$T=\sum_{\rm polarizations}\left[\bar{u}_L \left(i
\frac{1-\gamma_5}{2}\right) C \bar{v}_{N_1}  \right]^*
 \bigg[\bar{u}_L \left(i \frac{1-\gamma_5}{2}\right)
\bigg(-i [\Ksl-\Psl_H+m_{N_j}]  C\bigg)
\times
$$
\begin{equation}\label{eq:trace}
\times
\bigg(i \frac{1-\gamma_5}{2}\bigg) \bigg(-i \left[(1+a)
\Ksl + b \Usl \right]
\bigg) \left(i
\frac{1+\gamma_5}{2}\right) \bar{v}_{N_1} \bigg]\, ,
\end{equation}
where $C$ is the charge conjugation matrix with the
properties
\begin{equation}
C \bar{v}=u \ ,\ \   \{ C,\gamma^{\mu}  \}=0  \ , \ \
[C,\gamma^5]=0\, .
\end{equation}
Then eq.~(\ref{eq:trace}) becomes
\begin{eqnarray}\nonumber
T(K) &=& -i\  \hbox{Tr}\, \left[m_{N_1}  \frac{1+\gamma_5}{2}\Psl_L
m_{N_j}
[(1+a) \Ksl + b \Usl]    \right]\,\\
&=&-2i m_{N_1} m_{N_j}[ (1+a)\, P_L \cdot K +b\,  E_L]\, .
 \label{T(q)}
\end{eqnarray}
Note that the term $b \Usl$ is not put to zero
 by chirality projectors  unlike a usual mass term.


We have to perform the integral in $d^4\!K$  in
eq.~(\ref{integrale}).
The most convenient technique is to use polar coordinates
$\theta$ and $\varphi$
and
integrate first in $d\cos\theta$ using the
$\delta[(P_N-K)^2-m_H^2]$
and then integrate in $d k$ using the
$\delta[\delta_L(\omega,k)]$.
As we discussed in section~\ref{cpN1},
we can approximate the $L$ dispersion relation
with $\omega^2-k^2=m_L^2$, finding the
following solution
\begin{equation}
 \cos\theta=\frac{m_H^2
- m_L^2 - m_{N_1}^2+ 2 E_N \omega}{2 k{p}_N }\,
,  \qquad
         k=\sqrt{\omega^2 - m_L^2}  \,
.\label{costheta}
\end{equation}
Imposing that $|\cos\theta|\leq 1$ we find that
 $\omega$ must be comprised between two positive values
$\omega_{\rm min}$ and $\omega_{\rm max}$.
 This implies that in the arguments of the
$\theta$-functions
 in eq.~(\ref{integrale}) we have $\omega >0$ and
 $E_N-\omega>0$, so
 that expanding
 $D_H^+ (P_N-K) D_L^+(K)+D_H^-(P_N-K) D^-_L(K)$ gives
\begin{eqnarray}
N&\equiv&
[1+f_B(E_N-\omega)][1-f_F(\omega)]-
f_B(E_N-\omega) f_F(\omega)
\nonumber
\\
 &=&1+f_B(E_N-\omega)-f_F(\omega)-2 f_B(E_N-\omega)
f_F(\omega)    \, .
\label{distrib}
\end{eqnarray}
At this point we are left  with integrals in $d\omega$ and
$d\varphi$.
Adding also the contribution of the ``wave'' diagram
the result is
%
given by
\begin{equation}
\epsilon_A=\frac{\textrm{Im}[(Y^{\dagger}
Y)_{1j}^2]}
{2\pi(Y^{\dagger} Y)_{11}\ |I_0|^2
}
 \int_0^{2 \pi} \frac{d\varphi}{2\pi}
\int_{\omega_{\rm min}}^{\omega_{\rm max}}
d\omega \frac{ k m_{N_1} m_{N_j}
}{|\partial\delta_L/\partial k|\,p_N}
[(1 + a)P_L\cdot K + b E_L]\,N\, [ P_V +
2 P_W ]\, , \label{ResultEpsN}
\end{equation}
where
\begin{equation}
|I_0|^2=2 P_N \cdot P_L=
m_{N_1}^2-m_H^2+m_L^2.
\end{equation}
$P_V$ and $P_W$ arise from $N_j$ propagators
in vertex and wave diagrams respectively
\begin{equation}
P_V=\frac{1}{m_H^2+m_L^2-m_{N_j}^2-2 P_H\cdot K}\, ,
\qquad
P_W=\frac{1}{m_{N_1}^2-m_{N_j}^2}.
\end{equation}
$P_W$ appears in \eq{ResultEpsN} multiplied by a factor $2$
 since in the wave diagram also the charged states of Higgs and
lepton fields can propagate~\cite{Covi}.
Finally, the relevant quantity for the Boltzmann equations
is
the average of $\epsilon_A$ over the phase space and the
thermal distribution of $N_1$:
\begin{equation}
\epsilon_{N_1}(T) =
 \frac{\int_{m_{N_1}}^\infty dE_{N}\, p_N \  f_B(E_N)\int_{-1}^1
d\cos\theta_0\  \epsilon_A (E_N,\theta_0)}
 {2\int_{m_{N_1}}^\infty dE_{N}\,p_N\  f_B(E_N) } \, .
\end{equation}
The result is presented in fig.~\ref{figepsilon}.
In the limit $m_{N_j}\gg m_{N_1}$ one has $P_V\approx
P_W\approx - 1/m_{N_j}^2$
and the integrals in $\varphi$ and $\theta_0$ can be done
analytically.
Approximating $f_B(E_N)\approx e^{-E_N/T}$
the explicit result is
$$ {\epsilon_{N_1}(T)\over\epsilon_{N_1}(0)} =\int_{m_{N_1}}^\infty dE_{N} 
\int_{\omega_{\rm min}}^{\omega_{\rm max}}d\omega~
[(1+a) (m_{N_1}^2-m_H^2+m_L^2)+2b
E_N] \frac{2km_{N_1} N\,e^{-E_N/T}}
{T\,{\rm K}_1({m_{N_1}}/{T})|\partial\delta_L/\partial k|}.$$

\subsubsection{Computation of $\epsilon_{N_1}$ for $N_1$ at
rest}
We have then performed the computation for the $N_1$ decay
in
the simplified situation with $N_1$ at rest in the plasma.
In this case, the $\delta((P_N-K)^2-m_H^2)$
factor does
not contain the angle $\theta$ in the scalar product
$P_N\cdot K$.
So we can proceed by integrating in $dk$ and
in $d\omega$
 using the two $\delta$-functions.
Moreover there is no dependence on the angles $\varphi$ and
$\theta_0$. The
 integrals in $d\varphi$ and $d\cos\theta_0$ are trivial
and the integral in
$d\cos\theta$
can be done analytically.
In this way one obtains the result of
section \ref{cpN1}.
The difference with respect to our full computation
is not fully negligible, as shown in fig.~\ref{figepsilon}a.

\subsubsection{Computation of $\epsilon_H$}
Finally, the same technique of computation can be applied
to the case of
 $H \rightarrow L N_1$.
There are, though, remarkable differences between the two
decays.
In  $H$ decay (see fig.~\ref{Hdecay}) the particles in
the loop
could never go on-shell with the usual $T=0$ Feynman rules.
This is due
to the presence of the $\theta$-functions in the cut
propagators.
However, using the finite-$T$ rules we know that particles
with
negative energy can be absorbed from the bath so
the process can have a CP asymmetry.
We can make the cuts in fig.~\ref{Hdecay},
 where again cuts of very heavy particles have been
neglected.
The $\delta$-functions have solutions identical to
eq.~(\ref{costheta}), in terms of the external
momenta
(see the notation of fig.~\ref{Hdecay}). However
 the masses are such that $\omega_{\rm min}<\omega_{\rm max}<0$.
So imposing $|\cos\theta|\leq 1$, 
$\omega$
has to lie
between two \textit{negative} values
(it is exactly for this reason that with $T=0$ Feynman
rules $\epsilon_H$ would
be zero), while $E_N-\omega$ is positive.
Then the products of $\theta$-functions are different from the previous case
\begin{equation}
\begin{split}
D_H^+ (P_N-K) D_L^+(K)+D_H^-(P_N-K)D^-_L(K) \propto
\\
& \makebox[-8cm]{} \propto
[f_B(|E_N-\omega|)][1-f_F(|\omega|)]-
[1+f_B(|E_N-\omega|)] f_F(|\omega|)
\\
& \makebox[-8cm]{} =f_B(|E_N-\omega|)-f_F(|\omega|)-2
f_B(|E_N-\omega|) f_F(|\omega|) .
\end{split}
\label{distribH}
\end{equation}
Indeed \eq{distribH}
is equal to 0 (and not to $1$) at $T=0$.

The result in this case has the same form of eq.~(\ref{ResultEpsN}).
As already explained, since the final effect of $\epsilon_H$ in our
scenario is
very small,
 we computed it only taking the incoming $H$ at rest,
 in order to simplify the computations. However,
in this case
it is not possible to give an analytic result as opposed to the
case of $N_1$ decay
at rest. The reason is that this time the
$\delta((P_N-K)^2-m_H^2)$
factor contains always the dependence on the
angle $\theta$, since $N_1$ is the outgoing particle
and $P_N$ is not of the form $P_N=(m_{N_1},0,0,0)$. 

The most convenient order of integration is the same we
followed
in the $N_1$ decay. We obtain again a result in an implicit
form
(even if there is no dependence on $\alpha$ and $\varphi$)
\begin{eqnarray}
\epsilon_H(T)=\frac{\textrm{Im}[(Y^{\dagger}
Y)_{1j}^2]}
{(Y^{\dagger} Y)_{11}|I_0|^2
}\cdot 2\pi
\int_{q_m}^{q_M} d\omega \frac{m_{N_1} m_{N_j}}{
|\partial\delta_L/\partial k| (2 \pi)^2}\
\left[(1 + a)\left(E_L \omega - p_L k \cos{\theta}\right) +
b E_L \right]\nonumber \\
 \times [f_B(|E_N-\omega|)-f_F(|\omega|)-2
f_B(|E_N-\omega|) f_F(|\omega|)]
[P_V(\omega) +2 P_W(\omega)]
\label{resultEpsH}\, ,
\end{eqnarray}
where $\delta_L$ is given by eq.(\ref{deltaL}) and $I_0$ is
the tree level rate
for this decay
\begin{equation}
|I_0|^2=m_H^2-m_{N_1}^2-m_L^2.  \label{treeH}
\end{equation}
%
%
The on-shell conditions fix
\begin{equation}
  k=\sqrt{\omega^2-m_L^2} \, ,
 \qquad    \cos{\theta}=\frac{m_{N_1}^2 + m_L^2-m_H^2 - 2
E_N \omega }{2 p_N k}.
\nonumber
\end{equation}
Finally $P_V(\omega)$ and $P_W(\omega)$ in this case are
given by
\begin{equation}
P_V(\omega)=\frac{k}{\, p_N \,
    \left( {{m_H}}^2 + {{m_L}}^2 - {{m_{N_j}}}^2 -
      2\,{m_H}\,{\omega} \right) }\, ,
\end{equation}
\begin{equation}
\qquad P_W(\omega)=\frac{k}{p_N \,\left(
{{m_{N_1}}}^2 - {{m_{N_j}}}^2 \right) \,} \, .
\end{equation}
Note that the tree level rate for this process in
eq.(\ref{treeH})
is small at high $T$ as long as the value of $m_H$ is
near to $m_L$. For this reason $\epsilon_H$ is bigger than
$\epsilon_{N_1}$. In fact if $|I_0|^2$ were
 ${\cal O}(1)T^2$ at high $T$, then $\epsilon_H$ would be
as big
as
$\epsilon_{N_1}$. Instead,  it goes like $c T^2$, where $c$
is a small number: $c=(m_H^2-m_L^2)/T^2$, and so
$\epsilon_H$
becomes $c^{-1}$ times bigger than $\epsilon_{N_1}$. In
particular
for the SM thermal masses $c^{-1}$ is  about
$15$.

\section{Thermal correction to  MSSM CP-asymmetries}\label{CPMSSM}
In the MSSM the situation becomes more complicated than in the SM
because
we must study both $N_1$ and $\tilde{N_1}$ decays,
each having two possible decay channels,
with each channel having more diagrams.
Including thermal effects, the relevant masses satisfy
$$m_{N_{1,2,3}} = m_{\Nt_{1,2,3}},\qquad
a_\Ht\equiv \frac{m_\Ht^2}{m_{N_1}^2} =
\frac{a_H}{2},\qquad
a_\Lt\equiv \frac{m_\Lt^2}{m_{N_1}^2} = 2 a_L.$$
As previously discussed we use approximate dispersion relations for the fermions,
and assume $m_{N_j}\gg m_{N_1}$.
For simplicity,
in the MSSM case
we compute the decay rates
and the CP-asymmetries
neglecting the velocity of the decaying particle with
respect to the plasma.

The CP-asymmetries in $N_1$ and $\tilde{N}_1$ decays are
\begin{eqnarray*}
\epsilon_{\Nt_1}(T) &\equiv&
\frac{\Gamma(\Nt_1\rightarrow H\Lt)-\Gamma(\Nt_1\rightarrow
\bar{H} \bar{\Lt}) +
\Gamma(\Nt_1\rightarrow\tilde{H} L)-\Gamma(\Nt_1\rightarrow
\bar{\tilde{H}} \bar{{L}})}
{\Gamma(\Nt_1\rightarrow H\Lt)+\Gamma(\Nt_1\rightarrow
\bar{H} \bar{\Lt})+
\Gamma(\Nt_1\rightarrow\tilde{H} L)+\Gamma(\Nt_1\rightarrow
\bar{\tilde{H}} \bar{{L}})}\\
&=&\epsilon_{\Nt_1}(T=0)
\frac{R_\Gamma(\Nt_1\to \Ht L) R_\epsilon(\Nt_1\to \Ht L)
 + R_\Gamma(\Nt_1\to H \Lt) R_\epsilon(\Nt_1\to H \Lt)}
{R_\Gamma(\Nt_1\to \Ht L)  + R_\Gamma(\Nt_1\to H\Lt)}\\
\epsilon_{N_1}(T) &=&  \epsilon_{N_1}(T=0)
 R_\epsilon(N_1\to HL)= \epsilon_{N_1}(T=0)
 R_\epsilon(N_1\to \Ht\Lt)
\end{eqnarray*}
where
$$R_\Gamma(i\to f) \equiv \frac{\Gamma(i\to f\hbox{ at }
T)}{\Gamma(i\to f\hbox{ at } T=0)}\qquad
R_\epsilon(i\to f) \equiv \frac{\epsilon(i\to f\hbox{ at }
T)}{\epsilon(i\to f\hbox{ at } T=0)}.$$
We have used the fact that
at $T=0$ the two decay channels have equal zero-temperature
width and CP-asymmetries
$\epsilon_{N_1}(T=0) = \epsilon_{\Nt_1}(T=0)$ given in eq.~(\ref{epsss}).

We now give the explicit expressions for the $R_\epsilon$.
Consider first the $N_1$ decays.
Two more one loop diagrams  (not plotted) contribute to the CP-asymmetry:
a ``vertex'' and a ``wave'' diagram with sparticles in the loop.
The SM function $R_\epsilon(N_1\to HL)$ has been given in
section~\ref{cpN1}.
Using analogous self-explanatory notations we find, in the
MSSM
\begin{eqnarray*}\label{gpgs}
R_\epsilon(N_1\to HL) &=& 8 \frac{k_L^2}{m_{N_1}}
[\omega_L(1+a_L)+b_L]
[1+f_H-f_L-2 f_H f_L]
\left|\left|
\begin{matrix}
 \partial \delta_H/\partial\omega_L &\partial
\delta_L/\partial\omega_L \\
\partial \delta_H/\partial k_L &\partial \delta_L/\partial
k_L
\end{matrix}\right|\right|^{-1}+\\
&& 8 \frac{k^2_\Ht}{m_{N_1}}
[\omega_\Ht(1+a_\Ht)+b_\Ht]
[1+n_\Lt-n_\Ht-2 n_\Lt n_\Ht]
\left|\left|
\begin{matrix}
 \partial \delta_\Lt/\partial\omega_\Ht &\partial
\delta_\Ht/\partial\omega_\Ht \\
\partial \delta_\Lt/\partial k_\Ht &\partial
\delta_\Ht/\partial k_\Ht
\end{matrix}\right|\right|^{-1}
\end{eqnarray*}
where
$$\begin{array}{ll}
\delta_H \equiv (m_{N_1}-\omega_L)^2 - k_L^2 - m_H^2
\qquad&
\delta_L \equiv [(1+a_L)\omega_L +b_L]^2-(1+a_L)^2 k_L^2
 \cr
\delta_\Lt \equiv (m_{N_1}-\omega_\Ht)^2 - k^2_\Ht -m_\Lt^2
&
 \delta_\Ht \equiv
[(1+a_\Ht)\omega_\Ht+b_\Ht]^2-(1+a_\Ht)^2k_\Ht^2 .
 \end{array}$$
$R_\epsilon(N_1\to \Ht \Lt)$ has to be evaluated at the
values of $k_L$ and $\omega_L$ which solve
$\delta_H=\delta_L=0$, approximatively given by
eq.~(\ref{eq:omegak}),
and at the values of $k_\Ht$ and $\omega_\Ht$ which solve
$\delta_\Ht =\delta_\Lt = 0$,
approximatively given by
$\omega_\Ht = (m_{N_1}^2 + m_\Ht^2-m_\Lt^2)/2m_{N_1}$ and
$k_\Ht=(\omega^2_\Ht-m_\Ht^2)^{1/2}$.

Finally $R_\epsilon(N_1\to \Ht\Lt) =R_\epsilon(N_1\to HL)$
since the CP-asymmetry in the two $N_1$ decay modes
is due to loops with the same virtual particles.

\medskip

For $\Nt_1$ decays, the situation is different.
When $\Nt_1$ decays into two fermions
($\tilde{N}_1\to \tilde{H}L$) the imaginary part is
obtained
cutting two internal bosons, $H$ and $\tilde{L}$.
The decay rate is suppressed by two Pauli-blocking factors,
but its CP-asymmetry is enhanced by two stimulated-emission
factors.
\begin{equation}
R_\epsilon(\Nt_1 \to \Ht L) = 2 \frac{k_H}{m_{\Nt_1}}
 [1+f_B(\omega_\Lt)+f_B(\omega_H)+2 f_B(\omega_\Lt)
f_B(\omega_H)]
\end{equation}
where
$\omega_H = (m_{\Nt_1}^2 + m_H^2-m_\Lt^2)/2m_{\Nt_1} =
 m_{\Nt_1}-\omega_\Lt$
and $k_H=(\omega_H^2-m_H^2)^{1/2}$.

When $\Nt_1$ decays into two bosons
($\tilde{N}_1\to H\Lt$) the imaginary part is obtained
cutting two internal fermions, $\Ht$ and $L$,
and it is therefore given by a lengthy expression.
The CP-asymmetry is suppressed by two Pauli-blocking
factors,
but the decay rate is enhanced by two stimulated-emission
factors.
$$
R_\epsilon(\Nt_1 \to H\Lt)  = 16\frac{k^2}{m^2_{\Nt_1}}
\Big\{
[(1+a_L)\omega+b_L ] 
 [(1+a_\Ht)\omega_\Ht +b_{\tilde{H}} ]
 + k^2  (1+a_L) (1+a_\Ht)
 \Big\} \times
 $$
\begin{equation}\qquad\times
[1-f_F(m_{\Nt_1}-\omega)-f_F(\omega)+2
f_F(m_{\Nt_1}-\omega)f_F(\omega)] \left|\left|
\begin{matrix}
 \partial \delta_L/\partial\omega &\partial
\delta_{\tilde{H}}/\partial\omega \\
\partial \delta_L/\partial k &\partial
\delta_{\tilde{H}}/\partial k
\end{matrix}\right|\right|^{-1}
\end{equation}
where $(\omega,k)$ is the quadri-momentum of $L$,
$\omega_\Ht = m_{\Nt_1}-\omega$ is the energy of $\Ht$ and
$$
\delta_{\tilde{H}}\equiv [(1+a_{\tilde{H}})(m_{\Nt_1}-\omega)
+b_{\tilde{H}}]^2-(1+a_{\tilde{H}})^2 k^2
\qquad
\delta_L \equiv [(1+a_L)\omega +b_L]^2-(1+a_L)^2k^2.$$
The expression should be evaluated at the values of
$\omega$ and $k$ which
solve $\delta_L = \delta_\Ht = 0$.
Using the approximate on-shell condition, they are given by
$$\omega = (m_{\Nt_1}^2 +m_L^2 - m_\Ht^2)/2m_{\Nt_1},\qquad
k = (\omega^2-m_L^2)^{1/2}.$$

Finally, the thermal corrections to the decay rates are
given by
\begin{eqnarray*}
R_\Gamma (N_1\to HL) &=& (1+ f_B(E_H))(1-f_F(E_L))
\lambda^{1/2}(1,a_H,a_L)(1-a_H+a_L)\\
R_\Gamma (N_1\to \tilde{H}\tilde{L}) &=&
(1-f_F(E_\Ht))(1+f_B(E_\Lt))\lambda^{1/2}(1,a_\Ht,a_\Lt)(1+a_\Ht+a_\Lt)\\
R_\Gamma (\Nt_1\to \Ht L) &=& (1- f_F(E_\Ht))(1-f_F(E_L))
\lambda^{1/2}(1,a_\Ht,a_L)(1-a_\Ht-a_L)\\
R_\Gamma (\Nt_1\to H\Lt) &=& (1+ f_B(E_H))(1+f_B(E_\Lt))
\lambda^{1/2}(1,a_H,a_\Lt) .
\end{eqnarray*}
The $1\pm f_{B,F}$ factors take into account Pauli blocking
or stimulated emission
(the thermal distributions must be evaluated at the
energies of final state particles)
while the other factors arise from thermal corrections to
kinematics.

\paragraph{Note added}
Some of our preliminary results appeared in~\cite{ery}.  However, 
in that paper sneutrino reheating was not correctly included
(we also take into account thermal corrections and proper subtraction of on-shell scatterings).

\frenchspacing

\begin{multicols}{2}

\end{multicols}

\footnotesize
\begin{thebibliography}{99}

\bibitem{oscdata}
The atmospheric mixing parameters are extracted from 
the revised SK analysis presented by K. Nishikawa at the 2003
Lepton-Photon Conference (see the web site conferences.fnal.gov/lp2003)
and from
\art[hep-ex/0212007]{K2K collaboration}{\PRL}{90}{2003}{041801}.
For a recent global fit see
L. Fogli, E. Lisi, A. Marrone, D. Montanino, A. Palazzo, A.M. Rotunno, hep-ph/0310012.
The solar mixing parameters are taken from the global fit in
\art[hep-ph/0102234]{P. Creminelli et al.}{\Jhep}{05}{2001}{052}.
Its e-print version, has been updated
including the recent data from SNO
(SNO collaboration, nucl-ex/0309004), KamLAND
(KamLAND collaboration, hep-ex/0212021), etc.


\bibitem{seesaw}
M. Gell-Mann, P. Ramond and R. Slansky, proceedings of the supergravity Stony Brook
workshop, New York, 1979, ed.s P. Van Nieuwenhuizen and D. Freedman (North-Holland,
Amsterdam); T. Yanagida, proceedings of the workshop on unified theories and baryon
number in the universe, Tsukuba, Japan 1979 (ed.s. O. Sawada and A. Sugamoto, KEK
Report No. 79-18, Tsukuba).
 \art{R.N. Mohapatra, G. Senjanovi\'c}{\PRL}{44}{912}{1980}.



\bibitem{fuk}
M.~Fukugita and T.~Yanagida,
Phys.\ Lett.\ B {174}, 45 (1986).

\bibitem{reviewbau} For reviews, see 
A.~Riotto, hep-ph/9807454.
A.~Riotto, M.~Trodden,
Ann.\ Rev.\ Nucl.\ Part.\ Sci.\  {49}, 35 (1999).

\bibitem{sak}
A.~D.~Sakharov,
Pisma Zh.\ Eksp.\ Teor.\ Fiz.\  {5}, 32 (1967)
[JETP Lett.\  {5}, 24 (1967)].


\bibitem{k-sm}
M.~A.~Luty, Phys.\ Rev.\ D {45} (1992) 455;
M.~Pl\"umacher, Z.\ Phys.\ C {74} (1997) 549;
W.~Buchmuller, M.~Pl\"umacher,
Int.\ J.\ Mod.\ Phys.\ A {15}, 5047 (2000)
[hep-ph/0007176].


\bibitem{diagrun}
See e.g.\
J.~A.~Casas, J.~R.~Espinosa, A.~Ibarra, I.~Navarro,
Nucl.\ Phys.\ B {\bf 573} (2000) 652
[hep-ph/9910420]
and~\cite{SMRGE2}.

\bibitem{BRS} \art[hep-ph/9906470]{R. Barbieri et al.}{JHEP}{10}{20}{1999}.




\bibitem{vari}
T.~Hambye, E.~Ma, U.~Sarkar, Nucl.\ Phys.\ B {590} (2000) 429 [hep-ph/0006173];
S.~Davidson, A.~Ibarra, JHEP {0109} (2001) 013;
G.C.~Branco, T.~Morozumi, B.M.~Nobre, M.N.~Rebelo, Nucl.\ Phys.\ B {617} (2001) 475;
A.S.~Joshipura, E.~A.~Paschos, W.~Rodejohann, JHEP {0108} (2001) 029;
D. Falcone, Phys. Rev. D {66} (2002) 053001 [hep-ph/0204335];
G.~C.~Branco, R.~Gonzalez Felipe, F.R.~Joaquim, M.N.~Rebelo, Nucl.\ Phys.\ B {640} (2002) 202;
W.~Rodejohann, Phys.\ Lett.\ B {542} (2002) 100;
J.R.~Ellis, M.~Raidal, Nucl.\ Phys.\ B {643} (2002) 229 [hep-ph/0206174];
G.C.~Branco, R.~Gonzalez Felipe, F.R.~Joaquim, I.~Masina, M.N.~Rebelo, 
C.A.~Savoy, hep-ph/0211001;
S.F.~King, hep-ph/0211228;
S.~Pascoli, S.T.~Petcov, C.E.~Yaguna, hep-ph/0301095;
S.~Davidson, JHEP {0303} (2003) 037 [hep-ph/0302075];
E. Kh. Akhmedov, M. Frigerio, A. Yu. Smirnov, hep-ph/0305322.


\bibitem{mBound}
W.~Buchmuller, P.~Di Bari, M.~Pl\"umacher,
Phys.\ Lett.\ B {547} (2002) 128 [hep-ph/0209301];
W.~Buchmuller, P.~Di Bari, M.~Pl\"umacher,
Nucl.\ Phys.\ B {665} (2003) 445 [hep-ph/0302092].


\bibitem{softl2}
Y.~Grossman, T.~Kashti, Y.~Nir, E.~Roulet,
hep-ph/0307081.

\bibitem{softl}
G.~D'Ambrosio, G.~F.~Giudice, M.~Raidal,
hep-ph/0308031.


\bibitem{bellac} For a review see
M. Le Bellac, {\it Thermal Field Theory},
Cambridge University Press, Cambridge, England, 1996.


\bibitem{enqvist}
P.~Elmfors, K.~Enqvist, I.~Vilja, Nucl.\ Phys.\ B {\bf
412}, 459 (1994).


\bibitem{weldon}
V.V. Klimov, Sov. J. Nucl. Phys. 33, 934 (1981);
H.~A.~Weldon, Phys.\ Rev.\ D {26}, 2789 (1982)
and Phys.\ Rev.\ D {40}, 2410 (1989).

\bibitem{yaffe} D.~J.~Gross, R.~D.~Pisarski,
L.~G.~Yaffe,
Rev.\ Mod.\ Phys.\  {53}, 43 (1981).





\bibitem{Kin} T.~Kinoshita, J. Math. Phys. {3}, 650 (1962).

\bibitem{LN} T.~D.~Lee, M.~Nauenberg, Phys. Rev. {133}, 1549 (1964).


\bibitem{BPR} 
R.~D.~Pisarski, Phys. Rev. Lett. {63}, 1129 (1989); 
E.~Braaten, R.~D.~Pisarski, Nucl. Phys. {B337}, 569 (1990); 
{\it ibidem} {B339}, 310 (1990).

\bibitem{weldonir} H.~A.~Weldon,
Phys.\ Rev.\ D {44}, 3955 (1991).


\bibitem{mikko}
K.~Kajantie, M.~Laine, K.~Rummukainen,
M.~E.~Shaposhnikov,
Nucl.\ Phys.\ B {458} (1996) 90
[hep-ph/9508379]. 


\bibitem{epsilon}
\hepart[hep-ph/0312203]{T. Hambye, Yin Lin,
A. Notari, M. Papucci, A. Strumia}.


\bibitem{SMRGE2} 
\hepart[hep-ph/0305273]{S.~Antusch, J.~Kersten, M.~Lindner, M.~Ratz}.



\bibitem{heavym0} \art[hep-ph/9912301]{A. Romanino,
 A. Strumia}{\PL}{B487}{165}{2000}.

\bibitem{thermalmasses}
Thermal masses can be found in the several works,
after fixing various discrepancies.
See refs.~\cite{enqvist,weldon,Olive49} and
D.~Comelli, J.~R.~Espinosa,
Phys.\ Rev.\ D {55} (1997) 6253
[hep-ph/9606438].


\bibitem{CoviTh}
L.~Covi, N.~Rius, E.~Roulet, F.~Vissani,
Phys.\ Rev.\ D {57} (1998) 93
[hep-ph/9704366].


\bibitem{Kobes}
R.~L.~Kobes, G.~W.~Semenoff,
Nucl.\ Phys.\ B {260} (1985) 714;
R.~L.~Kobes, G.~W.~Semenoff,
Nucl.\ Phys.\ B {272} (1986) 329.



\bibitem{Covi}
L.~Covi, E.~Roulet, F.~Vissani,
Phys.\ Lett.\ B {384}, 169 (1996)
[hep-ph/9605319].


\bibitem{bcst} 
R.~Barbieri, P.~Creminelli, A.~Strumia, N.~Tetradis,
Nucl.\ Phys.\ B {575} (2000) 61 [hep-ph/9911315].
The hep-ph version will present revised
analytical approximations, in the light of the modified 
Boltzmann equations discussed here.

\bibitem{di2}
S.~Davidson, A.~Ibarra,
Phys.\ Lett.\ B {535} (2002) 25.

\bibitem{bbp}
W.~Buchmuller, P.~Di Bari, M.~Plumacher,
Nucl.\ Phys.\ B {643} (2002) 367
[hep-ph/0205349].

\bibitem{pilaf}
A.~Pilaftsis, T.~E.~Underwood,
hep-ph/0309342.


\bibitem{k-mssm}
M.~Pl\"umacher,
Nucl.\ Phys.\ B {530} (1998) 207
[hep-ph/9704231].

\bibitem{mueg}
F. Borzumati, A. Masiero, Phys. Rev. Lett. 57 (1986) 961;
\art{L.J.~Hall, V.A.~Kostelecky, S.~Raby}{\NP}{B267}{415}{1986};
\art[hep-ph/9501407]{J. Hisano, T. Moroi, K. Tobe, M. Yamaguchi, T. Yanagida}{\PL}{B357}{579}{1995};
J.~Hisano, T.~Moroi, K.~Tobe, M.~Yamaguchi, Phys.\ Rev.\ D {53} (1996) 2442.
For some recent analyses see e.g.\
J.A.~Casas, A.~Ibarra, Nucl.\ Phys.\ B {618} (2001) 171;
\hepart[hep-ph/0104076]{S.~Davidson, A.~Ibarra};
\art[hep-ph/0106245]{S.~Lavignac, I.~Masina, C.~A.~Savoy}{\PL}{B520}{269}{2001};
J.R.~Ellis, J.~Hisano, S.~Lola, M.~Raidal, Nucl.\ Phys.\ B {621} (2002) 208 [hep-ph/0109125];
\art[hep-ph/0108275]{A. Romanino, A. Strumia}{\NP}{B622}{73}{2002};
J.R.~Ellis, J.~Hisano, M.~Raidal, Y.~Shimizu, Phys.\ Rev.\ D {66} (2002) 115013 [hep-ph/0206110];
A.~Dedes, J.R.~Ellis, M.~Raidal, Phys.\ Lett.\ B {549}, 159 (2002) [hep-ph/0209207].


\bibitem{minimalseesaw}
M. Raidal, A. Strumia, Phys. Lett. B553 (2003) 72 [hep-ph/0210021].




\bibitem{reviewinf} For a review on inflation, see 
D.~H.~Lyth, A.~Riotto,
Phys.\ Rept.\  {314}, 1 (1999) 
[hep-ph/9807278].
\bibitem{Davidson:2000dw}
S.~Davidson, M.~Losada, A.~Riotto,
Phys.\ Rev.\ Lett.\  {84}, 4284 (2000)
[hep-ph/0001301].

\bibitem{Giudice:2000dp}
G.~F.~Giudice, E.~W.~Kolb, A.~Riotto, D.~V.~Semikoz,
I.~I.~Tkachev,
Phys.\ Rev.\ D {64}, 043512 (2001)
[hep-ph/0012317].
\bibitem{giudiceetal}
G.~F.~Giudice, E.~W.~Kolb, A.~Riotto,
Phys.\ Rev.\ D {64}, 023508 (2001)
[hep-ph/0005123].

\bibitem{fornengo}
N.~Fornengo, A.~Riotto, S.~Scopel,
Phys.\ Rev.\ D {67}, 023514 (2003)
[hep-ph/0208072].


\bibitem{book} E. W. Kolb, M. S. Turner, {\it The Early
Universe},
(Addison-Wesley, Menlo Park, Ca., 1990).


\bibitem{ckr} D.~J.~Chung, E.~W.~Kolb, A.~Riotto,
Phys.\ Rev.\ D {60}, 063504 (1999)
[hep-ph/9809453].

\bibitem{infla} K. Kumekawa, T. Moroi, T. Yanagida,
Prog. Theor. Phys. 92, 437 (1994).


\bibitem{np} E. W. Kolb, A. D. Linde, A. Riotto, Phys.
Rev. Lett.
77, 4290 (1996); E. W. Kolb, A.  Riotto,  I. I. Tkachev,
Phys. Lett. B423, 348 (1998).

\bibitem{np2} G. F. Giudice, M. Peloso, A. Riotto
and I. I. Tkachev, JHEP 9908, 014 (1999) [hep-ph/9905242].

\bibitem{knr} E.~W.~Kolb, A.~Notari, A.~Riotto,
hep-ph/0307241, to be published in Phys. Rev. {D}.

\bibitem{Treh}
P.~H.~Chankowski, K.~Turzynski,
Phys.\ Lett.\ B {570}, 198 (2003)
[hep-ph/0306059].


\bibitem{ellis} 
J. R. Ellis, J. Kim,  D. V. Nanopoulos, 
Phys. Lett. B145, 181 (1984);
L. M. Krauss, 
Nucl. Phys. B227, 556 (1983);
M. Yu. Khlopov, A. D. Linde,
Phys. Lett. 138B, 265 (1984);
J. R. Ellis, D. V. Nanopoulos, K. A. Olive, S.-J. Rey,
Astropart. Phys. 4, 371 (1996);
M. Bolz, A. Brandenburg, W. Buchmuller,
Nucl. Phys. B 606, 518 (2001).

\bibitem{nucleo}
R. H. Cyburt, J. Ellis, B. D. Fields, K. A. Olive,
Phys. Rev. D 67, 103521 (2003).
For a review see
M.Yu. Khlopov, `Cosmoparticle physics', World Scientific, 1999.

\bibitem{sn1} 
H.~Murayama, H.~Suzuki, T.~Yanagida, J.~Yokoyama,
Phys.\ Rev.\ Lett.\  {70} (1993) 1912;
H.~Murayama, H.~Suzuki, T.~Yanagida, J.~Yokoyama,
Phys.\ Rev.\ D {50} (1994) 2356
[hep-ph/9311326].


\bibitem{sn2} 
H.~Murayama, T.~Yanagida,
Phys.\ Lett.\ B {322} (1994) 349
[hep-ph/9310297];
K.~Hamaguchi, H.~Murayama, T.~Yanagida,
Phys.\ Rev.\ D {65} (2002) 043512
[hep-ph/0109030];
T.~Moroi, H.~Murayama,
Phys.\ Lett.\ B {553} (2003) 126
[hep-ph/0211019].

\bibitem{ery}
J.~R.~Ellis, M.~Raidal, T.~Yanagida, hep-ph/0303242. 

\bibitem{pilaf1}
A.~Pilaftsis, Phys.\ Rev.\ D {56} (1997) 5431 [hep-ph/9707235];
T.~Hambye, Nucl.\ Phys.\ B {633} (2002) 171 [hep-ph/0111089];
J.~R.~Ellis, M.~Raidal, T.~Yanagida, Phys.\ Lett.\ B {546}, 228 (2002) [hep-ph/0206300].


\bibitem{DolgovZeldovich}
I.~Vysotsky, A.~D.~Dolgov, Y.~B.~Zeldovich,
Pisma Zh.\ Eksp.\ Teor.\ Fiz.\  {26}, 200 (1977).


\bibitem{Wolfram}
E.~W.~Kolb, S.~Wolfram,
Nucl.\ Phys.\ B {172}, 224 (1980)
[Erratum-ibid.\ B {195}, 542 (1982)].

\bibitem{Olive49}
J.~M.~Cline, K.~Kainulainen, K.~A.~Olive,
Phys.\ Rev.\ D {49}, 6394 (1994).

\bibitem{RGESM}
\art{C.~Ford, D.R.~Jones, P.W.~Stephenson, M.B.~Einhorn}
{\NP}{B395}{17}{1993};
\art{A.~Sirlin, R.~Zucchini}{\NP}{B266}{389}{1986};
\art{R.~Hempfling, B.A.~Kniehl}{\PR}{D51}{1386}{1995}.


\bibitem{SMRGE}
\art[hep-ph/9306333]{P.H.~Chankowski,
Z.~Pluciennik}{\PL}{B316}{312}{1993};
\art[hep-ph/9309223]{K.S.~Babu, C.N.~Leung,
J.~Pantaleone}{\PL}{B319}{191}{1993}.
An error has been corrected
in \art[hep-ph/0108005]{S.~Antusch, M.~Drees, J.~Kersten,
M.~Lindner, M.~Ratz}{\PL}{B519}{238}{2001}.


\bibitem{Gamma(T)}
Thermal corrections to decay rates have been employed
in \art{D.A. Dicus et al.}{\PR}{D26}{2694}{1982} and formalized in
H.~A.~Weldon, Phys.\ Rev.\ D {28}, 2007 (1983).


\bibitem{russi}
I.~F.~Ginzburg,
Nucl.\ Phys.\ Proc.\ Suppl.\  {51A} (1996) 85
[hep-ph/9601272];
K.~Melnikov, G.~L.~Kotkin, V.~G.~Serbo,
Phys.\ Rev.\ D {54} (1996) 3289
[hep-ph/9603352].

\bibitem{diagrammar}
Diagrammar,
by G. 't Hooft, M.J.G. Veltman,  CERN report 73-9 (1973). 



\end{thebibliography}
\end{document}